\documentclass[11pt]{article}
\usepackage{lmodern}
\usepackage{jheppub}
\usepackage{caption} 
\usepackage{hyperref}
\usepackage{amsmath,amssymb,amsthm,amscd}
\usepackage{mathrsfs}
\usepackage{mathtools}
\usepackage[all,cmtip]{xy}
\usepackage{xcolor}
\usepackage{blkarray}
\usepackage{float} 
\usepackage{youngtab}
\usepackage{ytableau}
\usepackage[shortlabels]{enumitem}
\Ylinethick{0.3pt}
\Yboxdim{5pt}
\usepackage{tikz,textcomp}
\usetikzlibrary{decorations.pathreplacing,calc,fadings,fit,shapes,arrows,positioning,chains,matrix}
\usepackage{textgreek}
\DeclareFontFamily{U}{wncy}{}
\DeclareFontShape{U}{wncy}{m}{n}{<->wncyr10}{}
\DeclareSymbolFont{mcy}{U}{wncy}{m}{n}
\DeclareMathSymbol{\Sh}{\mathord}{mcy}{"58} 
\usepackage{silence}
\numberwithin{equation}{section}

\usepackage{multirow}
\usepackage{booktabs}
\def\bea{\begin{eqnarray}}
\def\eea{\end{eqnarray}}
\def\be{\begin{equation}}
\def\ee{\end{equation}}
\def\ba{\begin{align}}
\def\ea{\end{align}}
\def\bse{\begin{subequations}}
\def\ese{\end{subequations}}

\newcommand{\sfy}{\mathsf{y}}

\newcommand{\eps}{\epsilon}

\newcommand{\sfq}{\mathsf{q}}
\newcommand{\sfp}{\mathsf{p}}
\newcommand{\cN}{\mathcal{N}}
\newcommand{\cV}{\mathcal{V}}
\newcommand{\cW}{\mathcal{W}}

\newcommand{\cO}{\mathcal{O}}

\newcommand{\bC}{\mathbb{C}}
\newcommand{\bR}{\mathbb{R}}
\newcommand{\bP}{\mathbb{P}}

\newcommand{\bZ}{\mathbb{Z}}

\newcommand{\Tr}{{\rm Tr}}

\newcommand{\stab}{\mathsf{Stab}}
\newcommand{\attr}{\mathsf{Attr}}
\newcommand{\kstab}{\mathsf{KStab}}
\newcommand{\CI}{\mathcal{I}}
\newcommand{\SL}{\mathscr{L}}
\usepackage[export]{adjustbox} 

\usepackage{longtable}
\usetikzlibrary{arrows.meta}

\usepackage{graphicx}
\usepackage{subcaption}

\title{Stable envelopes as BPS defects in gauged linear sigma models }
\abstract{
Stable envelopes are central objects in the geometric representation theory of symplectic resolutions: they provide chamber-dependent bases of equivariant cohomology and K-theory whose transition matrices are R-matrices of quantum groups. We realize the stable envelopes of the cotangent bundle of the complex Grassmannian manifold $X_{k,n}\equiv T^*{\rm Gr}(k,n)$ as BPS defects in the 2d $\mathcal{N} = (2,2)^*$ and 3d $\mathcal{N}=2^*$ gauged linear sigma models (GLSMs) with target space $X_{k,n}$. In the 2d case, these are point defects defined in terms of 0d $\mathcal{N}=2$ quiver gauge theory coupled to the bulk, and in 3d, they are line defects defined in terms of an $\mathcal{N}=2$ quiver quantum mechanics coupled to the bulk. The defects are constructed such that they restrict the target geometry to be the conormal bundle classes supporting these stable envelopes. In addition, when computing the partition function of the 0d-2d coupled system (or the 1d flavored Witten index of the 1d-3d coupled system), we recover the explicit formulas representing the (K-theoretic) stable envelope classes. Dimensional reduction relates the 3d and 2d constructions, moving from K-theory down to cohomology. Another limit is where we integrate out the cotangent fibers of $X_{k,n}$. We show that, in this case, our construction reduces to that of the Schubert defects in Gr$(k,n)$ of~\cite{Closset:2023bdr}. Using Gröbner-basis techniques from computational algebraic geometry, we compute A-model correlation functions of the defects, obtaining products of stable envelopes in the equivariant quantum cohomology and quantum K-theory of $X_{k,n}$. With this, we reproduce known results in the cohomology case while producing novel ones in the K-theory case.}

\author[a]{Peng Cheng,}
\affiliation[a]{Mainz Institute for Theoretical Physics, Johannes Gutenberg-Universit\"at,\\ Staudinger Weg 7, 55128 Mainz, Germany}
\emailAdd{pcheng@uni-mainz.de}

\author[b]{Zhihao Duan,}
\affiliation[b]{Shanghai Institute for Mathematics and Interdisciplinary Sciences (SIMIS),\\ 657 Songhu Road, Shanghai 200433, China}
\emailAdd{xduanz@gmail.com}

\author[c]{Osama Khlaif,}
\affiliation[c]{Philippe Meyer Institute, Physics Department, École Normale Supérieure (ENS), Université PSL,\\ 24 rue Lhomond, F-75231 Paris, France}
\emailAdd{osama.khlaif@phys.ens.fr}

\begin{document}

\maketitle
\raggedbottom

\section{Introduction and Summary}

The gauge theory/quantum cohomology correspondence has been a very fruitful area of research for both theoretical high-energy physics and mathematical physics~\cite{Bullimore:2014awa,Koroteev:2017nab,Jockers:2018sfl,Bullimore:2018jlp,Jockers:2019wjh,Ueda:2019qhg,Bullimore:2019qnt,Bullimore:2020nhv,Gu:2020zpg,Gu:2021yek,Jockers:2021omw,Dedushenko:2021mds,Gu:2021beo,Gu:2022yvj,Gu:2023tcv,Dedushenko:2023qjq,Closset:2023bdr,Gu:2023fpw,Closset:2023izb,Sharpe:2024ujm,Huq-Kuruvilla:2024tsg,AminiHuqKuruvillaMihalceaOrrXu2025,Huq-Kuruvilla:2025nlf,GuMihalceaSharpeXuZhangZou2025,Closset:2025akk,Closset:2026bnk,Duan:2026mfq,Cao:2026qbt,Brunner:2026fxy}. This is a relation between gauged linear sigma models (GLSMs)~\cite{Witten:1993yc} and the quantum cohomology theory of some geometry $X$ realized as the GLSM target space (Higgs branch). More explicitly, the correspondence states that the correlation functions of certain BPS operators in the GLSM correspond to Gromov--Witten (GW) invariants (or rather, quasimaps invariants) of $X$. Moreover, the ring of these operators is isomorphic to a quantum deformation of the standard cohomology theory of $X$. For the case where the GLSM is two-dimensional with spacetime geometry $\mathbb{R}^2$, one obtains the quantum (equivariant) cohomology ring $QH_T^\bullet(X)$, deforming the usual de Rham cohomology ring of $X$. In contrast, if we take 3d GLSM on $\mathbb{R}^2\times\mathbb{S}^1_{\boldsymbol\beta}$, one gets the quantum deformation of the K-theory ring of $X$; $QK_T(X)$. Here, the equivariance is with respect to the maximal torus of the isometry group of the target.

For this, the GLSM needs to have a certain amount of supersymmetry, depending on the geometry of the target $X$ of interest. In this work, we will consider theories with four supercharges. To define the BPS operators of interest, one needs to perform a topological A-twist along $\mathbb{R}^2$ and define the physical operators as cohomology classes of the twisting supercharge~\cite{Witten:1988xj,Witten:1991zz}. In the 2d $\mathcal{N}=(2,2)$ case, one gets the twisted chiral ring $\mathcal{R}_{\rm 2d}$ consisting of BPS point operators along the 2d spacetime. In the 3d uplift, we have a 3d $\mathcal{N}=2$ theory compactified on $\mathbb{R}^2\times\mathbb{S}^1_{\boldsymbol\beta}$. In this case, the twisted chiral ring $\mathcal{R}_{\rm 3d}$ consists of BPS line operators (e.g., Wilson loops) wrapping the $\mathbb{S}^1_{\boldsymbol\beta}$ fiber at some point along $\bR^2$. In terms of these rings, the correspondence can be written as the following isomorphisms:
\begin{equation}\label{corresp}
    \mathcal{R}_{\rm 2d}\,\cong\,QH_T^\bullet(X)~, \qquad \mathcal{R}_{\rm 3d}\,\cong\,QK_T(X)~.
\end{equation}

An ongoing program has recently been to construct point/line operators (or defects) in the GLSM that correspond to certain cohomology/K-theory classes of $X$. This has been carried out for cases where $X$ is a complex partial flag manifold, and the cohomology/K-theory classes are the Schubert classes. For instance, this was carried out in~\cite{Closset:2023bdr} for the case $X = {\rm Gr}(k,n)$ is the complex Grassmannian manifold, in~\cite{Gu:2025tda} for $X$ a Lagrangian Grassmannian LGr$(k,n)$, in~\cite{Closset:2025akk}, and in~\cite{Closset:2026bnk} for any partial flag manifold. In this work, we generalize this construction to the Nakajima quiver variety $X = X_{k,n}\equiv T^*{\rm Gr}(k,n)$, where a geometrically-natural basis in this case is given in terms of the (K-theoretical) stable envelopes studied in~\cite{MaulikOkounkov2019,Okounkov:2015spn, RimanyiTarasovVarchenko2014,rimanyi2015partial,rimanyi2015trigonometric}.

In 2d, the hyperKähler manifold $X_{k,n}$ is realized in terms of the 2d $\mathcal{N}=(4,4)$ GLSM with a $U(k)$ vector multiplet and $n$ fundamental hypermultiplets. To see the correspondence~\eqref{corresp}, the supersymmetry of the model can be softly broken down to four supercharges by turning on a $U(1)_\hbar$ symmetry group. This, in turn, gives masses proportional to $\hbar$ to some of the fields in both sets of multiplets parameterizing the cotangent directions of $X_{k,n}$. In this case, the equivariance of the cohomology theory is with respect to $T = A\times\bC_\hbar^\times$; the maximal torus of the $GL(n,\bC)\times \bC^\times_\hbar$ isometry group of $X_{k,n}$ (the complexification of the GLSM flavor symmetry $U(n)\times U(1)_\hbar$) with $A = (\bC^\times)^n$.

This construction can be uplifted to 3d $\mathcal{N}=4$ theory on $\mathbb{R}^3$ or $\mathbb{R}^2\times\mathbb{S}^1_{\boldsymbol\beta}$. In this case, upon softly breaking the symmetry, we get 3d $\mathcal{N}=2$ theory with massive fields. In this case, one needs to be careful about the choice of the Chern--Simons (CS) levels of the gauge group. Since this theory originates from $\mathcal{N}=4$, we take these levels to be zero. Other choices of the CS levels might give rise to topological and/or strongly coupled vacua alongside the Higgs branch. See~\cite{Intriligator:2013lca,Bullimore:2020nhv,Closset:2023jiq,Khlaif:2025ccg} for a recent discussion on this.

\medskip
\noindent
\textbf{(K-theoretic) Stable envelopes}
Given a Nakajima variety with its symplectic form $(X,\omega)$, choose a generic embedding of $\bC^\times \hookrightarrow  {\rm Aut}(X,\omega)$. The corresponding $\bC^\times$ will have a set of fixed loci:
\begin{equation*}
    \bigsqcup_i F_i \,\subset\, X~.
\end{equation*}
Each fixed locus $F_i$ comes with an attractor manifold:
\begin{equation*}
    {\rm Attr}_{F_i}\, =\, \{x\,\in\, X~|~ {\rm lim}_{t\to 0 \in \bC^{\times}}\,t\cdot x \,\in\, F_i\}~.
\end{equation*}
There is also a partial ordering:
\begin{equation*}
    F_i\,>\, F_j~,  \qquad{\ \ \rm if \ \ }  F_j \,\subset\, \overline{\rm Attr}_{F_i}~.
\end{equation*}
Stable envelope associated with fixed locus $F_i$ is defined as the dual of cycles given by a linear combination of $\{\overline{\rm Attr}_{F_j}| F_j\leq F_i\}$ which is well defined in the deformation of $X$. As the definition of stable envelopes depends on a choice of embedding  $\bC^\times \hookrightarrow  {\rm Aut}(X,\omega)$, they have a chamber-dependent behavior in ${\rm Lie}({\rm Aut}(X,\omega))$.  Their existence and uniqueness are proven in~\cite{MaulikOkounkov2019}. They also admit K-theoretic and elliptic generalization\cite{Okounkov:2015spn,aganagic2021ellipticstableenvelopes}.  In this paper, we focus on cohomological and K-theoretic stable envelopes for $X_{k,n}$ with fixed locus isolated points, although our proposal can be generalized to a broader class of Nakajima varieties. Stable envelopes have already appeared in the physics literature~\cite{Bullimore:2017lwu,Bullimore:2021rnr,Dedushenko:2021mds,Galakhov:2022uyu,Dedushenko:2023qjq,Ishtiaque:2023acr} and are being interpreted as Janus walls of the corresponding $3d$ $\cN=4$ gauge theory, as well as their dimensional reduction. In this paper, we propose a different realization of stable envelopes as BPS defects. 
 
\medskip
\noindent
\textbf{BPS defects corresponding to stable envelopes.}
In this work, we follow the construction algorithm of~\cite{Closset:2023bdr,Gu:2025tda,Closset:2025lqt,Closset:2026bnk} to propose a set of BPS point/line defects in 2d/3d GLSM of $X_{k,n}$ that correspond to the stable envelope classes in the equivariant cohomology/K-theory ring. The algorithm is as follows. Motivated by the geometry of these stable envelopes, we propose the defect as a lower-dimensional quiver gauge theory coupled to the GLSM in the UV via superpotential terms coupling the defect fields with those of the bulk and by taking some of the defect fields to be charged under the bulk symmetries. We provide two supporting pieces of evidence in the paper. First, we analyze the defect vacuum structure and argue that it realizes the conormal bundle class (i.e., the support of the stable envelope class). Second, we compute the partition function of the point defect (or the 1d flavored Witten index of the line defect) and show that it reproduces the stable envelope (Chern character of the K-theoretic stable envelope).  As we will see later in this paper, the defect realization naturally makes contact with Chern--Schwartz--MacPherson (CSM) classes and Motivic Chern (MC) classes and provides one way to study their quantum group aspects.

\medskip
\noindent
\textbf{From the Bethe/Gauge correspondence perspective.} Another motivation of our work is the Bethe/Gauge correspondence of Nekrasov and Shatashvili~\cite{Nekrasov:2009uh,Nekrasov:2014xaa}. In that work, they observed that the supersymmetric vacuum equations of the 2d/3d $X_{k,n}$ GLSM are of the same form as the Bethe ansatz equations (BAEs) governing the Bethe eigenstates of the quantum spin-$1/2$ $XXX$/$XXZ$ model. In terms of the twisted chiral ring, Nekrasov~\cite{Nekrasov2013} (see also~\cite{Bullimore:2017lwu}) constructed a set of `orbifold defects' that correspond to these Bethe states. But another, more natural basis for the spin chain model is the \textit{spin basis}. In~\cite{MaulikOkounkov2019}, Maulik and Okounkov established that, in the equivariant cohomology ring of $X_{k,n}$, this spin basis corresponds to the basis of cohomological stable envelopes. The BPS point/line defects that we construct in this paper correspond to the spin states in the corresponding spin chain model.

\begin{figure}
    \centering
    \begin{tikzpicture}[
    font=\large,
    >={Stealth[length=6pt,width=5pt]},
    every node/.style={inner sep=5pt},
    every path/.style={line width=0.9pt}
]

\node (Kcot) at (0,0)
    {$QK_T\!\left(T^*\mathrm{Gr}(k,n)\right)$};

\node (Kgr) at (8,0)
    {$QK_A\!\left(\mathrm{Gr}(k,n)\right)$};

\node (Hcot) at (0,-3)
    {$QH_T^\bullet\!\left(T^*\mathrm{Gr}(k,n)\right)$};

\node (Hgr) at (8,-3)
    {$QH_A^\bullet\!\left(\mathrm{Gr}(k,n)\right)$};

\draw[->] (Kcot.east) --
    node[above] {$\hbar\to\infty$} (Kgr.west);

\draw[->] (Hcot.east) --
    node[above] {$\hbar\to\infty$} (Hgr.west);

\draw[->] (Kcot.south) --
    node[left] {$\beta\to0$} (Hcot.north);

\draw[->] (Kgr.south) --
    node[right] {$\beta\to0$} (Hgr.north);

\end{tikzpicture}

    \caption{Commuting diagram of moving between quantum K-theory and quantum cohomology of $X_{k,n}$ and Gr$(k,n)$.}
    \label{fig:XknGrknLimits}
\end{figure}

\medskip
\noindent
\textbf{Connection with the base Grassmannian.} One interesting limit that we will apply in different places in this work is the $\hbar\rightarrow \infty$ limit. From the bulk perspective, in this limit, the charged fields become very massive and are integrated out. This is equivalent to `integrating out' the cotangent fibers of the $X_{k,n}$, leaving us with the compact base Gr$(k,n)$. At the level of the BPS defect, we will show that in this limit, some of the defect fields also get integrated out. The remaining defect coincides exactly with the one proposed in~\cite{Closset:2023bdr} for the Schubert classes in the Grassmannian manifold. Another way of stating this is that, in this limit, the stable envelopes of $X_{k,n}$ reduce to the Schubert classes (i.e., stable envelopes) of Gr$(k,n)$. This limit works both in the 2d and 3d theories and is compatible with the reduction limit between the two theories. This is summarized in Figure~\ref{fig:XknGrknLimits}.

\medskip
\noindent
\textbf{Extending the Gr\"obner algorithm.}
As already mentioned, the twisted chiral rings of the 2d and 3d GLSMs encode quantum
cohomology and quantum K-theory, respectively. Gauge-invariant
observables are represented by symmetric functions of the
Coulomb-branch variables, subject to the Bethe
ansatz equations. This turns the quantum products into classical multiplication in a quotient ring. We use computational algebraic geometry to carry out the computation efficiently.

In the meantime, we adapt the standard Gr\"obner basis method to account for the denominators appearing in the stable envelopes.
We first saturate the Bethe ideal with respect to a common
denominator of the product and the basis representatives.
Then we can remove the denominator and reduce the resulting polynomials to their Gr\"obner normal forms. Comparing with the normal forms of basis representatives then yields a finite linear system, which determines the linear combination in terms of basis. The same procedure applies to both quantum cohomological products and quantum K-products, with the respective Bethe ideal and stable envelopes as input.

\medskip
\noindent
\textbf{Plan of the paper.} The article is structured as follows. In Section~\ref{sec:2dGLSM}, we give a lightning review of the 2d GLSM realization of $X_{k,n} = T^*{\rm Gr}(k,n)$ as the Higgs branch. We also discuss the corresponding 2d A-model description, giving us the quantum cohomology ring (the quasimap theory) of $X_{k,n}$. We uplift this discussion to 3d GLSM and quantum K-theory in Section~\ref{sec:3dGLSM}. 

In Section~\ref{sec:stabs}, we review the axiomatic definition of the cohomological stable envelopes and the explicit form of their images in the equivariant cohomology ring of $X_{k,n}$. Moreover, we provide the K-theoretic version of these formulas, where now we have the equivariant Chern characters of the resulting classes in the equivariant K-theory ring of $X_{k,n}$. In Section~\ref{sec:BPSdefects}, we propose our BPS defect construction corresponding to the stable envelope classes in both the cohomology and the K-theory cases. We argue that these defects restrict $X_{k,n}$ to the conormal bundle that supports the stable envelopes. In addition, we show in detail how the formulas for these stable envelopes can be reproduced as the partition function (or the flavored index) of these defects.

Provided with the explicit form of these defects, in Section~\ref{sec:quantumProd} we extend the Gr\"obner basis algorithm to explicitly compute quantum products involving stable envelope classes. For the cohomological case, we compare our results with those in the literature.
As for the K-theoretic case, to the best of our knowledge, our results are novel.

In Appendix~\ref{append:TastP2}, we provide some explicit formulas for the quantum products in both cohomology and K-theory involving the stable envelopes in $T^*\bP^2$.

\begin{center}
    \large \textit{No part of this paper was written by AI}
\end{center}

\section{2d GLSM and quantum cohomology of \texorpdfstring{$X_{k,n}$}{Xkn}}\label{sec:2dGLSM}
In this section, we give a brief review of how to realize the cotangent bundle of the Grassmannian manifold in terms of 2d GLSMs. Moreover, we discuss the genus-$0$ quasimaps theory and how to compute intersection numbers on it in terms of the 2d A-model correlation functions.
\subsection{A gentle review of 2d GLSM to \texorpdfstring{$X_{k,n}$}{Xkn}}

\begin{figure}
    \centering
    \includegraphics[width=0.5\linewidth]{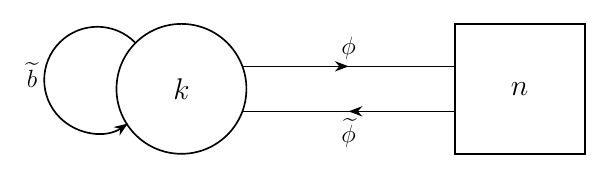}
    \caption{Quiver representation of the Nakajima variety $T^*{\rm Gr}(k,n)$. The circle node stands for the gauge group $U(k)$, and the square node stands for the flavor symmetry group $SU(n)$. Moreover, we have the pair $(\phi, \widetilde\phi)$ living in $(\square_k,\overline{\square}_k)$ representation, and $b$ living in adjoint representation of $U(k)$.} 
    \label{fig:cotGrkn}
\end{figure}

Let us start with a lightning review of the 2d GLSM realization of the symplectic variety $X_{k,n}\equiv T^*{\rm Gr}(k,n)$. As is well known~\cite{Nekrasov:2009uh}, this is realized as the Higgs branch of the 2d $\mathcal{N}=(2,2)^*$ gauge theory depicted in Figure~\ref{fig:cotGrkn}. We have a $G = U(k)$ gauge theory with $n$ fundamental chiral matter multiplets $\Phi_i = (\phi_i, \cdots)$, $n$ antifundamental chiral matter multiplets $\widetilde{\Phi}^i= (\widetilde{\phi}^i,\cdots)$, and an adjoint chiral multiplet $\widetilde{B} = (\widetilde{b}, \cdots)$, along with the superpotential:
\begin{equation}\label{Wbulk}
    W_{\rm bulk}\,=\,\Tr\,\left(\phi\,\widetilde{b}\,\widetilde{\phi}\right)~.
\end{equation}
The global symmetry group of the theory is $SU(n)\times U(1)_\hbar$
and the charges of these matter fields are summarized in Table~\ref{tab:charges}. Note that, for the $R$-charges, we are taking the fields $\phi, \widetilde{\phi}$ to be neutral; meanwhile, we take the $R$-charge of the adjoint to be $2$ so that the total $R$-charge of the superpotential~\eqref{Wbulk} is $+2$.

\begin{table}[t]
    \centering
    \begin{tabular}{|c||c|c|c|c|}
    \hline
        &$U(k)$ & $SU(n)$ & $U(1)_\hbar$ & $U(1)_R$\\
        \hline
\hline
$\phi$ & $\square$ & $\overline{\square}$ & $0$ & $0$\\ 
$\widetilde{\phi}$ & $\overline{\square}$ & $\square$ & $+1 $ & $0$\\
$\widetilde{b}$ & $\mathsf{Adj}$ & \textbf{1} & $-1$ & $2$\\
\hline
    \end{tabular}
    \caption{The charges of the matter fields appearing in the quiver in Figure~\ref{fig:cotGrkn} under the symmetry group $U(k)\times SU(n)\times U(1)_\hbar\times U(1)_R$.}
    \label{tab:charges}
\end{table}

\medskip
\noindent
\textbf{The Higgs branch as a symplectic quotient}
The Higgs branch $X_{k,n}$ of the theory is given as the following symplectic quotient:
\begin{equation}\label{Xkn}
    X_{k,n}\,=\,\{\mu_{\mathbb R}\,=\,r\,1_{k}~,\quad \mu_{\mathbb C}\,=\,0\}/G~,
\end{equation}
where the real and complex moment maps are given by:
\begin{equation}\label{muRmuC}
    \mu_{\mathbb R}\,\coloneqq\,\phi\cdot\,\phi^\dagger\,-\,\widetilde{\phi}^\dagger \,\cdot\,\widetilde{\phi}~, \qquad \mu_\mathbb{C}\,\coloneqq\,\phi\,\cdot\,\widetilde{\phi}~,
\end{equation}
and $G_\bC \equiv GL(k;\bC)$ is the complexification of the gauge group.
Moreover, the parameter $r$ is the real 2d FI parameter associated with $U(1)\subseteq U(k)$ and for the case $r>0$, the Higgs branch $X_{k,n}$ is none other than the cotangent bundle of the Grassmannian manifold Gr$(k,n)$ where $\phi_i$ span the base, and $\widetilde{\phi}^i$ span the cotangent fibers.\footnote{For the case $r<0$, the roles of $\phi$ and $\widetilde{\phi}$ get reversed.} 

\medskip
\noindent
\textbf{The Higgs branch as a Nakajima quiver variety.} Given $V\cong \bC^{k}, \,W\cong \bC^{n},\,G_\bC\cong GL(V,\bC)$. Take $ \phi \in {\sf Hom}(V,W),\widetilde\phi \in {\sf Hom}(W,V)$. Then, $X_{k,n}$ can be realized by the following GIT quotient:
\begin{equation}
    X_{k,n}\, =\,  \mu^{-1}(0)^{\rm stable}/G_\bC~,
\end{equation}
where
\begin{equation}
    \mu\,:\, {\sf Hom}(V,W)\oplus{\sf Hom}(W,V) \,\longrightarrow \,{\sf Hom(V,V)} ~\qquad \mu(\widetilde\phi,\phi) \,\coloneqq\, \widetilde \phi\circ \phi~,
\end{equation}
which is $\mu_\bC$ defined in~\eqref{muRmuC}. The stability condition is:
\begin{equation*}
    {\sf Ker} (\phi) \,=\, 0~, 
\end{equation*}
i.e., it is injective. This is equivalent to the $\mu_\bR = r>0$ condition in the symplectic quotient~\eqref{Xkn}.

At a point $(\phi, \widetilde \phi) \in X_{k,n}$, the tangent space $T_{(\phi, \widetilde \phi)}X_{k,n}$ is given by the virtual linear space of the following complex: 
\begin{equation}
    \label{complex chracterization of TGr}
    0 \to {\sf Hom}(V,V) \xrightarrow{\rm gauge\ trans} {\sf Hom}(V,W)\oplus{\sf Hom}(W,V) \xrightarrow{(\widetilde \phi\circ -)+(-\circ \phi ) }{\sf Hom}(V,V) \to 0~.
\end{equation}
Or, more explicitly, it is given by:
\begin{equation}
    \label{tangent space as virtual bundle}
    T_{(\phi, \widetilde \phi)} X_{k,n}\,=\, {\sf Hom}(V,W)\,\oplus\,{\sf Hom}(W,V)\,\ominus\,  {\sf Hom}(V,V)\, \ominus\,  {\sf Hom}(V,V)~.
\end{equation}
Taking the $\bC_{\hbar}^{\times}$ weights into account, we have:
\begin{equation}
    \label{refined tangent space as virtual bundle}
    T_{(\phi, \widetilde \phi)}X_{k,n} \,=\, {\sf Hom}(V,W)\,\oplus\, \hbar\,{\sf Hom}(W,V)\,\ominus \, {\sf Hom}(V,V) \,\ominus\,  \hbar\,{\sf Hom}(V,V)~.
\end{equation}
This will play an important role in the quasimaps theory of $X_{k,n}$, as we will come back to later.

\subsubsection{On the Coulomb branch and the BAEs} 

Let us now look at the Coulomb branch description of the theory. Recall that, for the gauge group, we have a 2d $\mathcal{N}=(2,2)$ vector multiplet $\mathcal{V}^{\rm 2d} = (\sigma, A, \cdots)$ with $\sigma$ being a complex scalar field and $A$ is the $U(k)$ gauge potential. All the components of this multiplet are in the adjoint representation of $U(k)$. On the Coulomb branch of the theory, the complex scalar field $\sigma$ gets diagonalized into the form $\sigma = {\rm diag}(\sigma_1, \cdots, \sigma_k)$. Additionally, associated with each component of the flavor symmetry group $SU(n)\times U(1)_\hbar$, we can turn on background vector multiplets with fixed VEVs of their corresponding complex scalars. In particular, from the complex scalar field associated with the $SU(n)$ vector multiplet, we get the twisted masses $m_1, \cdots, m_n$\footnote{Throughout this work, we will take the $SU(n)$ flavor symmetry group as $U(n)$ with the tracelessness condition. For instance, we will take the maximal torus to be $U(1)^n$ accompanied with the condition that the twisted masses satisfy the relation $\sum_{i=1}^n m_i = 0$.}, and, from that associated with the $U(1)_\hbar$ symmetry group, we get the twisted mass $\hbar$.

The effective twisted superpotential of the theory is then written as follows~\cite{Witten:1993yc,Morrison:1994fr,Nekrasov:2009uh,Closset:2015rna}:
\begin{equation}\label{effTwsited}
    \mathcal{W}^{\rm 2d}\,=\,\mathcal{W}^{\rm 2d}_\Phi\,+\,\mathcal{W}^{\rm 2d}_{\widetilde\Phi}\,+\,\mathcal{W}^{\rm 2d}_{\widetilde B}\,+\,\mathcal{W}^{\rm 2d}_{\rm FI}~,
\end{equation}
where:
\begin{equation}
    \begin{split}
        &\mathcal{W}^{\rm 2d}_\Phi\,\coloneqq\,-\,\frac{1}{2\pi i}\,\sum_{a=1}^k \sum_{i=1}^n\left(\sigma_a \,-\, m_i\right)\,\left(\log\left(\sigma_a\,-\,m_i\right)\,-\,1\right)~,\\
        &\mathcal{W}^{\rm 2d}_{\widetilde\Phi}\,\coloneqq\,\frac{1}{2\pi i}\,\sum_{a=1}^k \sum_{i=1}^n\left(\sigma_a \,-\, m_i\,-\,\hbar\right)\,\left(\log\left(-\,\sigma_a\,+\,m_i\,+\,\hbar\right)\,-\,1\right)~,\\
        &\mathcal{W}^{\rm 2d}_{\widetilde B}\,\coloneqq\,-\,\frac{1}{2\pi i}\,\sum_{\substack{a,b=1\\a\neq b}}^k(\sigma_a\,-\,\sigma_b\,-\,\hbar)\left(\log(\sigma_a\,-\,\sigma_b\,-\,\hbar)\,-\,1\right)~,\\
        &\mathcal{W}^{\rm 2d}_{\rm FI}\,\coloneqq\,\xi\,\sum_{a=1}^k\sigma_a~,
    \end{split}
\end{equation}
with $\xi\equiv ir+\frac{\theta}{2\pi}$ being the complexified 2d FI parameter associated with $U(1)\subseteq U(k)$. From this, we find the explicit form of the BAEs:
\begin{equation}\label{2dBAEs}
    \prod_{i=1}^n\frac{\sigma_a\,-\,m_i}{\sigma_a\,-\,m_i\,-\,\hbar}\,=\,(-1)^n\,\sfq\,\prod_{b\neq a}\,\frac{\sigma_b\,-\,\sigma_a\,-\,\hbar}{\sigma_a\,-\,\sigma_b\,-\,\hbar}~, \qquad a\,=\,1,\cdots, k~,
\end{equation}
with $\sfq\equiv e^{2\pi i \xi}$. These are obtained as the critical locus of the effective twisted superpotential~\eqref{effTwsited}, namely $\exp(2\pi i\partial\mathcal{W}^{\rm 2d}/\partial\sigma_a) = 1$.

\medskip
\noindent 
\textbf{Geometric interpretation: connecting back to $X_{k,n}$.}
The flavor symmetry group $SU(n)$ corresponds to the isometry group of $X_{k,n}$ acting both on the base and the fiber, and $U(1)_\hbar$ corresponds to the isometry group scaling the cotangent directions. The associated twisted masses correspond to the equivariant parameters of these isometries. As for the Coulomb branch parameters $\sigma_a$, they are interpreted as the Chern roots of the tautological vector bundle $\mathcal{S}\rightarrow {\rm Gr}(k,n)$~\cite{Witten:1993xi}. With this dictionary, the BAEs given in~\eqref{2dBAEs} are interpreted as the defining relations for the equivariant quantum cohomology ring of the Higgs branch:
\begin{equation}\label{QH=R2d}
    QH^\bullet_{T}(X_{k,n})\,\cong\,\mathcal{R}^{\rm 2d}\,\cong\,\frac{\mathbb{Z}(\sfq,m,\hbar)[\sigma_1,\cdots, \sigma_k]^{S_k}}{{\mathcal I}^{\rm 2d}_{\rm BAE}}~, 
\end{equation}
where $T = (\mathbb{C}^\times)^n\,\times\,\mathbb{C}_\hbar^\times$ is the complexification of the maximal torus of $SU(n)\times U(1)_\hbar$, the symmetric group $S_k$ is the residual gauge symmetry on the Coulomb branch and $\mathcal{I}^{\rm 2d}_{\rm BAE}$ is the ideal generated by the BAEs~\eqref{2dBAEs}. 

From this, one can obtain the quantum cohomology ring relations of the Grassmannian manifold Gr$(k,n)$ simply by taking the double scaling limit:
\begin{equation}
    \hbar\,\rightarrow\, \infty~, \quad \sfq\,\rightarrow\, 0~, \quad \sfq_\infty\,\equiv\, \hbar^n\,\sfq\,=\,{\rm finite} ~.
\end{equation}
Applying this limit to~\eqref{2dBAEs}, we end up with:
\begin{equation}
    \prod_{i=1}^n(\sigma_a\,-\,m_i)\,=\,\sfq_\infty~\qquad a\,=\,1\,,\,\cdots\,,\,k~,
\end{equation}
which are none other than those defining the $(\bC^\times)^n$-equivariant quantum cohomology ring relations of Gr$(k,n)$~\cite{Witten:1993xi}.

\subsection{2d A-model and quantum cohomology}
Let us now put our 2d theory defined in Figure~\ref{fig:cotGrkn} on $\mathbb{P}^1$ and perform a topological A-twist~\cite{Witten:1988xj,Witten:1991zz,Closset:2015rna}. The ring of half-BPS point operators (a.k.a. the 2d twisted chiral ring) $\mathcal{R}_{\rm 2d}$ is then isomorphic to the quantum cohomology ring of the Higgs branch $X_{k,n}$~\cite{Witten:1991zz,Witten:1993yc,Morrison:1994fr}. BPS point operators correspond to differential forms on $X_{k,n}$, and the fusion of these operators is translated into the quantum cup product between those differential forms. In particular, taking $\{\mathcal{O}_w\}_{w\in\mathrm{W}}$ to be a basis for $\mathcal{R}_{\rm 2d}$, then, we can write the fusion rules as:
\begin{equation}\label{Ow.Ow'}
    \mathcal{O}_w\,\cup_\sfq\,\mathcal{O}_{w'}\,=\,{\mathsf{C}_{w\,w'}}^v(\sfq,m)\,\mathcal{O}_v~,
\end{equation}
with repeated indices being summed over. The OPE coefficients are defined in terms of the 2- and 3-point correlation functions:
\begin{equation}
    {\mathsf{C}_{w\,w'}}^{v}(\sfq,m)\,\coloneqq\,\eta^{v\,v'}(\sfq,m)\,\mathsf{C}_{w\,w'\,v'}(\sfq,m)~, 
\end{equation}
where,
\begin{equation}
    \eta_{w\,w'}(\sfq,m)\,\coloneqq\,\langle\mathcal{O}_w\,\mathcal{O}_{w'}\rangle_{\mathbb{P}^1}~, \qquad  \mathsf{C}_{w\,w'\,w''}(\sfq,m)\,\coloneqq\,\langle\mathcal{O}_w\,\mathcal{O}_{w'}\,\mathcal{O}_{w''}\rangle_{\mathbb{P}^1}~,
\end{equation}
and, $\eta^{v\,v'}\eta_{v'\,w} = \delta_{w}^v$.

\subsubsection{Computing the correlation functions}
Upon topological twisting, the 2d theory becomes a 2d TQFT and the correlation functions of BPS operators can be worked out exactly using the following Vafa--Intriligator-type formula~\cite{Vafa:1990mu,Intriligator:1991nk,Nekrasov:2014xaa}:
\begin{equation}\label{<Ow...>}
\langle\mathcal{O}_w\,\cdots\rangle_{\mathbb{P}^1}\,=\,\sum_{\hat{\sigma}\in\mathcal{S}^{\rm 2d}_{\rm BE}}\,\mathcal{H}^{\rm 2d}(\hat{\sigma},m,\hbar)^{-1}\,\mathcal{O}_w(\hat{\sigma},m,\hbar)\,\cdots~,
\end{equation}
where $\mathcal{S}^{\rm 2d}_{\rm BE}$ are the solutions to the BAEs~\eqref{2dBAEs}:
\begin{equation}\label{2dSBE}
    \mathcal{S}^{\rm 2d}_{\rm BE}\,\coloneqq\,\{\hat{\sigma}~:~\exp(2\pi i\partial\mathcal{W}^{\rm 2d}/\partial\sigma)\mid_{\hat{\sigma}} \,=\, 1~, \quad \hat{\sigma}_a \,\neq\,\hat{\sigma}_b~,\forall a\neq b \}/S_k~.
\end{equation}
Moreover, the handle-gluing operator is defined as follows:
\begin{equation}
    \mathcal{H}^{\rm 2d}(\sigma, m, \hbar)\,=\,\exp(2\pi i \Omega^{\rm 2d})\,\det_{a,b}\left(2\pi i\frac{\partial^2\mathcal{W}^{\rm 2d}}{\partial\sigma_a\,\partial\sigma_b}\right)~.
\end{equation}
Here, the second term on the r.h.s. is the Hessian of the effective twisted superpotential given in~\eqref{effTwsited}. As for the first term, we have the dilaton potential $\Omega^{\rm 2d}$ which is given by:
\begin{equation}
    \Omega^{\rm 2d}\,=\,\Omega^{\rm 2d}_\Phi\,+\,\Omega^{\rm 2d}_{\widetilde\Phi}\,+\,\Omega^{\rm 2d}_{\widetilde B}\,+\,\Omega^{\rm 2d}_{\rm W}~,
\end{equation}
where, keeping in mind the $R$-charge assignments in Table~\ref{tab:charges}:
\begin{equation}
    \begin{split}
        &\Omega^{\rm 2d}_\Phi\,\coloneqq\,\frac{1}{2\pi i}\,\sum_{a=1}^k\sum_{i=1}^n\,\log(\sigma_a\,-\,m_i)~,\\
        &\Omega^{\rm 2d}_{\widetilde\Phi}\,\coloneqq\,\frac{1}{2\pi i}\,\sum_{a=1}^k\sum_{i=1}^n\,\log(-\sigma_a\,+\,m_i\,+\,\hbar)~,\\
        &\Omega^{\rm 2d}_{\widetilde B}\,\coloneqq\,-\,\frac{1}{2\pi i}\,\sum_{\substack{a,b=1}}^k\log(\sigma_a\,-\,\sigma_b\,-\,\hbar)~,\\
        &\Omega^{\rm 2d}_{\rm W}\,\coloneqq\,-\,\frac{1}{2\pi i}\,\sum_{\substack{a,b=1\\a\neq b}}^k\log(\sigma_a\,-\,\sigma_b)~.\\
    \end{split}
\end{equation}
The last contribution comes from the 2d W-bosons in the gauge vector multiplet. Hence, 
\begin{equation}
    \exp(2\pi i\Omega^{\rm 2d})\,=\,\frac{\prod_{a=1}^k\prod_{i=1}^n(\sigma_a\,-\,m_i)\,(-\sigma_a\,+\,m_i\,+\,\hbar)}{(-\hbar)^k\,\prod_{a\neq b}(\sigma_a\,-\,\sigma_b)\,(\sigma_a\,-\,\sigma_b\,-\,\hbar)}~.
\end{equation}

\medskip
\noindent
\textbf{Example: the 2d flavored twisted index with no insertion.} As an example of the above construction, let us compute explicitly the correlation function of the trivial point operator. Doing so for several choices of $k$ and $n$, we find that it takes the following form:
\begin{equation}\label{2d<1>Xkn}
    \left\langle1\right\rangle_{\mathbb{P}^1}\,=\,\,\sum_{\substack{I\subset\{1,\cdots,n\}\\|I|=k}}\,\prod_{\substack{i\in I\\j\notin I}}\,\frac{1}{(m_i\,-\,m_j)\,(m_j\,-\,m_i\,+\,\hbar)}~.
\end{equation}
The final expression coincides with the equivariant localization result of $\int_{X_{k,n}}1$. We will come back to this point momentarily. 

\medskip
\noindent
\textbf{Correlation functions as JK residues.} Another way of computing the correlation functions~\eqref{<Ow...>} comes from the Coulomb branch localization of the 2d GLSM in the UV (see~\cite{Closset:2015rna} and references therein for the details). These results were applied in~\cite {KimOhUedaYoshida2020} extensively for the Grassmannian manifold Gr$(k,n)$ case, and we will not discuss it here. Instead, in the next section, we will discuss its 3d uplift and its connection with K-theoretic quasimaps theory.   

\subsection{Quasimaps to \texorpdfstring{$X_{k,n}$}{Xkn} and quantum cohomology}

From an enumerative geometry perspective, the moduli space problem that appears from 2d GLSM is that of the moduli space of quasimaps.\footnote{These were first discussed in physics in~\cite{Witten:1993yc,Morrison:1994fr} where they used the terminology of small worldsheet instantons to refer to points on this moduli space.} Taking the 2d geometry to be $\Sigma_g$ (compact genus-$g$ Riemann surface), and $[\mu^{-1}(0)/G]$ to be the stack containing $X_{k,n} = \mu^{-1}(0)^{\rm stable}/G$, a quasimap is a map~\cite{Kim2016StableQuasimaps,CiocanFontanineKimMaulik2014}:
\begin{equation}
    f\,:\, \Sigma_g \,\longrightarrow\, [\mu^{-1}(0)/G]~,
\end{equation}
satisfying the following set of conditions:
\begin{itemize}
    \item The preimage of unstable locus $f^{-1}(\mu^{-1}(0)^{\rm unstable}/G)$ is a finite number of points on $\Sigma_g$. We call these points \textit{base points} of quasimap $f$.\footnote{This justifies the name quasimap; $f$ is not an honest map from $\Sigma_g$ to $X_{k,n}$, it is an honest map up to these base points.}
    \item $\Sigma_g$ is allowed to have nodal points. When $\Sigma_g$ has nodal points, then the base points are distinct from the nodal points.
    \item When $\Sigma_g$ has marked points $\{p_1,..,p_n\}$, then the marked points are different from both nodal points of $\Sigma_g$ and the defect points of the map $f$.
\end{itemize}

\subsubsection{Genus-\texorpdfstring{$0$}{0} quasimaps}
From now on, we set $\Sigma_g = \bP^1$ with one marked point\footnote{Using the terminology of~\cite{Okounkov:2015spn}, this point is a non-singular point.} as this is the most relevant case for quantum cohomology and quantum K-theory.\footnote{See~\cite{Bullimore:2018jlp} for some calculations in higher genus cases.} The quasimap moduli space naturally provides a hypergeometric series ${\rm Z} ^{k,n}$ valued in $H_T^\bullet(X_{k,n})[[\sfq]]$ ($K_T(X_{k,n})[[q]])$ in $T$-equivariant cohomology (K-theory). Instead of presenting details on calculations in quasimap theory, we only sketch the general ideas here. Some detailed calculations can be found in, e.g.~\cite{Bullimore:2018jlp,Dedushenko:2023qjq}. A generalization of this hypergeometric series valued in equivariant elliptic cohomology has been proposed in \cite{Brunner:2026fxy}.  

We will denote this quasimap moduli space by $\mathsf{QM}^{k,n}_{1}$. It admits a grading structure given by:
\begin{equation*}
    \int_{\bP^1}f^*c_{1}({\rm \mathcal{S}^{*}}) \in \bZ_{\geq 0}~,\qquad \forall f\in \mathsf{QM}^{k,n}_{1}~,
\end{equation*}
i.e.,
\begin{equation}\label{gradQM}
    \mathsf{QM}^{k,n}_{1} \,=\, \bigsqcup_{d\geq 0} \mathsf{QM}^{k,n}_{1,d}~.
\end{equation}
 Moreover, the virtual fundamental cycle (or the virtual structure sheaf in the K-theory case) can be defined on $\mathsf{QM}^{k,n}_{1,d}$~\cite{CiocanFontanineKimMaulik2014}. We denote it by $[\mathsf{QM}^{k,n}_{1,d}]^{\rm vir}$ ($\cO_{\mathsf{QM}^{k,n}_{1,d}}^{\rm vir}$). The non-singular marked point gives a map:
\begin{equation}\label{ev map}
    {\sf ev}_d\,:\, \mathsf{QM}^{k,n}_{1,d}\, \longrightarrow \,X_{k,n}~.
\end{equation}

\paragraph{Symmetries of $\mathsf{QM}^{k,n}_{1}$.} The symmetry group of $\mathsf{QM}^{k,n}_{1}$ comes from both the domain curve $\bP^1$ and the target space $X_{k,n}$. The symmetry group of $X_{k,n}$ is $T$ and the symmetry group of $\bP^1$ is $\bC_{{q_\epsilon}}^\times$ (i.e., the $\Omega$-background):
\begin{equation*}
    q_{\epsilon}\,\in \,\bC^\times_{q_{\epsilon}}\,:\, [z:w] \,\longrightarrow\, [q_{\epsilon}z\,:\,w]~.
\end{equation*}
 The $\widetilde{T}\equiv T\times \bC_{q_{\epsilon}}^\times$ fixed locus of $\mathsf{QM}^{k,n}_{1,d}$ are isolated fixed points, see~\cite[Section 4.3]{Brunner:2026fxy} for more details.

\paragraph{Why $\mathsf{QM}^{k,n}_{1}$?} The reason we focus on genus $0$ quasimap with one marked point $\mathsf{QM}^{k,n}$ is not by coincidence. For a general algebraic variety $X$, based on the reconstruction theorem~\cite{Givental2016}, the hypergeometric series based on $\mathsf{QM}_{0,1}(X)$ that we review below contains information of all genus $0$ Gromov--Witten (GW) invariants of $X$. As a result, it contains the GW invariants with three marked points, which are used to define quantum cohomology. Although $T^*{\rm Gr}(k,n)$, as a hyperk\"ahler manifold, is not ``general", its quasimap moduli space $\mathsf{QM}^{k,n}_{1}$ still plays an essential role in its enumerative geometry, see e.g.~\cite{Okounkov2020}.

The hypergeometric function mentioned above is defined as:
\begin{equation}\label{def of hypergeo}
    {\rm Z}^{k,n} \,\coloneqq\ \sum_{d\geq 0}q^d \,{\sf ev}_{d,*} ([\mathsf{QM}^{k,n}_{1,d}]^{\rm vir}) \,\in\, H_{T\times \bC_{q_{\epsilon}}^\times }^\bullet(X_{k,n})[[\sfq]] \,\cong\,H_{T}^\bullet(X_{k,n})({\epsilon})[[\sfq ]]~,
\end{equation}
in the cohomological case, and, 
\begin{equation}\label{def of K hypergeo}
    {\rm Z}^{k,n} \,\coloneqq\, \sum_{d\geq 0}q^d \,{\sf ev}_{d,*} (\cO_{\mathsf{QM}^{k,n}_{1,d}}^{\rm vir}) \,\in\, K_{T\times \bC_{q_{\epsilon}}^\times }(X_{k,n})[[q]] \,\cong\, K_{T}(X_{k,n}) (q_\eps)[[q]]~,
\end{equation} in K-theory.\footnote{In many cases it is better to replace $\cO_{\mathsf{QM}^{k,n}_{1,d}}^{\rm vir}$ by $\cO_{\mathsf{QM}^{k,n}_{1,d}}^{\rm vir}\otimes K_{\mathsf{QM}^{k,n}_{1,d}}^{1/2}$, where $K_{\mathsf{QM}^{k,n}_{1,d}}^{1/2}$ is the canonical line bundle.}

In both cohomology theories, the hypergeometric functions above can be calculated using equivariant localization. In fact, as we choose the marked point to be non-singular, these functions can only be well-defined with generic equivariant parameters turned on. More explicitly, cohomological equivariant localization gives:
\begin{align} \label{equi loc of hypergeo}
    & {\rm Z}^{k,n} \,=\, 
 \,\sum_{d\geq 0}\,\sum_{p\, \in \,(\mathsf{QM}^{k,n}_{1,d})^{\widetilde T}}\,\sfq^d\, \frac{1}{e(\mathcal{N}^{\rm vir}_p)}~,
 \end{align}
where $\mathcal{N}^{\rm vir}_p$ is the virtual normal bundle to the fixed point $p\in\mathsf{QM}^{k,n}_{1,d}$. The calculation is the same in K-theory, with $e(\mathcal{N}^{\rm vir}_p)$ replaced by the K-theoretic Euler class $e^K(\mathcal{N}^{\rm vir}_p) = \wedge^\bullet \mathcal{N}^{\rm vir,*}_p$.  

  The normal bundle $\mathcal{N}^{\rm vir}_p$ can be calculated using~\eqref{refined tangent space as virtual bundle}. The tangent space of a $T$-fixed point in $X_{k,n}$ can be described using~\eqref{refined tangent space as virtual bundle}. A point in $\mathsf{QM}^{k,n}_{1,d}$ describes a map from the domain curve $\bP^1$ to $X_{k,n}$, hence, the tangent space to this point in $\mathsf{QM}^{k,n}_{1,d}$ is described by the same complex~\eqref{refined tangent space as virtual bundle}, with the vector space $V$ and $W$ replaced by vector bundles $\cV$ and $\cW$ over the domain curve $\bP^1$.
    
   Those bundles will have non-trivial topological properties as vector bundles over $\bP^1$. Here we consider untwisted quasimaps where $\cW$ is a topologically trivial vector bundle. The topological properties of $\cV$ are tied with the grading structure of $\mathsf{QM}^{k,n}_{1}$~\eqref{gradQM}. In particular, in the $d$-sector, the first Chern class of $\cV$ satisfies:
  \begin{equation*}
      \int_{\bP^1}c_1(\cV) \,=\, d~.
  \end{equation*}

  With this at hand, the virtual bundle $\mathcal{N}^{\rm vir}_p$ is related to the following complex of sheaves:
  \begin{equation}  
    \label{complex chracterization of quasimap to TGr}
    0 \,\longrightarrow\, {\mathcal Hom}(\cV,\cV) \,\longrightarrow\, {\mathcal Hom}(\cV,\cW)\,\oplus\, {\mathcal Hom}(\cW,\cV)\, \longrightarrow\,{\mathcal Hom}(\cV,\cV)\, \longrightarrow\, 0~.
\end{equation}
  Taking the global section of~\eqref{complex chracterization of quasimap to TGr} gives:
  \begin{equation}
      0 \,\longrightarrow\, {\sf Ext^{*}}(\cV,\cV) \,\longrightarrow\, {\sf Ext^*}(\cV,\cW)\,\oplus\,{\sf Ext^*}(\cW,\cV) \longrightarrow \,{\sf Ext^*}(\cV,\cV)\,\longrightarrow\, 0~,
  \end{equation}
  where ${\sf Ext^{*}}(\cV,\cV) ={\sf Ext^{0}}(\cV,\cV)\ominus {\sf Ext^{1}}(\cV,\cV)$, etc. The virtual normal bundle $\mathcal{N}^{\rm vir}_p$ is given by:
  \begin{equation*}
      \mathcal{N}^{\rm vir}_p\,=\,{\sf Ext^*}(\cV,\cW)\,\oplus\,{\sf Ext^*}(\cW,\cV)\, \ominus\, {\sf Ext^{*}}(\cV,\cV) \,\ominus\, {\sf Ext^{*}}(\cV,\cV)~.
  \end{equation*}
  Substituting it into~\eqref{equi loc of hypergeo} gives the hypergeometric series Z$^{k,n}$ which plays an important role in quantum cohomology and quantum K-theory.

Quasimap is naturally preferred by GLSM with $4$ supercharges.  The field configuration naturally localizes to the BPS vortex configuration, with the first Chern class of the gauge bundle non-vanishing at the location of BPS vortex. As a result, each BPS vortex configuration gives a quasimap with the core as the base point. These vortices are called point-like instantons in~\cite{Witten:1991zz}.

\paragraph{Quantum cohomology based on quasimaps} The usual quantum cohomology is defined using stable maps, which do not necessarily coincide with the gauge theoretic chiral ring relations~\eqref{Ow.Ow'}. As gauge theory prefers quasimaps, it is natural to guess that the chiral ring relations~\eqref{Ow.Ow'} are provided by three-(relative) points quasimap invariants:
\begin{equation}
    \Omega_w\,\cup_\sfq\,\Omega_{w'}\,=\,{\mathsf{C}_{w\,w'}}^{w''}(\sfq, m,\hbar)\,\Omega_{w''}\,=\,\Omega_w\,\cup\,\Omega_{w'}\,+\,\sum_{d\geq 1} \sfq^d\,(\Omega_w\,\cup_d\,\Omega_{w'})~,
\end{equation}
with:
\begin{equation*}
 (\Omega_w\,\cup_d\,\Omega_{w'}) \,\coloneqq\, \sum_{{w''}\in {\rm W}}\left(\int_{[\mathsf{QM}_{3,d}^{k,n}]^{\rm vir}} {\sf ev}_1^{*}\Omega_w \,\cup\, {\sf ev}_2^*\Omega_{w'}\,\cup\, {\sf ev}_3^*\Omega_{w''}^{\vee}\right)\,  \Omega_{w''}~,
\end{equation*}
where $\Omega_{w''}^{\vee}$ is the dual of $\Omega_{w''}$ w.r.t. the Poincar\'e pairing on the cohomology ring. This ring structure could be different from the one defined by stable maps, see~\cite[Section 2]{CiocanFontanineKim2017} for some technical subtleties. Some relations between quasimap invariants and stable map invariants are given by the mirror map. In particular, for ${\rm Gr}(k,n)$, the chiral ring relations~\eqref{Ow.Ow'} coincide with the one defined by stable maps. In the next section, we use GLSM to illustrate why the mirror map is trivial in $\bP^n$ case. For $T^*{\rm Gr}(k,n)$ that we study in this paper,  the mirror map is also trivial. The reason is that supersymmetry in this case is enhanced to $\mathcal{N}=(4,4)$ and suppresses the instanton corrections. To have non-trivial chiral ring relations, we need to turn on $U(1)_{\hbar}$ background to softly break supersymmetry from ${\cal N}=(4,4)$ $T^*{\rm Gr}(k,n)$ GLSM to ${\cal N}=(2,2)$ ${\rm Gr}(k,n)$ GLSM, which receives non trivial instanton corrections but the mirror map is trivial. Mathematically, it corresponds to studying equivariant, or reduced GW invariants instead of the standard ones. It is natural, therefore, to expect that the chiral ring of $T^*{\rm Gr}(k,n)$ GLSM with  $U(1)_{\hbar}$ background is the same as the quantum cohomology defined by stable maps using equivariant (or reduced) GW theory.

\subsubsection{Stable maps vs. quasimaps}
Conventionally, quantum cohomology is formulated using the moduli space of stable curves. A stable map $f$ maps the entire domain curve into the target space, in contrast to quasimaps, which are well defined on the domain curve up to a finite number of base points. Both moduli spaces naturally appear in physics. As mentioned above, quasimap is closely related to $\mathcal{N}=(2,2)$ GLSM, and the base points are naturally identified with the location of the BPS vortices. Stable maps show up in $\mathcal{N}=(2,2)$ NLSM. In many cases, the UV GLSM flows to an NLSM in the IR, with the target space the classical Higgs branch of the UV theory. Hence, physically these two moduli spaces, as well as topological quantities on them, are related by renormalization group flow. Mathematically, the map provided by such an RG flow is called \textit{the mirror map} \footnote{Mathematically, it is a map between algebraic coordinates and flat coordinates of complexified K\"ahler moduli space associated with the classical Higgs branch. This map is important for counting curves using mirror symmetry. See~\cite{Candelas:1990rm,Chiang:1999tz,Bourget:2026hsg} for more details as well as its applications. } . Since the beta function of the 2d GLSM is proportional to the first Chern class of the target~\cite[Chapter 15]{Hori:MirrorSymmetry}, the details of the mirror map depend on whether the target is a Fano or a Calabi--Yau (CY).

 \medskip
 \noindent
  \textbf{Example I: ${\cal N} = (2,2)$ $\bP^{n-1}$ GLSM.} We have the following field contents:
   \begin{itemize}
       \item $U(1)$ vector multiplet $(A,\lambda_+,\lambda_-,\sigma)$.
       \item $n$ chiral multiplet $(\phi,\psi_{+},\psi_{-})_i$ with vector $U(1)_V$ and axial $U(1)_A$ charges $0$ and gauge charge $+1$.
   \end{itemize}
In this theory, $U(1)_A$ symmetry is broken due to the ABJ anomaly. This can be seen from the fermionic measure of the path integral:
\begin{equation*}
    \int d\psi_+d\bar{\psi}_+d\psi_-d\bar{\psi}_- \xrightarrow{\psi_{\pm}\to e^{\pm i\alpha/2}\psi_{\pm}}  \int e^{i(n_+-n_-)\alpha/2} d\psi_+d\bar{\psi}_+d\psi_-d\bar{\psi}_- = \int e^{in\alpha  } d\psi_+d\bar{\psi}_+d\psi_-d\bar{\psi}_- ~.
\end{equation*}
   However, this anomalous $U(1)_A$ symmetry can give a superselection rule by assigning a $U(1)_A$ charge to the background field. Note that there is a topological term $\theta \int f_{U(1)}$ in the action of the theory. Hence, the phase $ e^{in\alpha f_{U(1)}}$ from the path integral measure can be interpreted as $e^{i\alpha} \in U(1)_A$ acting on the background coefficient $\theta$.
\begin{equation*}
    \theta \xrightarrow{e^{i\alpha}}\theta +n\alpha.
\end{equation*}
As a result, the exponentiated FI term:
\begin{equation*}
    \sfq\,=\, \exp\left(2\pi i \left(i r \,+\,\frac{\theta}{2\pi}\right)\right)
\end{equation*}
carries a $U(1)_A$ charge $n$. Hence, the relevant symmetry of this theory is in Table~\ref{U(1)A}.
  
  \begin{table}[t]
    \centering
        \renewcommand{\arraystretch}{1.2}
    \begin{tabular}{|c||c|c|c|c|}
    \hline
        &$\sigma$ & $\psi_-$ & $\psi_+$ &$q_{UV}$\\
        \hline
        \hline
        $U(1)_A$&~$1$~&~$-1/2$~&~$+1/2$~&~$n$ \\
\hline
    \end{tabular}
    \caption{$U(1)_A$ charge}
    \label{U(1)A}
\end{table}
In the low energy limit with $r>0$, the theory flows to $\bP^{n-1}$ NLSM with $\sigma$ identified as the first Chern class of tautological bundle. Holomorphicity tells us that the quantum cohomology ring is of the following form\footnote{Recall quantum cohomology ring of $\bP^{n-1}$ is $\bC[\sigma][[\sfq_{\rm IR}]]/(\sigma^{n}-\sfq_{\rm IR})$.}:
      \begin{equation}\label{mirror map for Pn}
          \sigma^n  \, =\,\sfq_{\rm IR} \,=\, f(\sfq)~,
      \end{equation}
    where the mirror map $f(\sfq)=\sfq_{\rm IR}$ is a holomorphic function of the UV parameter $\sfq$. We can choose the following ansatz:
    \begin{equation}
        f(\sfq) \,=\, \sum_{i\geq 0}\,c_i\, \sfq^i~. 
    \end{equation}
  The superselection rule from $U(1)_A$ requires both sides of~\eqref{mirror map for Pn} to have the same $U(1)_A$ quantum number. $\sigma^n$ has $U(1)_A$ quantum number $n$ and $\sfq^i$ has $U(1)_A$ quantum number $ni$. Hence, up to a rescaling, the only possibility is\footnote{The case $c_1 =0$ can be excluded by a one-loop calculation.}:
  \begin{equation*}
      \sigma^n \,=\,\sfq~.
  \end{equation*}
We conclude that the mirror map for $\bP^{n-1}$ is trivial.

To summarize, the anomalous $U(1)_A$ symmetry leads to a nontrivial $U(1)_A$ quantum number for $\sfq$ and the superselection rule associated with $U(1)_A$ strongly constrains the mirror map. This also explains why the mirror map for generic CY cases is more complicated. For CY cases, $U(1)_A$ is not anomalous\footnote{A quick way to see this is by noticing CY NLSM leads to a  $\mathcal{N}=(2,2)$ SCFT. The superconformal algebra contains both left and right-moving $U(1)$ $R$ symmetries coming from the $U(1)_V$ and $U(1)_A$ symmetries in the UV GLSM.}. As a result, $\sfq$ has $U(1)_A$ quantum number $0$ so there is no $U(1)_A$ superselection rule.

\section{3d GLSM and quasimaps quantum K-theory}\label{sec:3dGLSM}
One can refine the discussion above by instead viewing the quiver theory in Figure~\ref{fig:cotGrkn} as a 3d $\mathcal{N}=2$ gauge theory on $\mathbb{P}^1\times\mathbb{S}^1_{\boldsymbol\beta}$ with a partial A-twist along the Riemann sphere. In this construction, correlation functions in the resulting 3d A-model compute (twisted) virtual holomorphic Euler characteristics of the quasimaps moduli space rather than intersection numbers as in the 2d case. In this subsection, we will review these ideas in more detail.

\medskip
\noindent
\textbf{A comment of the 3d CS terms.} Let us comment here on the choice of CS levels we will be following in this work. For the gauge group $U(k)_{\mathsf{k}, \mathsf{k}+{\sf l} k}$, there are two CS levels that one needs to fix. Since the theory we are interested in comes from softly breaking 3d $\mathcal{N}=4$ theory, these two CS levels are fixed to $({\sf k}, {\sf l}) = (0,0)$. Therefore, the effective CS levels that appear at the level of the effective twisted superpotential are of the form:
\begin{equation}\label{KandL}
    K\,\equiv\,k\,+\,n~, \qquad L\,\equiv\,-1~.
\end{equation}
For more details on this see~\cite[Subsection 2.2]{Closset:2023vos}. One can generalize the discussion below by considering other choices of ${\sf k}$ and ${\sf l}$. For more discussion on the geometric interpretation of these choices, see~\cite{Jockers:2019lwe,Jockers:2021omw,Closset:2023jiq,Closset:2023izb,Huq-Kuruvilla:2025nlf}.

\subsection{3d twisted chiral ring and correlation functions}
Physical operators in the 3d twisted theory are half-BPS line operators wrapping the circle fiber at a point along $\mathbb{P}^1$. The ring of these line operators, with the product operation being the fusion of parallel lines, is the 3d twisted chiral ring, which we denote by $\mathcal{R}_{\rm 3d}$. Taking $\{\SL_w\}_{w\in{\rm W}}$ to be a set of generators for $\mathcal{R}_{\rm 3d}$, one can write the fusion relations as follows:
\begin{equation}\label{Lw.Lw'}
    \SL_{w}\,\otimes_q\,\SL_{w'}\,=\,{\mathsf{N}_{w\,w'}}^{w''}(q,y,\mathsf{y})\,\SL_{w''}~.
\end{equation}
Here, $q\equiv e^{2\pi i \zeta}$ is the exponential of the 3d complexified FI parameter $\zeta$, $y_i$ stand for the 3d equivariant masses associated with the flavor symmetry $U(n)$. These are related to the 2d twisted masses via $y_i\equiv e^{\boldsymbol{\beta}m_i}$. Similarly, $\mathsf{y}\equiv e^{\boldsymbol{\beta}\hbar}$ is the 3d mass parameter associated with the $U(1)_\hbar$ symmetry group.

The explicit form for the OPE coefficients in~\eqref{Lw.Lw'} can be worked out using the 2 and 3-point functions of the line operators $\SL_w$:
\begin{equation}
    g_{w\,w'}(q,y,{\sf y})\,\coloneqq\,\left\langle\SL_w\,\SL_{w'}\right\rangle_{\mathbb{P}^1\times\mathbb{S}^1_{\boldsymbol \beta}}~, \qquad \mathsf{N}_{w\,w'\,w''}(q,y,{\sf y})\,\coloneqq\,\left\langle\SL_w\,\SL_{w'}\,\SL_{w''}\right\rangle_{\mathbb{P}^1\times\mathbb{S}^1_{\boldsymbol \beta}}~.
\end{equation}
In terms of these, we have:
\begin{equation}\label{quantum k from correlator}
    {\mathsf{N}_{w\,w'}}^{w''}(q,y,{\sf y})\,=\,g^{w''\,v}(q,y,{\sf y})\,\mathsf{N}_{w\,w'\,v}(q,y,{\sf y})~.
\end{equation}

\subsubsection{Correlation functions as sums over Bethe vacua}
In the 3d A-model, the correlation functions of half-BPS line operators can be computed simply as~\cite{Nekrasov:2009uh,Nekrasov:2014xaa,Benini:2015noa,Benini:2016hjo, Closset:2016arn,Closset:2017zgf}:\footnote{This can be viewed as the K-theoretic version of the Vafa--Intriligator-type formula~\eqref{<Ow...>}.}
\begin{equation}\label{<Lw...>}
\left\langle\SL\,\cdots\right\rangle_{\mathbb{P}^1\times\mathbb{S}^1_{\boldsymbol\beta}}\,=\,\sum_{\hat{x}\in\mathcal{S}_{\rm BE}^{\rm 3d}}\mathcal{H}(\hat{x},y,{\sf y})^{-1}\,\SL(\hat{x},y,{\sf y})\,\cdots~,
\end{equation}
where $x_a \equiv e^{ \boldsymbol{\beta}\sigma_a}$ are the 3d Coulomb branch parameters defined as the exponential of the 3d complexified scalar fields $\sigma_a$ living in the 3d gauge vector multiplet. The sum is over the Bethe vacua of the 3d theory defined as:\footnote{This formalism has been recently generalized beyond 3d theories with simply-connected/unitary gauge groups in~\cite{Closset:2024sle,Closset:2025lqt}.}
\begin{equation}\label{eq:SBE3d}
    \mathcal{S}_{\rm BE}^{\rm 3d}\,\coloneqq\,\left\{\hat{x} \,:\,\exp\left(\frac{\partial\mathcal{W}}{\partial\log x_a}\right)\mid_{\hat{x}} \,=\, 1~,\quad \hat{x}_a\neq\hat{x}_b~,~\forall a\neq b \right\}/S_k~.
\end{equation}

The Bethe vacua are defined in term of the 3d effective twisted superpotential $\mathcal{W}^{\rm 3d}$. This is given by:
\begin{equation}
    \mathcal{W}^{\rm 3d}(x,y,{\sf y},q)\,=\, \mathcal{W}^{\rm 3d}_{\Phi}\,+\, \mathcal{W}^{\rm 3d}_{\widetilde{\Phi}}\,+\, \mathcal{W}^{\rm 3d}_{\widetilde{B}}\,+\, \mathcal{W}^{\rm 3d}_{\rm FI}\,+\,\mathcal{W}_{\rm CS}^{\rm 3d}~,
\end{equation}
with the following explicit different contributions:
\begin{equation}
    \begin{split}
        & \mathcal{W}^{\rm 3d}_{\Phi}\,\coloneqq\,\sum_{a=1}^k\sum_{i=1}^n\,{\rm Li}_2(x_a\,y_i^{-1})~,\\
        & \mathcal{W}^{\rm 3d}_{\widetilde\Phi}\,\coloneqq\,\sum_{a=1}^{k}\sum_{i=1}^n\,{\rm Li}_2(x_a^{-1}\,y_i\,{\sf y})~,\\
        & \mathcal{W}^{\rm 3d}_{\widetilde B}\,\coloneqq\,\sum_{a,b=1}^k\,{\rm Li}_2(x_a\,x_b^{-1}\,{\sf y}^{-1})~,\\
        & \mathcal{W}^{\rm 3d}_{\rm FI}\,\coloneqq\,2\pi i\zeta\,\sum_{a=1}^k\log x_a~,\\
        &\mathcal{W}_{\rm CS}^{\rm 3d}\,\coloneqq\,\frac{K}{2}\,\sum_{a=1}^k \log x_a\,(\log x_a\,+\,2\pi i)\,+\,\frac{L}{2}\,\left(\sum_{a=1}^k\log x_a\right)\,\left(\sum_{a=1}^k\log x_a\,+\,2\pi i \right)~,
    \end{split}
\end{equation}
where the effective CS levels $K$ and $L$ are defined in~\eqref{KandL}. In terms of these, the BAEs have the following form~\cite[Equation (9)]{Pushkar:2016qvw},~\cite[Equation (2.47)]{Nekrasov:2009uh}:
\begin{equation}\label{3dBAEs}
\prod_{i=1}^n\,\frac{x_a\,-\,y_i}{x_a\,-\,{\sf y}\,y_i}\,=\,q\,\prod_{\substack{b=1\\b\neq a}}^k \frac{{\sf y}\,x_a\,-\,x_b}{x_a\,-\,{\sf y}\,x_b}~,\qquad a\,=\,1\,,\,\cdots\,,\,k~.
\end{equation}
Here, we used the fact that the symmetry group is $SU(n)$, thus, $\prod_{i=1}^ny_i=1$.

As for the 3d handle gluing operator appearing on the r.h.s. of~\eqref{<Lw...>}, it is given by:
\begin{equation}\label{3dH}
    \mathcal{H}^{\rm 3d}(x,y,{\sf y})\,=\,e^{2\pi i \Omega^{\rm 3d}}\,\det_{a,b}\left(\frac{\partial^2\mathcal{W}^{\rm 3d}}{\partial\log x_a\,\partial\log x_b}\right)~,
\end{equation}
with the dilaton potential of the form:\footnote{As part of defining the theory, we choose the bare $U(1)_R$ CS levels such that there is a constant term $\frac{k}{2}$ contributing to the dilaton potential.}
\begin{equation}
    \Omega^{\rm 3d}(x,y,{\sf y})\,=\,\Omega_\Phi^{\rm 3d}\,+\,\Omega_{\widetilde\Phi}^{\rm 3d}\,+\,\Omega_{\widetilde B}^{\rm 3d}\,+\,\Omega_{\rm W}^{\rm 3d}\,+\,\frac{k}{2}~, 
\end{equation}
where, 
\begin{equation}
\begin{split}
    & \Omega_\Phi^{\rm 3d}\,\coloneqq\,\frac{1}{2\pi i}\,\sum_{a=1}^k\sum_{i=1}^n\,\log\left(1\,-\,x_a\,y_i^{-1}\right)~,\\
    &\Omega_{\widetilde\Phi}^{\rm 3d}\,\coloneqq\,\frac{1}{2\pi i}\,\sum_{a=1}^k\sum_{i=1}^n\,\log\left(1\,-\,x_a^{-1}\,y_i\,{\sf y}\right)~,\\
    &\Omega_{\widetilde B}^{\rm 3d}\,\coloneqq\,-\,\frac{1}{2\pi i}\sum_{a,b=1}^{k}\,\log\left(1\,-\,x_a\,x_b^{-1}\,{\sf y}^{-1}\right)~,\\
    &\Omega_{\rm W}^{\rm 3d}\,\coloneqq\,-\,\frac{1}{2\pi i }\,\sum_{\substack{a,b=1\\a\neq b}}^k\,\log\left(1\,-\,x_a\,x_b^{-1}\right)~,
\end{split}
\end{equation}
where we recall the $R$-charge assignments for the different chiral multiplets as given in Table~\ref{tab:charges}.

\medskip
\noindent
\textbf{Example: the 3d twisted index.} As an application of the above construction, one can compute the correlation function~\eqref{<Lw...>} with the trivial insertion. In this case, we find:
\begin{equation}\label{<1>Xkn}
    \left\langle1\right\rangle_{\mathbb{P}^1\times\mathbb{S}^1_{\boldsymbol \beta}}\,=\,{\sf y}^{-k^2}\,\sum_{\substack{I\subset\{1,\cdots,n\}\\|I|=k}}\,\prod_{\substack{i\in I\\j\notin I}}\,\frac{1}{(1\,-\,y_{i}\,y_j^{-1})\,(1\,-\,{\sf y}\,y_i^{-1}\,y_j)}~.
\end{equation}
Note that there is no dependence on the FI parameter $q$ in the final expression. The final expression on the r.h.s. is related to the $T$-equivariant holomorphic Euler characteristic of the Higgs branch $X_{k,n}$:
\begin{equation}
    \left\langle1\right\rangle_{\mathbb{P}^1\times\mathbb{S}^1_{\boldsymbol\beta}}\,=\,{\sf y}^{-k^2}\,\chi_{T}(X_{k,n}\,,\,\mathcal{O}_{X_{k,n}})~.
\end{equation}
In the small radius limit (i.e. $\boldsymbol{\beta}\rightarrow 0$), the expression on the r.h.s. reduces to that in~\eqref{2d<1>Xkn}.

\subsubsection{Correlation functions as JK residues}
Another way of computing the correlation functions of the half-BPS line operators~\eqref{<Lw...>} comes from Coulomb branch localization~\cite{Benini:2015noa,Benini:2016hjo,Closset:2016arn,Closset:2017zgf,Bullimore:2018jlp}. In this case, the path integral reduces to the following JK residue:
\begin{equation}
    \left\langle \SL_w\,\cdots \right\rangle_{\mathbb{P}^1\times\mathbb{S}^1_{\boldsymbol \beta}}\,=\,\sum_{d\geq 0}\,q^d\,\left\langle \SL_w\,\cdots\right\rangle_{\mathbb{P}^1\times\mathbb{S}^1_{\boldsymbol\beta}}^d~,
\end{equation}
where the sum is over the magnetic fluxes, i.e., the first Chern class of the principal $U(k)$ gauge bundle, and the contribution of each flux sector is given by:
\begin{equation}\label{<Lw...>d}
    \left\langle\SL_w\,\cdots\right\rangle_{\mathbb{P}^1\times\mathbb{S}^1_{\boldsymbol\beta}}^d\,=\,(-1)^{(k+n-1)d}\,\sum_{\substack{\boldsymbol{d}\in\mathbb{Z}_{\geq 0}^k\\|\boldsymbol{d}|=d}}\,\oint_{\rm JK}\,\left(\prod_{a=1}^k\frac{-\,dx_a}{2\pi i \,x_a}\right)\,\frac{1}{k!}\,Z_{\boldsymbol d}(x,y,{\sf y})\,\SL_w(x,y,{\sf y})\,\cdots~.
\end{equation}
The sum on the r.h.s. is over flux vectors of the form $\boldsymbol{d} = (d_1, \cdots, d_k)$, and: 
\begin{equation}
    Z_{\boldsymbol d}(x,y,{\sf y})\,=\,\,Z^{\rm CS}_{\boldsymbol d}\,Z_{\boldsymbol d}^{\rm W}\,Z_{\boldsymbol{d}}^{\widetilde B}\,Z_{\boldsymbol d}^{\Phi}\,Z_{\boldsymbol d}^{\widetilde\Phi}~, 
\end{equation}
with: 
\begin{equation}
    \begin{split}
        &Z_{\boldsymbol d}^{\rm CS}\,\coloneqq\,(-1)^k\,\prod_{a=1}^k\,x_a^{(k+n)d_a-d}~, \\
        &Z_{\boldsymbol d}^{\rm W}\,\coloneqq\,\prod_{\substack{a,b=1\\a\neq b}}^k\,\left(1\,-\,\frac{x_a}{x_b}\right)~,\\
        &Z_{\boldsymbol d}^{\widetilde B}\,\coloneqq\,(1\,-\,{\sf y}^{-1})^k\,\prod_{\substack{a,b=1\\a\neq b}}^k\,\left(1\,-\,\frac{x_a}{{x_b}\,{\sf y}}\right)^{1-d_a+d_b}~,\\
        &Z_{\boldsymbol d}^{\Phi}\,\coloneqq\,\prod_{a=1}^k\,\prod_{i=1}^n\,\left(1\,-\,\frac{x_a}{y_i}\right)^{-d_a-1}~, \\
        &Z_{\boldsymbol d}^{\widetilde \Phi}\,\coloneqq\,\prod_{a=1}^k\,\prod_{i=1}^n\,\left(1\,-\,\frac{{\sf y}\,y_i}{x_a}\right)^{d_a-1}~.
    \end{split}
\end{equation}

To perform the JK residue above, we follow the analysis done in~\cite[Appendix A]{Closset:2023bdr}. For a positive 3d real FI parameter, we pick the poles coming from the chiral fields spanning the compact core of the Higgs branch $X_{k,n}$. Doing so,  we find that the degree-$d$ contribution to the correlation function is of the form:
\begin{equation}
    \left\langle\SL_w\,\cdots\,\right\rangle_{\mathbb{P}^1\times\mathbb{S}^1_{\boldsymbol\beta}}^d\,=\,(-1)^{(k+n-1)d-k}\,\sum_{\substack{\boldsymbol{d}\in\mathbb{Z}_{\geq 0}^k\\|\boldsymbol{d}|=d}}\,\sum_{\substack{I\subset \{1,\cdots,n\}\\|I|=k}}\,\operatorname*{\rm Res}_{x_a=y_{i_a}}\left[\frac{Z_{\boldsymbol{d}}(x,y,{\sf y})\,\SL_w(x,y,{\sf y})\,\cdots}{\det x}\right]~.
\end{equation}
Here, the second sum is over subsets of the form $I = \{i_1, \cdots, i_k\}\subset\{1,\cdots,n\}$.

\medskip
\noindent
\textbf{Example: the 3d twisted index with no insertions.}
As an explicit application for this, let us look at the case with trivial insertion. For the degree-$0$ term, we have:
\begin{equation}
    \left\langle 1\right\rangle_{\mathbb{P}^1\times\mathbb{S}^1_{\boldsymbol \beta}}^0\,=\,\sum_{\substack{I\subset\{1,\cdots,n\}\\|I|=k}}\,\operatorname*{\rm Res}_{x_a=y_{i_a}}\left[\frac{(1\,-\,{\sf y}^{-1})^k\,\prod_{a\neq b}\left(1\,-\,\frac{x_a}{x_b}\right)\,\left(1\,-\,\frac{x_a}{x_b\,{\sf y}}\right)}{\det x\,\prod_{a=1}^k\prod_{i=1}^n\left(1\,-\,\frac{x_a}{y_i}\right)\left(1\,-\,\frac{{\sf y}\,y_i}{x_a}\right)}\right]~.
\end{equation}
The poles in this case are simple, and it is straightforward to perform the residue. We end up with:
\begin{equation}
    \left\langle 1\right\rangle_{\mathbb{P}^1\times\mathbb{S}^1_{\boldsymbol \beta}}^0\,=\,{\sf y}^{-k^2}\,\sum_{\substack{I\subset\{1,\cdots,n\}\\|I|=k}}\,\frac{1}{\prod_{\substack{i\in I\\j\notin I}}\left(1\,-\,\frac{y_i}{y_j}\right)\,\left(1\,-\,\frac{y_j\,{\sf y}}{y_i}\right)}~,
\end{equation}
Moreover, for $d\geq 1$, we find that the final form of the residue calculation is trivial. In total, the JK residue calculation is in agreement with our earlier result~\eqref{<1>Xkn}.

\subsection{Quasimaps quantum K-theory of \texorpdfstring{$X_{k,n}$}{Xkn}}
The BPS line operators $\SL_w$ in $\mathcal{R}_{\rm 3d}$ are in one-to-one correspondence with vector bundles (or coherent sheaves in general) $\mathcal{E}_w\in K(X_{k,n})$. Moreover, correlation functions of these line operators are interpreted as twisted virtual holomorphic Euler characteristics of the genus-0 quasimaps moduli space $\mathsf{QM}^{k,n}_{m,d}$ with $m$ being the number of insertions and $d$ the degree of the map. For this, one uses the evaluation maps at the points of insertion to pull back the sheaf $\mathcal{E}_w$ to the quasimaps moduli space. More explicitly:\footnote{Here we want to address the different appearance of $\SL_w$ in the quasimap calculation. It can be understood as a pullback of the K-theoretic class of stack $[V/G]$, i.e., from $f: \mathsf{QM}^{k,n}_{m,d} \to [V/G] $. As a result, $\SL_w$ is a K-theoretic class of the stack $[V/G]$ and the inserted point of $\SL_w$ is not a marked point, neither relative nor non-singular. In~\eqref{K-theoretic correlator}, we choose $\SL_w\in K_T(T^*{\rm Gr}(k,n))$. The corresponding inserted point is a marked point and ${\sf ev}$ is the associated evaluation map. Moreover, as $\bP^1$ is compact,  the marked points are relative. One hint is that~\eqref{K-theoretic correlator} is well defined without turning on the Omega background $\bP^1_\eps \times {\mathbb S}^1_{\boldsymbol\beta}$. BPS ring structure from~\eqref{quantum k from correlator},~\eqref{K-theoretic correlator} should be closely related to the quasimap quantum K-theory proposed in~\cite{Pushkar:2016qvw}.}
\begin{equation}\label{K-theoretic correlator}
\left\langle\SL_w\,\cdots\right\rangle_{\mathbb{P}^1\times\mathbb{S}^1_{\boldsymbol\beta}}\,=\,\sum_{d\geq 0}q^d\,\chi_{T}\left({{\sf QM}^{k,n}_{m,d}}\,,\,\widehat{\mathcal{O}}^{\rm vir}_{{\sf QM}^{k,n}_{m,d}}\,\otimes\,{\sf ev}^{*}[\mathcal{E}_w]\,\otimes\,\cdots\right)~,
\end{equation}
where $T = (\mathbb{C}^\times)^n\times \mathbb{C}_{\hbar}^\times$ is the maximal torus of the isometry group of $X_{k,n}$ uplifted to the quasimaps moduli space. Moreover, 
\begin{equation}
    \widehat{\cal O}^{\rm vir}_{{\sf QM}^{k,n}_{m,d}}\,\coloneqq\,  {\cal O}^{\rm vir}_{{\sf QM}^{k,n}_{m,d}}\,\otimes\,K^{\frac{1}{2}}_{{\sf QM}^{k,n}_{m,d}}~.
\end{equation}

In the earlier work~\cite{Bullimore:2018jlp}, the A-twist we are following here corresponds to what they refer to as the C-twist.\footnote{This follows from our choice of the $R$-charges in Table~\ref{tab:charges}. Instead, one can take these charges to be $(0,0,2)$ for the fields $(\Phi,\widetilde{\Phi},\widetilde{B})$, which corresponds to the H-twist in~\cite{Bullimore:2018jlp}. In this case, the 3d twisted index with no insertions computes the virtual $\chi_{{\sf y}}$-genus of the moduli space of \textit{twisted} quasimaps studied in~\cite{Kim2016StableQuasimaps}.} In the limit ${\sf y}\rightarrow 1$, the chiral multiplets $\widetilde{B}$ and $\widetilde{\Phi}_i$ get `integrated in' enhancing the supersymmetry of the theory to $\mathcal{N}=4$. The topological twist in this case becomes Rozansky--Witten~\cite{Rozansky:1996bq,GukovPutrovVafa2017}, and the 3d twisted index computes the holomorphic Euler characteristic of the Higgs branch:
\begin{equation}
    \lim_{{\sf y}\rightarrow 1}\,\left\langle 1\right\rangle_{\mathbb{P}^1\times\mathbb{S}^1_{\boldsymbol\beta}}\,=\,\lim_{{\sf y}\rightarrow 1}\,\sum_{d\geq 0} q^d\,\chi_{T}\left({\sf QM}^{k,n}_{0,d}\,,\,\widehat{\mathcal{O}}^{\rm vir}_{{\sf QM}^{k,n}_{0,d}}\right)\,=\,\chi_{A}(X_{k,n}, \widehat{\mathcal{O}}_{X_{k,n}})~,
\end{equation}
where $A = (\mathbb{C}^\times)^n$. That is, in this limit, we receive contributions only from the degree-$0$ sector, which consists of constant maps into the Higgs branch. This is consistent with our earlier result in~\eqref{<1>Xkn}. More generally, in the presence of the half-BPS line operators in the correlation function, we have:
\begin{align}
\lim_{{\sf y}\rightarrow 1}\,\left\langle\SL_w\,\cdots\right\rangle_{\mathbb{P}^1\times\mathbb{S}^1_{\boldsymbol\beta}}\,=\,\chi_A(X_{k,n},\widehat{\mathcal{O}}_{X_{k,n}}\,\otimes\,\mathcal{E}_w\,\otimes\,\cdots)~.
\end{align}

\medskip
\noindent
\textbf{Further comments on the $m=0$ moduli space.}
Let us examine the case with zero insertions in more detail. The $T$-fixed locus of the quasimaps moduli space in this case is well known, and it is of the following form~\cite[Subsection 3.2]{MarianOprea2007},~\cite[Equation (5.99)]{Bullimore:2018jlp}:\footnote{When comparing with~\cite{Bullimore:2018jlp}, we take $r= 0$, $g=0$, and use the fact that Sym$^d(\mathbb{P}^1) = \mathbb{P}^d$.}
\begin{equation}\label{QMTfixed}
    ({\sf QM}^{k,n}_{0,d})^T\,=\,\bigsqcup_{\substack{I\subset\{1, \cdots, n\}\\|I|=k}}\,\bigsqcup_{\substack{d_1,\cdots,d_k\geq 0\\d_1+\cdots+d_k=d}}\,\prod_{i
    \in I}\mathbb{P}^{d_i}~.
\end{equation}
From this expression, one can explicitly compute the generating function of the Euler numbers of the quasimaps moduli spaces. Recall that, the Euler number of $\mathbb{P}^m$ is  $m+1$, therefore:
\begin{equation}
    e({{\sf QM}^{k,n}_{0,d}})\,=\,\binom{n}{k}\,\sum_{\substack{d_1,\cdots\,d_k\geq 0\\d_1+\cdots+d_k=d}}\,\prod_{a=1}^k(d_a+1)\,=\,\binom{n}{k}\,\binom{d+2k-1}{2k-1}~.
\end{equation}
Taking the sum over all non-negative degrees $d$, we get:
\begin{align}\label{sum_qe(QM)}
    \sum_{d\geq 0} q^d\,e({{\sf QM}^{k,n}_{0,d}})\,=\,\binom{n}{k}\,\sum_{d\geq 0} \binom{d+2k-1}{2k-1} \,q^d\,=\,\binom{n}{k}\,(1\,-\,q)^{-2k}~.
\end{align}
For instance, we see that the coefficient of $q^0$ is $\binom{n}{k}$ in agreement with our earlier statement that in this case, the quasimaps moduli space coincides with the Higgs branch $X_{k,n}$.

\medskip
\noindent
\textbf{Quasimaps quantum K-theory.} This leads us to the 3d gauge theory/quantum K-theory correspondence for the example we are studying here. The higher-degree terms appearing above define a quantum deformation of the classical tensor product of the K-theory ring of $X_{k,n}$. For two K-theory classes $[\mathcal{E}_w]$ and $[\mathcal{E}_{w'}]$, we have:
\begin{equation}
    [\mathcal{E}_{w}]\,\otimes_q\,[\mathcal{E}_{w'}]\,=\,{\mathsf{N}_{w\,w'}}^{w''}(y,{\sf y},q)\,[\mathcal{E}_{w''}]\,=\,[\mathcal{E}_w]\,\otimes\,[\mathcal{E}_{w'}]\,+\,\sum_{d\geq 1}\,q^d\,[\mathcal{E}_w]\,\otimes_d\,[\mathcal{E}_{w'}]~,
\end{equation}
where $\otimes_d$ encodes the contribution of the twisted virtual holomorphic Euler characteristics:
\begin{align}
   \chi_T\left({\sf QM}^{k,n}_{3,d}\,,\,\widehat{\mathcal{O}}^{\rm vir}_{{\sf QM}^{k,n}_{3,d}}\,\otimes\,{\sf ev}_1^*[\mathcal{E}_w]\,\otimes\,{\sf ev}_2^*[\mathcal{E}_{w'}]\,\otimes\,{\sf ev}_3^*[\mathcal{E}_{w''}]\right)~.
\end{align}

The K-theory ring of $X_{k,n}$ equipped with the deformed tensor product given above is usually referred to as the $T$-equivariant quasimaps quantum K-theory ring which we will denote here by QK$_T^{\rm qm}(X_{k,n})$ and it is isomorphic to the 3d twisted chiral $\mathcal{R}^{\rm 3d}$:
\begin{equation}\label{QK=R3d}
    {\rm QK}_T^{\rm qm}(X_{k,n})\,\cong\,\mathcal{R}^{\rm 3d}\,\cong\,\frac{\bZ(y,{\sf y})[x_1, \cdots, x_k]^{S_k}[[q]]}{\mathcal{I}^{\rm 3d}_{\rm BAE}}~,
\end{equation}
with the symmetric group $S_k$ being the residual gauge symmetry on the Coulomb branch and $\mathcal{I}_{\rm BAE}^{\rm 3d}$ is the ideal generated by the BAEs~\eqref{3dBAEs}.

\medskip
\noindent
\textbf{Connection with the K-theoretic GW theory.} Quantum K-theory based on stable maps has been developed~\cite{Givental2017, Lee2004}. What plays an important role there is determining the `input' from the hypergeometric series~\eqref{def of K hypergeo} derived by quasimap calculations. Relations between quantum K-theory based on quasimap and stable maps are under current investigation. Some progress can be found in~\cite{Liu2019, GiventalYan2021}. But more related to our work, for targets being the cotangent bundles of partial flag manifolds, it was conjectured in~\cite[Conjecture 13.15]{rimanyi2015trigonometric} that the corresponding quasimaps quantum K-theory that we get from 3d GLSM coincides with that defined in terms of stable maps theory.

\medskip
\noindent
\textbf{Another limit for ${\sf y}$.} Above, we discussed one important limit for the fugacity $\mathsf{y}$ where the supersymmetry of the theory gets enhanced to $\mathcal{N}=4$. Another important limit is where we take $\mathsf{y}\rightarrow \infty$. This amounts to integrating out the massive adjoint multiplet $\widetilde{B}$ and the antifundamental chirals $\widetilde{\Phi}_i$. Geometrically, this limit amounts to integrating out the cotangent fibers of the Higgs branch, and we end up with Gr$(k,n)$. At the level of the BAEs~\eqref{3dBAEs}, we get:
\begin{equation}
    \prod_{i=1}^n\,\left(1\,-\,x_a\,y_i^{-1}\right)\,=\,(-1)^{k-1}\,q_{\infty}\,x_a^k\,(\det x)^{-1}~,
\end{equation}
with $q_\infty\equiv q\,{\sf y}^n$. These are none other than the defining relations of the quantum K-theory ring of the Grassmannian manifold~\cite{Ueda:2019qhg,Jockers:2019lwe,Gu:2022yvj,Closset:2023bdr}.

\section{Stable envelopes in \texorpdfstring{$X_{k,n}$}{Xkn}}\label{sec:stabs}
In this section, we review the axiomatic definition of the cohomological stable envelopes in $X_{k,n}$ following~\cite{MaulikOkounkov2019}. As part of the definition, one needs to fix a chamber in the Lie algebra of the torus $A = (\mathbb{C}^\times)^n$ acting on $X_{k,n}$. After reviewing the definition of these chambers and the attractor mechanism, we provide arguments on how one can work out the explicit form of these stable envelopes using the abelian/non-abelian correspondence. Moreover, we give a brief review of how the concept of $\mathsf{R}$-matrix enters when moving across the chambers.

In the second part of the section, we discuss the K-theoretic uplift of these concepts and give the final formula for the K-theoretic stable envelopes. Finally, we revisit the product formulas in both quantum cohomology and quantum K-theory rings of $X_{k,n}$ involving the stable envelopes.

\medskip
\noindent
\textbf{The torus-fixed locus of $X_{k,n}$.}
Recall from our earlier discussion that the cotangent bundle of the Grassmannian manifold $X_{k,n}$ has an isometry group whose maximal torus is $T = A\times \bC^\times_\hbar$ with $A = \left(\mathbb{C}^{\times}\right)^n$. In the GIT description of $X_{k,n}$, $\operatorname{diag}(e^{m_1},\ldots,e^{m_n})\in A$ acts diagonally on $\mathbb C^n$ and
$e^{-\hbar} \in \mathbb C^\times_\hbar$ scales the fiber. The fixed locus of the $A$-action on $X_{k,n}$ is given by: 
\begin{equation}
    X_{k,n}^A\,=\,\bigsqcup_{w\in{\rm W}^{(k;n)}}\,\sfp_w~.
\end{equation}
Here, the fixed points are indexed by Grassmannian permutations $w =(w_1\,\cdots\,w_k\,\cdots)\in {\rm W}^{(k;n)}\subset S_n$.\footnote{Here, we are following the notation introduced in~\cite[Equation (3.8)]{Closset:2026bnk}.} We remind the reader that these permutations are such that they have a descent at most at the $k$-th location. Therefore, they are fully determined by their first $k$ entries.

For a permutation $w\in {\rm W}^{(k;n)}$, the associated fixed point $\sfp_w\in X_{k,n}^A$ is given by:
\begin{equation}
    \sfp_w\,\coloneqq\,\bigoplus_{a=1}^k\,\mathbb{C}e_{w_{a}}~, 
\end{equation}
with $\{e_1, \cdots, e_n\}$ denoting the standard basis of $\mathbb{C}^n$. There are $\binom{n}{k}$ of such points, which is none other than the topological Euler number of $X_{k,n}$. From the GLSM perspective, they correspond to the massive vacua of the theory.

\begin{figure}[t!]
    \centering

    \begin{subfigure}[c]{0.41\textwidth}
        \centering
        \includegraphics[width=\linewidth,scale=0.7]{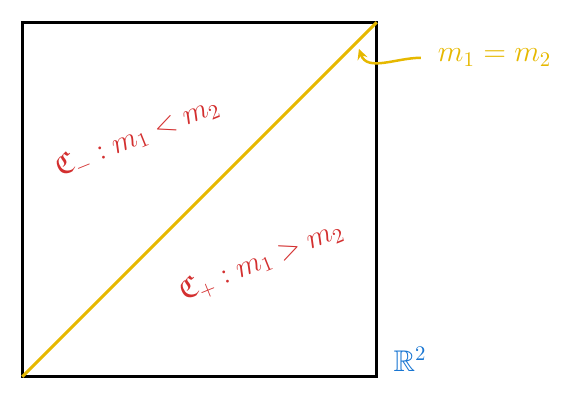}

    \end{subfigure}
    \hfill
    \begin{subfigure}[c]{0.55\textwidth}
        \centering
        \includegraphics[width=\linewidth, scale=2]{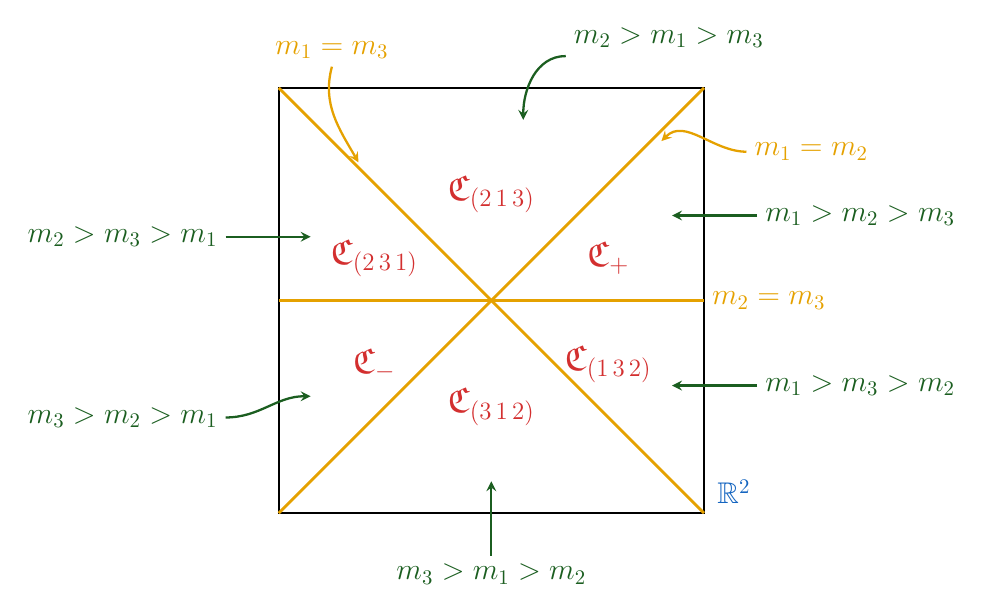}
    \end{subfigure}

    \caption{\textsc{Left:} Chamber decomposition for $n=2$. \textsc{Right:} Chamber decomposition for $n=3$ where we look at an $\bR^2$ slice inside $\bR^3$.}
    \label{fig:chambern23}
\end{figure}

\medskip
\noindent
\textbf{Normal directions and chamber decomposition.}\footnote{Throughout this work, when discussing the chamber decomposition, we will denote the real part of the equivariant parameters/twisted masses using the same notation $m_i$. This is an abuse of notation, and we hope that it will not cause any confusion.}As an essential ingredient that enters into the definition of the stable envelope, which we will introduce below, one needs to fix the choice of chamber $\mathfrak{C}$ inside the Lie algebra of the torus $A$. The chamber decomposition comes about by looking at the set of hyperplanes in $\mathbb{R}^n = {\rm Span}_{\mathbb R}(m_1, \cdots, m_n)$ defined by setting $m_i=m_j$ for $i\neq j\in\{1,\cdots, n\}$. There are therefore $n!$ possible chambers each indexed by a permutation $\pi\in S_n$:
\begin{equation}
    \mathfrak{C}_\pi\,\coloneqq\,\{(m_1, \cdots, m_n)\in\mathbb{R}^n\,:\,m_{\pi(1)}\,>\,m_{\pi(2)}\,>\,\cdots\,>\,m_{\pi(n)}\}~.
\end{equation}
 In particular, looking at the roots $\alpha_{ij}\equiv m_i-m_j$, the choice of a chamber amounts to decomposing these roots into positive and negative ones.  In what follows, we will use the notation $\mathfrak{C}_+\equiv\mathfrak{C}_{\rm id}$ where ${\rm id} = (1\,2\,\cdots\,n)$ being the trivial permutation. In contrast, we will denote the opposite chamber by $\mathfrak{C}_{-} \coloneqq \mathfrak{C}_{w_0} = \{m_1<m_2<\cdots<m_n\}$ where $w_0=(n\,\cdots\,2\,1)$ is the longest permutation in $S_n$. 
 
To illustrate this more, let us look at the example of $n=2$. In this case, we have the hyperplane $m_1=m_2$ that cuts $\mathbb{R}^2$ into two chambers $\mathfrak{C}_+ =\{m_1>m_2\}\subset \mathbb{R}^2$ and $\mathfrak{C}_-=\{m_1<m_2\}\subset \mathbb{R}^2$. This decomposition is shown in the left part of Figure~\ref{fig:chambern23}.  As a second example, let us consider $n=3$. In this case, we get $3$ hyperplanes $\alpha_{i,j} = 0$ that cut $\mathbb{R}^3$ into $6$ chambers as shown in the right part of Figure~\ref{fig:chambern23}.

At every fixed point $\sfp_w$, the normal bundle is none other than the fiber of the tangent bundle $\mathcal{N}_w \,\equiv\, T_{\sfp_w}X_{k,n}$ 
whose weights are given in terms of the roots $\alpha_{ij}$ introduced above. Therefore, the choice of a chamber $\mathfrak{C}_\pi$ decomposes the normal bundle as:
\begin{equation}
    \mathcal{N}_w\,=\,\mathcal{N}^+_{\pi,w}\,\oplus\,\mathcal{N}^-_{\pi,w}~,\qquad 
\end{equation}
which are usually referred to as attracting and repelling directions, respectively. Moreover, $\dim_\bC\mathcal{N}_{\pi,w}^+ = \dim_{\bC}\mathcal{N}_{\pi,w}^- = k(n-k)$.

\subsection{Cohomological stable envelopes}
Let us fix a chamber $\mathfrak{C}_\pi$ inside Lie$(A)$. The equivariant action on $X_{k,n}$ generates a gradient flow and for each point $\sfp_w\in X_{k,n}^A$ one can define the attractor manifold:
\begin{equation}\label{attrpiw}
    \mathsf{Attr}_{\pi,w}\,\equiv\,\mathsf{Attr}_{\mathfrak{C}_\pi,w}\,\coloneqq\,\{x\in X_{k,n}\,:\,\lim_{t\rightarrow+\infty} e^{-t\xi}\cdot x \,=\, \mathsf{p}_w~, \forall \xi\in\mathfrak{C}_\pi\}\subseteq X_{k,n}~.
\end{equation}
This is a Lagrangian submanifold of $X_{k,n}$. We also define the full attractor of the point $\sfp_w$ to be the closure of ${\sf Attr}_{\pi,w}$, where one includes the attractors of other fixed points that flow to $\sfp_w$. More explicitly, let us write $\sfp_{w'}\leq\sfp_w$ if $\sfp_{w'}\in{\sf Attr}_{\pi,w}$, then, the full attractor of $\sfp_w$ is given by:
\begin{align}\label{attrcfull}
    {\sf Attr}_{\pi,w}^{\rm f}\,=\,\bigsqcup_{\mathsf{p}_{w'}\leq \mathsf{p}_w}\, \mathsf{Attr}_{\pi,w'}~,
\end{align}
which is a Lagrangian subvariety of $X_{k,n}$.

\medskip
\noindent
\textbf{Attractor manifolds are the conormal bundles.} Let us focus on the Grassmannian manifold Gr$(k,n)$ at the base of our Higgs branch $X_{k,n}$. In this case, the attractor manifold associated with each fixed point $\sfp_w$ is none other than the Schubert cell $C_{\pi,w}$. See~\cite{Anderson_Fulton_2023} for a pedagogical discussion. See also~\cite{Closset:2023bdr,Gu:2025tda,Closset:2025akk,Closset:2026bnk} for some recent reviews. Including the cotangent direction, then, the attractor manifold $\attr_{\pi,w}$ given in~\eqref{attrpiw} becomes the conormal bundle over $C_{\pi,w}$:
\begin{equation}\label{attr=conorm}
    \attr_{\pi,w}\,=\,\mathcal{C}_{\pi,w}\,\coloneqq\,\{(c,v)\,:\,c\,\in\,C_{\pi,w}~,~v\,\in\,T^*_c{\rm Gr}(k,n)\,:\,v\mid_{T_cC_{\pi,w}}\,=\,0\}\,\subset\,X_{k,n}~,
\end{equation}
where, in particular, the fiber at a point $c$ is $(T_cC_{\pi,w})^\perp =\mathcal{N}^*_{\pi,w}\mid_c $ is the dual of the normal bundle of $C_{\pi,w}$ restricted at that point.

\subsubsection{Matrix representation of attractors in the negative chamber}
Looking at the chamber $\mathfrak{C}_-$, the attractor on the Grassmannian base is given by the Schubert cell $C_w\equiv C_{-,w}$ which, following the notation in~\cite{Closset:2023bdr}, has the following tangent space:
\begin{equation}\label{TpwCw}
    T_{\sfp_w}C_{w} \,=\, \bigoplus_{a=1}^{k}{\sf Hom}\left(\bC e_{w_a}\,,\, \bigoplus_{i=w_{a}+1}^{n}\bC e_{i}\right)\,\ominus\, \bigoplus_{a=1}^{k} {\sf Hom}\left(\bC e_{w_a}\,,\, \bigoplus_{b=a+1}^{k}\bC e_{w_b}\right)~.
\end{equation}
Where the second factor is the redundancy in rotating the $\bC^{k}$ subspace inside $\bC^n$. The Schubert cell is represented by the $k\times n$ matrix $\phi_w$~\cite[Equation (3.56)]{Closset:2023bdr}:
\begin{equation}\label{phiwMatrix}
    (\phi_w)_{a,i}\,=\,\begin{cases}
        0~, \qquad &i\,<\,w_a~,\\
        1~, \qquad &i \,= \,w_a~,\\
        0~, \qquad &i\, =\, w_b~, b\,\neq\, a~,\\
        \star~,\qquad & i\,>\,w_a~,\,i\notin\,\{w_1,\cdots,w_k\}~.
    \end{cases}\,=\,\left(\begin{array}{c|c|c|c|c|c|c|c|c|c}
0 & 1 & \star & 0 & \cdots & 0 & \star & \cdots & 0 & \star\\ \hline
0 & 0 & 0 & 1 & \cdots & 0 & \star & \cdots & 0 & \star\\ \hline
\vdots & \vdots & \vdots & \vdots & \ddots & \vdots & \vdots & \ddots & \vdots & \vdots\\ \hline
0 & 0 & 0 & 0 & \cdots & 1 & \star & \cdots & 0 & \star\\ \hline
\vdots & \vdots & \vdots & \vdots & \ddots & \vdots & \vdots & \ddots & \vdots & \vdots\\ \hline
0 & 0 & 0 & 0 & \cdots & 0 & 0 & \cdots & 1 & \star
\end{array}\right)_{a,i}~.
\end{equation}
Here, the $\star$ entries stand for the tangent directions to the Schubert cell, and their number is none other than the dimension of the cell:
\begin{equation}
    \dim C_w\,=\,\sum_{a=1}^k \underbrace{(n\,-\,w_a)}_{i>w_a}\,-\,\underbrace{(k\,-\,a)}_{i\neq w_{a+1},\cdots, w_k}\,=\,k(n\,-\,k)\,-\,\ell(w)~,
\end{equation}
where $\ell(w)\coloneqq \sum_{a=1}^k(w_a-a)$ is the length of the Grassmannian permutation $w$.

Similarly, we see that the normal bundle of $C_w$ in ${\rm Gr}(k,n)$ at a fixed point $\sfp_w$, which we denote by $\widetilde{\mathcal{N}}_w\mid_{\sfp_w}$, takes the form:
\begin{equation}\label{tildeNwMatrix}
\big(\widetilde{\mathcal N}_w\,\big|_{\mathsf p_w}\big)_{a,i}=
\begin{cases}
\star\,, & i<w_a\,,\ i\notin\{w_1,\cdots,w_k\}\,,\\[2pt]
0\,, & i=w_b\,,\ b=1,\cdots,k\,,\\[2pt]
0\,, & i>w_a\,.
\end{cases}
\qquad=\quad
\left(\begin{array}{c|c|c|c|c|c|c|c|c|c}
\star & 0 & 0     & 0 & \cdots & 0 & 0     & \cdots & 0 & 0\\ \hline
\star & 0 & \star & 0 & \cdots & 0 & 0     & \cdots & 0 & 0\\ \hline
\vdots & \vdots & \vdots & \vdots & \ddots & \vdots & \vdots & \ddots & \vdots & \vdots\\ \hline
\star & 0 & \star & 0 & \cdots & 0 & 0     & \cdots & 0 & 0\\ \hline
\vdots & \vdots & \vdots & \vdots & \ddots & \vdots & \vdots & \ddots & \vdots & \vdots\\ \hline
\star & 0 & \star & 0 & \cdots & 0 & \star & \cdots & 0 & 0
\end{array}\right)_{a,i}\,.
\end{equation}
As a complex vector bundle, we see that the normal bundle has the following rank:
\begin{equation}
    {\rm rank}\,\widetilde{\mathcal{N}}_w\,=\,\sum_{a=1}^k\underbrace{(w_a\,-\,1)}_{i<w_a}\,-\,\underbrace{(a\,-\,1)}_{i\neq w_1, \cdots, w_{a-1}}\,=\,\ell(w)\,=\,{\rm codim}\,C_w ~.
\end{equation}
So, in total, the normal vector bundle has dimension $k(n-k)$.

As defined in~\eqref{attr=conorm}, the conormal bundle $\mathcal{C}_{-,w}$ is subspace of $X_{k,n}$ such that it kills the tangent directions in $\phi_w$~\eqref{phiwMatrix}.\footnote{Said differently, the conormality constraint translates into $\widetilde\phi_w\cdot \phi_w$ is strictly upper triangular.} As we discussed in the previous section, this is represented by $\widetilde\phi_w$. And it has the following explicit form:
\begin{equation}\label{tildePhiwMatrix}
(\widetilde\phi_w)_{i,a}\,=\,
\begin{cases}
\star\,, & i<w_a\,,\ i\notin\{w_1,\cdots,w_k\}\,,\\[2pt]
0\,, & i=w_b\,,\ b=1,\cdots,k\,,\\[2pt]
0\,, & i>w_a\,.
\end{cases}
\qquad=\quad
\left(\begin{array}{c|c|c|c|c|c}
\star & \star & \cdots & \star & \cdots & \star\\ \hline
0 & 0 & \cdots & 0 & \cdots & 0\\ \hline
0 & \star & \cdots & \star & \cdots & \star\\ \hline
0 & 0 & \cdots & 0 & \cdots & 0\\ \hline
\vdots & \vdots & \ddots & \vdots & \ddots & \vdots\\ \hline
0 & 0 & \cdots & 0 & \cdots & 0\\ \hline
0 & 0 & \cdots & 0 & \cdots & \star\\ \hline
\vdots & \vdots & \ddots & \vdots & \ddots & \vdots\\ \hline
0 & 0 & \cdots & 0 & \cdots & 0\\ \hline
0 & 0 & \cdots & 0 & \cdots & 0
\end{array}\right)_{i,a}~.
\end{equation}
This is none other than the matrix transpose of the normal bundle~\eqref{tildeNwMatrix}, therefore, its dimension is $k(n-k)$ as anticipated from the fact that it is a Lagrange submanifold of $X_{k,n}$.

These explicit forms of the matrices will be of great importance to us when we construct the BPS defects of interest in Section~\ref{sec:BPSdefects}. As an explicit example, in Table~\ref{tab:Gr24-cells}, we list the form of these matrices for all Schubert cells in the Gr$(2,4)$ case.

\begin{table}[ht]
\centering
\renewcommand{\arraystretch}{1.25}
\setlength{\arraycolsep}{2.5pt}
\setlength{\tabcolsep}{5pt}
\small
\begin{tabular}{cccccc}
\toprule
$w$ & $\ell(w)$ & $\dim C_w$ & $\phi_w$ & $\widetilde{\mathcal N}_w\big|_{\mathsf p_w}$ & $\widetilde\phi_w$\\
\midrule
\addlinespace[6pt]
$(1\,2\,3\,4)$ & $0$ & $4$ & $\begin{pmatrix}1&0&\star&\star\\0&1&\star&\star\end{pmatrix}$ & $\begin{pmatrix}0&0&0&0\\0&0&0&0\end{pmatrix}$ & $\begin{pmatrix}0&0\\0&0\\0&0\\0&0\end{pmatrix}$ \\ \addlinespace[10pt]
$(1\,3\,2\,4)$ & $1$ & $3$ & $\begin{pmatrix}1&\star&0&\star\\0&0&1&\star\end{pmatrix}$ & $\begin{pmatrix}0&0&0&0\\0&\star&0&0\end{pmatrix}$ & $\begin{pmatrix}0&0\\0&\star\\0&0\\0&0\end{pmatrix}$ \\ \addlinespace[10pt]
$(1\,4\,2\,3)$ & $2$ & $2$ & $\begin{pmatrix}1&\star&\star&0\\0&0&0&1\end{pmatrix}$ & $\begin{pmatrix}0&0&0&0\\0&\star&\star&0\end{pmatrix}$ & $\begin{pmatrix}0&0\\0&\star\\0&\star\\0&0\end{pmatrix}$ \\ \addlinespace[10pt]
$(2\,3\,1\,4)$ & $2$ & $2$ & $\begin{pmatrix}0&1&0&\star\\0&0&1&\star\end{pmatrix}$ & $\begin{pmatrix}\star&0&0&0\\\star&0&0&0\end{pmatrix}$ & $\begin{pmatrix}\star&\star\\0&0\\0&0\\0&0\end{pmatrix}$ \\ \addlinespace[10pt]
$(2\,4\,1\,3)$ & $3$ & $1$ & $\begin{pmatrix}0&1&\star&0\\0&0&0&1\end{pmatrix}$ & $\begin{pmatrix}\star&0&0&0\\\star&0&\star&0\end{pmatrix}$ & $\begin{pmatrix}\star&\star\\0&0\\0&\star\\0&0\end{pmatrix}$ \\ \addlinespace[10pt]
$(3\,4\,1\,2)$ & $4$ & $0$ & $\begin{pmatrix}0&0&1&0\\0&0&0&1\end{pmatrix}$ & $\begin{pmatrix}\star&\star&0&0\\\star&\star&0&0\end{pmatrix}$ & $\begin{pmatrix}\star&\star\\\star&\star\\0&0\\0&0\end{pmatrix}$ \\
\addlinespace[4pt]
\bottomrule
\end{tabular}
\caption{Schubert cells $C_w\subset\mathrm{Gr}(2,4)$ for all Grassmannian permutations $w$.
Columns: the cell representative $\phi_w$~\eqref{phiwMatrix},
the normal space at the fixed point $\mathsf p_w$~\eqref{tildeNwMatrix}, and the conormal fiber at $\mathsf p_w$ $\widetilde{\phi}_w$~\eqref{tildePhiwMatrix}. In each row, $\phi_w$ has $\dim C_w=4-\ell(w)$ entries $\star$,
while $\widetilde{\mathcal N}_w|_{\mathsf p_w}$ and $\widetilde\phi$ each have $\ell(w)$ entries $\star$. One can check that, for every $w$, $\phi_w\cdot\widetilde{\phi}_w = 0$ and $ \widetilde{\phi}_w\cdot \phi_w$ is upper triangular.}
\label{tab:Gr24-cells}
\end{table}

\subsubsection{Definition of the stable envelope}
For each fixed point $\sfp_w\in X_{k,n}^A$, an equivariant cohomology class of $X_{k,n}$ can be associated, referred to as the stable envelope map:
\begin{equation}
    \stab_\pi\,:\,H_T^\bullet(X_{k,n}^A)\,\longrightarrow \,H_T^\bullet(X_{k,n})~.
\end{equation}
In what follows, we will denote the image of $\sfp_w$ under this map by $\stab_{\pi,w}\in H_T^{2k(n-k)}(X_{k,n})$. This is defined by the following three axioms~\cite[Theorem 3.3.4]{MaulikOkounkov2019}:
\begin{itemize}
    \item \textbf{Support:} the support of the image of a fixed point $\sfp_w$ under the stable envelope map is given by the full attractor of that point:
    \begin{equation}\label{suppAxiom}
        {\rm Supp}[\stab_{\pi,w}]\,\subseteq\,\attr_{\pi,w}^{\rm f}~.
    \end{equation}
    \item \textbf{Normalization:} when restricted to the fixed point $\sfp_w$, it gives the Euler class of the repelling directions $\mathcal{N}_{\pi,w}^-$:
    \begin{equation}\label{normAxiom}
        \stab_{\pi,w}\mid_{\sfp_w}\,=\,\pm\,e(\mathcal{N}_{\pi,w}^-)~.
    \end{equation}
    Here, the $\pm$ sign is related to the polarization of the cotangent bundle of $X_{k,n}$. In our work, we will fix the sign in explicit examples.
    \item \textbf{Degree:} when evaluated at other fixed points, the degree of the final expression with respect to the $A$-equivariant parameters is strictly less than $k(n-k)$ (i.e., half the cohomological degree of $\stab_{\pi,w}$.):
    \begin{equation}\label{degAxiom}
    \mathrm{deg}_A\,\Big(\stab_{\pi,w}\mid_{\sfp_{w'}}\Big)\,<\,k\,(n\,-\,k)~,\qquad \forall w'\neq w~.
    \end{equation}
\end{itemize}

\medskip
\noindent
\textbf{The explicit form of the stable envelope.} Let us now fix the chamber to be $\mathfrak{C}_+ = \{m_1>\cdots>m_n\}$. As an element of the $T$-equivariant cohomology ring of $X_{k,n}$, the stable envelope $\stab_{+,w}$ is given explicitly as follows~\cite[Theorem 4.1]{rimanyi2015partial}:
\begin{equation}\label{stab=VT}
    \mathsf{Stab}_{+,w}\,=\,c_n^{-1}\,\mathsf{W}_{+,w}(\sigma,m,\hbar)\,\in\,H_T^\bullet(T^*{\rm Gr}(k,n))~,
\end{equation}
with $c_n$ being the following equivariant Euler class~\cite[Equation (15)]{rimanyi2015partial}:
\begin{equation}
    c_n\,\coloneqq\,\prod_{a=1}^k\,\prod_{b=1}^k\,(\sigma_a\,-\,\sigma_b\,-\,\hbar)~.
\end{equation}
Meanwhile, ${\sf W}_{+,w}(\sigma,m,\hbar)$ is the \textit{Tarasov--Varchenko} (TV) weight function~\cite[Subsection 3.1]{rimanyi2015partial} (see also~\cite{Tarasov:1994bb,RimanyiTarasovVarchenko2014,tarasov2013combinatorial}):
\begin{multline}\label{VTwdefn}
        \mathsf{W}_{+,w}(\sigma; m, \hbar) \,\coloneqq\,\\ (-\hbar)^k\,\operatorname*{\mathsf{Sym}}_{\sigma_\bullet} \left[ \prod_{a=1}^k \left( \prod_{\alpha_a=1}^{w_a - 1} ( \sigma_a \,-\,m_{\alpha_a}\,-\,\hbar) \cdot \prod_{\gamma_a=w_a + 1}^n (\sigma_a \,-\, m_{\gamma_a}) \right) \cdot \prod_{1 \le a < b \le k} \frac{\sigma_a - \sigma_b - \hbar}{\sigma_a - \sigma_b} \right]~.
\end{multline}
Here, we introduced the symmetrization operator:
\begin{equation}\label{sym-op}
   \operatorname*{\mathsf{Sym}}_{\sigma_\bullet} \left[ f(\sigma_1, \dots, \sigma_k) \right] \coloneqq \sum_{\pi \in S_k} f\left(\sigma_{\pi(1)}, \sigma_{\pi(2)}, \dots, \sigma_{\pi(k)}\right)~.
\end{equation}
With these explicit expressions, we have:
\begin{equation}\label{stab+w}
    \stab_{+,w}(\sigma, m,\hbar)\,=\,(-1)^{\eps(w)}\,\operatorname*{\mathsf{Sym}}_{\sigma_\bullet}\left[\frac{\prod_{a=1}^k\left(\prod_{\alpha_a=1}^{w_a-1}(\sigma_a\,-\,m_{\alpha_a}\,-\,\hbar)\,\prod_{\gamma_a=w_a+1}^n(\sigma_a\,-\,m_{\gamma_a})\right)}{\prod_{{1\leq a<b\leq k}}(\sigma_a\,-\,\sigma_b)\,(\sigma_b\,-\,\sigma_a\,-\,\hbar)}\right]~.
\end{equation}
Here, we are fixing an overall sign with:
\begin{equation}\label{eps(w)}
    \eps(w)\,\coloneqq\,\ell(w)\,-\,k\,w_1~.
\end{equation}
Recall here that the length of a Grassmannian permutation $w_a$ can be written as $\ell(w)\,=\,\sum_{a}(w_a\,-\,a)$. This sign is fixed to get a matching with our defect calculations below. The term $(-1)^{\eps(w)}$ is related to a choice of polarization in~\cite{MaulikOkounkov2019}.

Similarly, one can write the expression for the stable envelope in the opposite chamber $\mathfrak{C}_-$. This can be done simply by applying the permutation $w_0$ to the indices of the $A$ equivariant parameters in~\eqref{stab+w}:
\begin{equation}\label{stab-w}
    \stab_{-,w}(\sigma, m,\hbar)\,=\,(-1)^{\eps(w)}\,\operatorname*{\mathsf{Sym}}_{\sigma_\bullet}\left[\frac{\prod_{a=1}^k\left(\prod_{\alpha_a=1}^{w_a-1}(\sigma_a\,-\,m_{\alpha_a})\,\prod_{\gamma_a=w_a+1}^n(\sigma_a\,-\,m_{\gamma_a}\,-\,\hbar)\right)}{\prod_{{1\leq a<b\leq k}}(\sigma_b\,-\,\sigma_a)\,(\sigma_a\,-\,\sigma_b\,-\,\hbar)}\right]~.
\end{equation}

For example, in $T^*\bP^{n-1}$ case, for any permutation $w = (w_1\,\cdots)$, these expressions for the stable envelopes simplify into:
\begin{equation}\label{stab+w1}
     \stab_{+,w_1}(\sigma, m, \hbar)\,\equiv\,\stab_{+,w}(\sigma, m, \hbar)\,=\,-\, \prod_{\alpha=1}^{w_1-1}(\sigma\,-\,m_\alpha\,-\,\hbar)\,\prod_{\gamma=w_1+1}^n(\sigma\,-\,m_\gamma)~,
\end{equation}
in $\mathfrak{C}_+$ chamber, and:
\begin{equation}\label{stab-w1}
    \stab_{-,w_1}(\sigma, m, \hbar)\,\equiv\,\stab_{-,w}(\sigma, m, \hbar)\,=\, -\,\prod_{\alpha=1}^{w_1-1}(\sigma\,-\,m_\alpha)\,\prod_{\gamma=w_1+1}^n(\sigma\,-\,m_\gamma\,-\,\hbar)~,
\end{equation}
in the chamber $\mathfrak{C}_-$.

\medskip
\noindent
\textbf{A comment on the $\hbar\rightarrow \infty$ limit.} Let us now examine what happens to these stable envelopes in the limit where we take $\hbar\rightarrow\infty$. As we discussed earlier, this limit amounts to integrating out the cotangent fibers of the Higgs branch $X_{k,n}$ and, hence, one expects the stable envelopes to reduce to the usual Schubert classes in the base Gr$(k,n)$. Indeed, performing this limit to~\eqref{stab-w}, we get:
\begin{equation}
    (-1)^{\eps(w)}\,\stab_{-,w}(\sigma,m,\hbar)\,=\,(-\hbar)^{{\rm codim}(C_w)}\,\mathfrak{S}_w(\sigma,m)\,+\,\cdots~,
\end{equation}
where, codim$(C_w) = k(n-k)-\ell(w)$ with $\ell(w)$ being the length of the permutation $w$, and, $\mathfrak{S}_w(\sigma, m)\in H^{2\ell(w)}_A({\rm Gr}(k,n))$ is the double Schubert polynomial. This has the following explicit form:\footnote{This can also be written as a determinant formula as in~\cite[Equation (5.4)]{Closset:2023bdr} or in terms of difference operators as in~\cite[Equation (A.8)]{Closset:2025akk}.}
\begin{equation}\label{doubleSchubert}
    \mathfrak{S}_w(\sigma, m)\,=\,\operatorname*{\mathsf{Sym}}_{\sigma_\bullet}\left[\frac{\prod_{a=1}^k\prod_{\alpha_a=1}^{w_a-1}(\sigma_a\,-\,m_{\alpha_a})}{\prod_{{1\leq a<b\leq k}}(\sigma_a\,-\,\sigma_b)}\right]~.
\end{equation}
This polynomial represents the Schubert class $[C_w]$ in the equivariant cohomology ring of the base manifold.\footnote{From this discussion, one can deduce that, when discussing the equivariant cohomology ring of Gr$(k,n)$, it is standard to study things in the chamber $\mathfrak{C}_-$. Therefore, what is usually denoted by $C_w$ in the related literature is what we denoted in this section by $C_{-,w}$.}

\subsubsection{Moving across the walls: the $\sf R$-matrix}
In our earlier discussion, we gave the expression for the stable envelopes in the two chambers $\mathfrak{C}_{+}$ and $\mathfrak{C}_-$ corresponding to the permutations ${\rm id} = (1\,\cdots\,n)$ and $w_0 = (n\,\cdots\,1)$. But, one can work out the form of these stable envelopes for any chamber $\mathfrak{C}_\pi$ inside Lie$(A)$. Starting with $\stab_{+,w}$, in~\cite[Subsection 3.2]{rimanyi2015partial}, the authors define the form of $\stab_{\pi,w}$. For our taste here, we can rewrite their expression such that $\mathfrak{C}_-$ is the starting point:
\begin{equation}\label{stabpi=stab-}
    \stab_{\pi, w}(\sigma, m_1,\cdots,m_n, \hbar)\,=\,\stab_{-,w^\pi}(\sigma, m_{\pi w_0(1)},\cdots, m_{\pi w_0(n)}, \hbar)~,
\end{equation}
where $w^\pi \in {\rm W}^{(k;n)}$ is the Grassmannian representative of the permutation $w_0\pi^{-1}w$.

More generally, starting with the set of stable envelopes $\{\stab_{\pi,w}\,:\,w\in{\rm W}^{(k;n)}\}$ in the chamber $\mathfrak{C}_\pi$, one can work out the form of these stable envelopes in any other chamber $\mathfrak{C}_{\pi'}$ using the $\sf R$-matrix which is defined as~\cite[Definition 4.1.1]{MaulikOkounkov2019}:
\begin{equation}\label{RmatDefn}
    {\sf R}_{\pi,\pi'}\,\coloneqq\,\stab_{\pi}^{-1}\,\circ\,\stab_{\pi'}\,\in\,\mathsf{End}(H_T(X_{k,n}^A))\,\otimes\,\mathbb{C}(\hbar, m)~.
\end{equation}
This can be viewed as a square matrix of size $\binom{n}{k}$.

The $\mathsf{R}$-matrix has the following three properties:
\begin{itemize}
    \item Going in the opposite direction between two chambers amounts to taking the inverse of the $\mathsf{R}$-matrix:
    \begin{equation}
    \mathsf{R}_{\pi,\pi'}\,=\,\mathsf{R}_{\pi',\pi}^{-1}~.
    \end{equation}
    \item For three chambers $\mathfrak{C}_\pi$, $\mathfrak{C}_{\pi'}$, and $\mathfrak{C}_{\pi''}$, we have:
    \begin{equation}\label{composR}
        \mathsf{R}_{\pi'',\pi}\,=\,\mathsf{R}_{\pi'',\pi'}\,\circ\,\mathsf{R}_{\pi',\pi}~.
    \end{equation}
    It is understood from our discussion here that the decomposition of the $\mathsf{R}$-matrices goes from left to right. This is the opposite of our convention for the multiplication of permutations.
    \item It satisfies the Yang--Baxter equation. See~\cite[Subsection 4.1.8]{MaulikOkounkov2019}. We will see an explicit example of this below.
\end{itemize}

\medskip
\noindent
\textbf{Connection to integrable models.}  The ${\sf R}$-matrix provides one way of defining quantum groups. Essentially, the Yang Baxter equation (see, for example ,~\eqref{YBE} below)
  defines a quantum group.  
  On the other hand, the geometric $\mathsf{R}$-matrix above arises from $X_{k,n}$ also appears in the algebraic Bethe ansatz of the $XXX$ spin chain. 

\subsubsection{An illustrative example: cohomological stable envelopes in \texorpdfstring{$T^*\bP^{n-1}$}{T*P1}}\label{sssec:examples}
Let us now pedagogically discuss how one computes the stable envelopes as written in~\eqref{stab-w}. Here, we will look at the two cases for $T^*\bP^{n-1}$ with $n=2,3$.

\medskip
\noindent
\underline{\textbf{Example I: Stable envelopes in $T^*\bP^1$.}}\footnote{This example is a slight elaboration upon~\cite[Example 4.1.2]{MaulikOkounkov2019}.} Let us look at the case $n=2$. Recall that the equivariant classical cohomology ring is given by:\footnote{Note that classically, $H^*_T(T^*{\rm Gr}(k,n)) \cong H^*_T({\rm Gr}(k,n))$. This isomorphism follows from the equivariant homotopy equivalence $T^*X\rightarrow X$, where one can shrink the cotangent fibers to a point. This isomorphism is lost in the world of quantum cohomology. $T^*{\rm Gr}(k,n)$ is hyperk\"ahler and its quantum cohomology is the same as the classical one, while the quantum cohomology ring of ${\rm Gr}(k,n)$ is not~\cite{Witten:1988xj}. To have an interesting quantum cohomology ring structure from $T^*{\rm Gr}(k,n)$, it is necessary to use equivariant, or reduced Gromov--Witten theory. One incarnation of this fact is that the quantum multiplication~\eqref{eq:Su_T^astGr} becomes trivial when the equivariant parameter $\hbar$ is turned off.}
\begin{equation}\label{calssical equi coho of cotang P1}
    H_T^{\bullet}(T^{*}\bP^1) \,= \,\frac{\bZ[\sigma,m_1,m_2,\hbar]}{\left\langle(\sigma\,-\,m_1)(\sigma\,-\,m_2)\right\rangle}~. 
\end{equation}
We recall here that $\sigma = c_1(\mathcal{S})$ is the first Chern class of the tautological line bundle $\mathcal{S} = \mathcal{O}(-1)$ over the base $\mathbb{P}^1$. 

From the action of the torus $A= (\bC^\times)^2$, there are two fixed points characterized by $\sigma = m_1,m_2$\footnote{By this we mean the restriction of the first Chern class of $\mathcal{O}_{\bP^1}(-1)$ at each fixed point.} corresponding to the two fixed points on the base $\sfp_1\equiv0 =[0:1]$ and $\sfp_2 = \infty \equiv [1:0]$, respectively. Therefore, we have the following Euler classes:
\begin{equation}\label{tangentP2weights}
    e_T(T_{\sfp_1}\mathbb{P}^1)\,=\,m_2\,-\,m_1~, \qquad e_T(T_{\sfp_2}\mathbb{P}^1)\,=\,m_1\,-\,m_2~.
\end{equation}
This follows directly from:\footnote{Moreover, for two vector bundles $E$ and $F$ with Chern roots $\alpha_i$ and $\beta_j$ respectively, the Euler class of their tensor product is given by $e(E\otimes F) = \prod_{i,j} (\alpha_i+\beta_j)$.}
\begin{equation}
    T_{\sfp}\mathbb{P}^1\,=\,{\sf Hom}(\mathcal{S}_\sfp,\mathbb{C}^2/\mathcal{S}_\sfp) \,=\,\mathcal{S}^\vee_\sfp\,\otimes\,\bC^2/\mathcal{S}_\sfp~.
\end{equation}

Since $T^*\mathbb{P}^1 = (T\mathbb{P}^1)^\vee$, we have:
\begin{equation}
\begin{split}
     e_T(T_{\sfp_1}^*\mathbb{P}^1) \,=\, -\,e_T(T_{\sfp_1}\mathbb{P}^1)\, -\, \hbar \,=\, m_1\,-\,m_2\,-\,\hbar~,\\
     e_T(T_{\sfp_2}^*\mathbb{P}^1) \,=\, -\,e_T(T_{\sfp_2}\mathbb{P}^1)\, -\, \hbar \,=\, m_2\,-\,m_1\,-\,\hbar~.\\
\end{split}
\end{equation}

\begin{figure}[t!]
    \centering

    \begin{subfigure}{0.48\textwidth}
        \centering
        \includegraphics[width=\linewidth,scale=0.7]{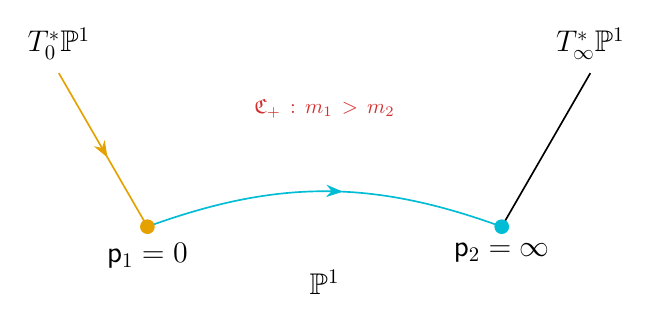}
    \end{subfigure}
    \hfill
    \begin{subfigure}{0.48\textwidth}
        \centering
        \includegraphics[width=\linewidth, scale=2]{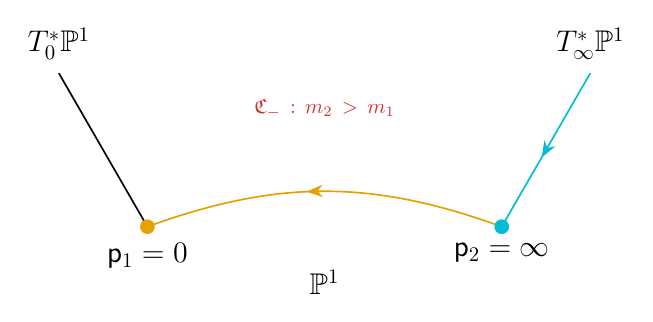}
    \end{subfigure}

    \caption{The attractors for the two fixed points of the torus $A$ action on $T^*\mathbb{P}^1$ in both possible chambers $\mathfrak{C}_+$ and $\mathfrak{C}_-$.}
    \label{fig:attrP1}
\end{figure}

Looking at the chamber $\mathfrak{C}_+ = \{m_1>m_2\}$, we have that $e_T(T_{\sfp_1}\mathbb{P}^1)<0$ and hence it is in the repelling direction of $\sfp_1$. Similarly, $e_{T}(T_{\sfp_2}\bP^1)>0$, hence, it is in the attracting direction of $\sfp_2$. To sum up, we have:
\begin{equation}
\begin{split}
    &\mathcal{N}_{+,1}^+\,=\,T^*_{\sfp_1}\bP^1~,\qquad \mathcal{N}_{+,1}^-\,=\,T_{\sfp_1}\mathbb{P}^1~,\\
    &\mathcal{N}_{+,2}^+\,=\,T_{\sfp_2}\bP^1~,\qquad \mathcal{N}_{+,2}^-\,=\,T^*_{\sfp_2}\bP^1~.\\
\end{split}
\end{equation}
In particular, we see that $\sfp_1\leq \sfp_2$. Thus, the full attractors are given by~\eqref{attrcfull}:
\begin{equation}
    \attr^{\rm f}_{+,1}\,=\,[T^*_{\sfp_1}\mathbb{P}^1]~,\qquad \attr_{+,2}^{\rm f}\,=\,[\mathbb{P}^1]\,+\,[T^*_{\sfp_1}\mathbb{P}^1]~.
\end{equation}
These results are summarized in Figure~\ref{fig:attrP1}.

Since the stable envelope $\stab_{+,w}$ is of complex degree $1$, let us write it to be of the form $a\sigma+b$. From the axioms, $a$ and $b$ can be fixed as follows:
\begin{equation}
    \stab_{+,1}(\sigma, m_1, m_2,\hbar)\,=\,\mp (\sigma\,-\,m_2)~,\qquad \stab_{+,2}(\sigma,m_1,m_2,\hbar)\,=\,\pm(\sigma\,-\,m_1\,-\,\hbar)~.
\end{equation}
Note that, $\stab_{+,1}\mid_{\sfp_2} = 0$ and $\stab_{+,2}\mid_{\sfp_1} = -\hbar$. These both are of complex degree $0$, which is consistent with the degree axiom~\eqref{degAxiom}. With the right choice of the overall signs, the final expressions for the stable envelopes match those given earlier in~\eqref{stab+w1}.

One can repeat the same calculation in the opposite chamber $\mathfrak{C}_-$. In this case, the attracting and repelling manifolds are summarized in Figure~\ref{fig:attrP1}. Following the same steps as above, we end up with the following expressions for the corresponding stable envelopes:
\begin{equation}
    \stab_{-,1}(\sigma,m_1,m_2,\hbar)\,=\,\pm\,(\sigma\,-\,m_2\,-\,\hbar)~, \qquad \stab_{-,2}(\sigma,m_1,m_2,\hbar)\,=\,\mp\,(\sigma\,-\,m_1)~,
\end{equation}
which satisfy the degree axiom~\eqref{degAxiom} and match the expressions provided in~\eqref{stab-w1} upon choosing the right overall sign.

From these explicit forms, one finds that the $\sf R$-matrix for this model is given by~\eqref{RmatDefn}:
\begin{equation}\label{RmatT*P1}
    \mathsf{R}_{+,-}(u,\hbar)\,=\,\frac{1}{u\,+\,\hbar}\,\begin{pmatrix}
        u&\hbar\\\hbar&u
    \end{pmatrix}~, \qquad \mathsf{R}_{-,+}(u,\hbar)\,=\,{\sf R}_{+,-}(u,\hbar)^{-1}\,=\,\frac{1}{u\,-\,\hbar}\,\begin{pmatrix}
        u&-\,\hbar\\-\,\hbar&u
    \end{pmatrix}~,
\end{equation}
where $u\equiv m_1-m_2$.

\medskip
\noindent
\underline{\textbf{Example II: Stable envelopes in $T^*\bP^2$.}} Let us now look at the case of $T^*\bP^2$. In this case, we have the torus action $T = A\times\mathbb{C}^\times_\hbar$ with $A = (\bC^\times)^3$. Upon acting with the torus $A$, we get three fixed points on the base $\bP^2$ which are given by:
\begin{equation}
    \sfp_1\,\equiv\,[1\,:\,0\,:\,0]~, \qquad \sfp_2\,\equiv\,[0\,:\,1\,:\,0]~, \qquad \sfp_3\,\equiv\,[0\,:\,0\,:\,1]~.
\end{equation}
The equivariant cohomology ring of this space is defined in terms of the relations:
\begin{equation}
    (\sigma\,-\,m_1)\,(\sigma\,-\,m_2)\,(\sigma\,-\,m_3)\,=\,0~.
\end{equation}
The three fixed points above correspond to the three zeros of this polynomial equation, respectively. That is, the fixed point $\sfp_i$ corresponds to $\sigma = m_i$ for $i=1, 2, 3$.
The tangent and cotangent fibers at each point are two-dimensional, and the corresponding weights are listed in Table~\ref{tab:weightsP2}.
\begin{table}[t!]
    \centering
    \begin{tabular}{|c||c|c|c|c|}
    \hline
        $i$ & $T_{\sfp_i}\mathbb{P}^2$ & $T_{\sfp_i}^*\bP^2$ \\
        \hline
\hline
$1$ & $m_2\,-\,m_1~,~m_3\,-\,m_1$ & $m_1\,-\,m_2\,-\,\hbar~,~m_1\,-\,m_3\,-\,\hbar$ \\ 
$2$ &  $m_1\,-\,m_2~,~m_3\,-\,m_2$ & $m_2\,-\,m_1\,-\,\hbar~,~m_2\,-\,m_3\,-\,\hbar$ \\
$3$ &  $m_1\,-\,m_3~,~m_2\,-\,m_3$ & $m_3\,-\,m_1\,-\,\hbar~,~m_3\,-\,m_2\,-\,\hbar$ \\
\hline
    \end{tabular}
    \caption{For each $A$-fixed point $\sfp_i\in\mathbb{P}^2$, we list the corresponding weights of the tangent and cotangent fibers.}
    \label{tab:weightsP2}
\end{table}

Let us now consider all the possible six chambers, and the corresponding stable envelopes are summarized in the following table:
\renewcommand{\arraystretch}{1.3}
\begin{longtable}{|c|| c | c |}
\hline
$\mathfrak{C}$ & Attractors & $\stab_\mathfrak{C}$\\ 
\hline
$m_1 > m_2 > m_3$ & 
\includegraphics[valign=c, scale=0.45]{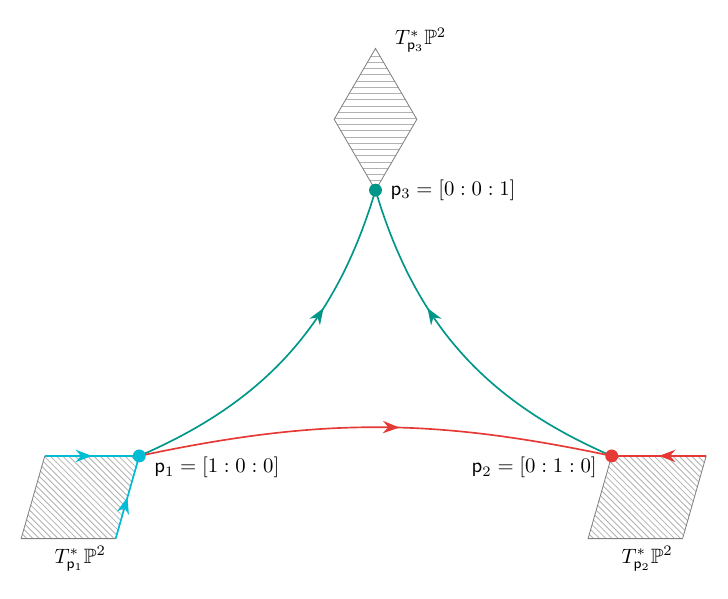} & 
$\begin{aligned}
   &\stab_{\mathfrak{C},1} \,=\, (\sigma\,-\,m_2 )\,(\sigma\,-\,m_3) \\
  &\stab_{\mathfrak{C},2} \,=\, (\sigma\,-\,m_1\,-\,\hbar)\,(\sigma\,-\,m_3)\\
  &\stab_{\mathfrak{C},3} \,=\, (\sigma\,-\,m_1\,-\,\hbar)\,(\sigma\,-\,m_2\,-\,\hbar)
\end{aligned}$ \\
\hline
$m_3 > m_2 > m_1$ & 
\includegraphics[valign=c, scale=0.5]{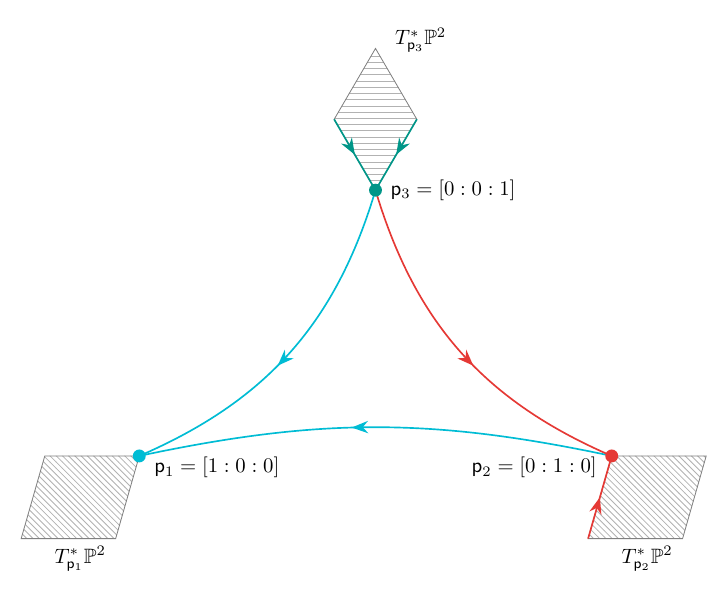} & 
$\begin{aligned}
   &\stab_{\mathfrak{C},1} \,=\, (\sigma\,-\,m_2\,-\,\hbar )\,(\sigma\,-\,m_3\,-\,\hbar) \\
  &\stab_{\mathfrak{C},2}\,=\, (\sigma\,-\,m_1)\,(\sigma\,-\,m_3\,-\,\hbar )\\
  &\stab_{\mathfrak{C},3}\,=\, (\sigma\,-\,m_1)\,(\sigma\,-\,m_2)
\end{aligned}$ \\
\hline
$m_1> m_3 > m_2$ & 
\includegraphics[valign=c, scale=0.5]{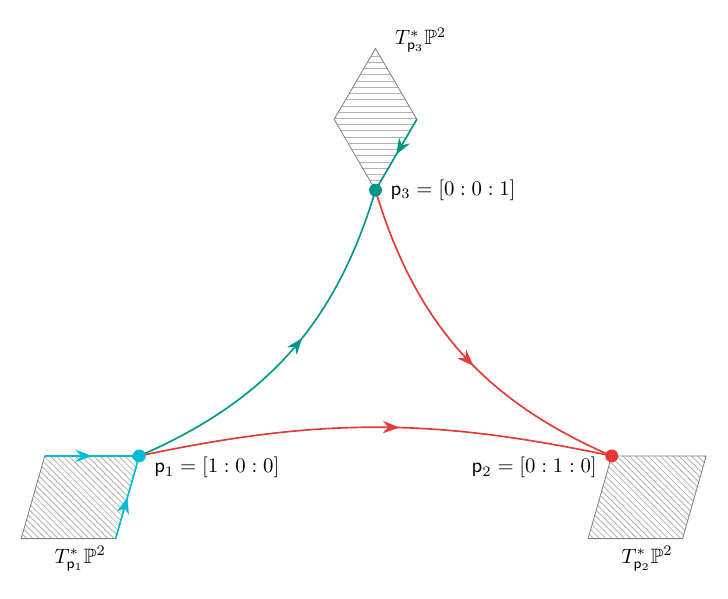} & 
$\begin{aligned}
   &\stab_{\mathfrak{C},1} \,=\, (\sigma\,-\,m_2 )\,(\sigma\,-\,m_3) \\
  &\stab_{\mathfrak{C},2} \,=\, (\sigma\,-\,m_1\,-\,\hbar)\,(\sigma\,-\,m_3\,-\,\hbar )\\
  &\stab_{\mathfrak{C},3} \,=\, (\sigma\,-\,m_1\,-\,\hbar)\,(\sigma\,-\,m_2)
\end{aligned}$ \\
\hline
$m_2> m_3 > m_1$ & 
\includegraphics[valign=c, scale=0.5]{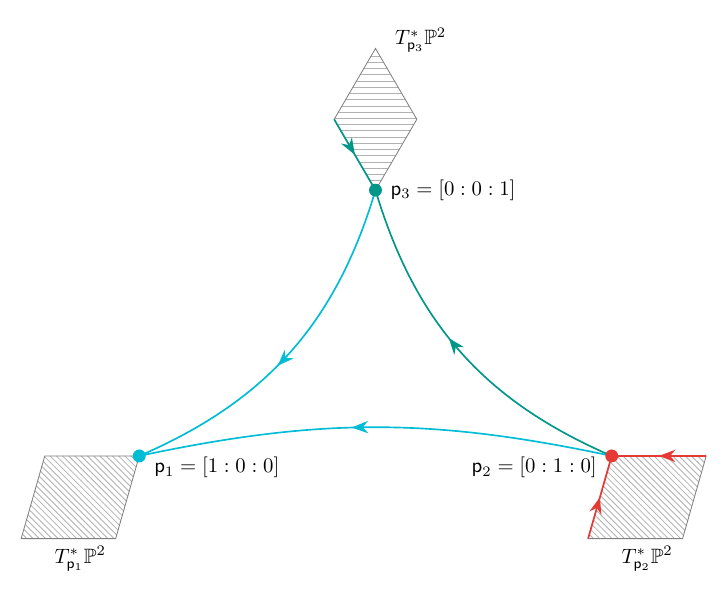} & 
$\begin{aligned}
   &\stab_{\mathfrak{C},1} \,=\, (\sigma\,-\,m_2\,-\,\hbar)\,(\sigma\,-\,m_3\,-\,\hbar) \\
  &\stab_{\mathfrak{C},2} \,=\, (\sigma\,-\,m_1)\,(\sigma\,-\,m_3 )\\
  &\stab_{\mathfrak{C},3} \,=\, (\sigma\,-\,m_1)\,(\sigma\,-\,m_2\,-\,\hbar)
\end{aligned}$ \\
\hline
$m_2> m_1 > m_3$ & 
\includegraphics[valign=c, scale=0.5]{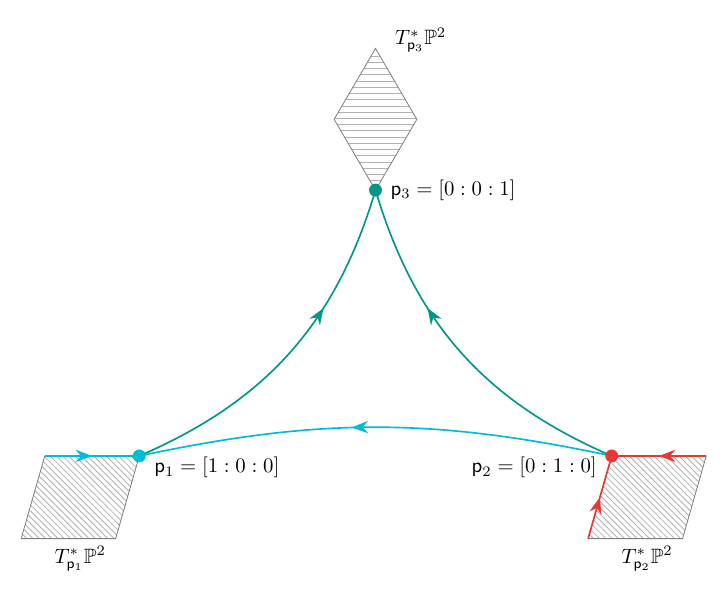} & 
$\begin{aligned}
   &\stab_{\mathfrak{C},1}\,=\, (\sigma\,-\,m_2 \,-\,\hbar)\,(\sigma\,-\,m_3) \\
  &\stab_{\mathfrak{C},2} \,=\, (\sigma\,-\,m_1)\,(\sigma\,-\,m_3 )\\
  &\stab_{\mathfrak{C},3} \,=\, (\sigma\,-\,m_1\,-\,\hbar)\,(\sigma\,-\,m_2\,-\,\hbar)
\end{aligned}$ \\
\hline
$m_3> m_1 > m_2$ & 
\includegraphics[valign=c, scale=0.5]{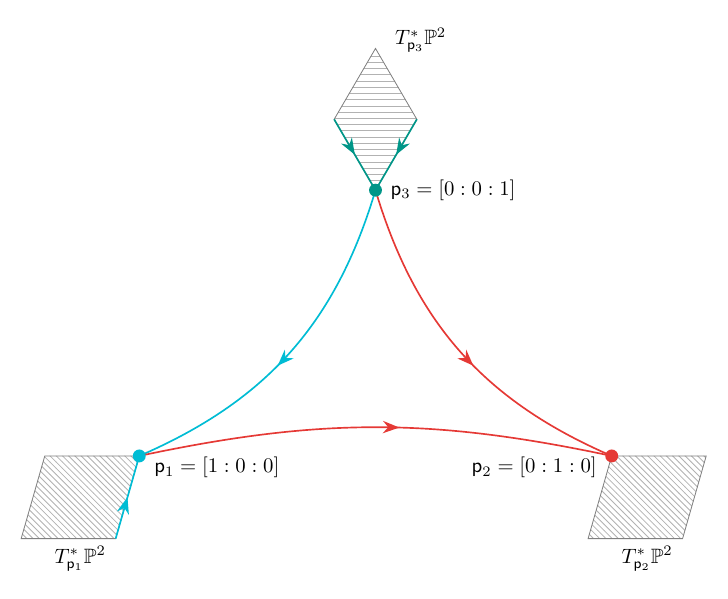} & 
$\begin{aligned}
   &\stab_{\mathfrak{C},1} \,=\, (\sigma\,-\,m_2 )\,(\sigma\,-\,m_3\,-\,\hbar) \\
  &\stab_{\mathfrak{C},2} \,=\, (\sigma\,-\,m_1\,-\,\hbar)\,(\sigma\,-\,m_3\,-\,\hbar )\\
  &\stab_{\mathfrak{C},3} \,=\, (\sigma\,-\,m_1)\,(\sigma\,-\,m_2)
\end{aligned}$ \\
\hline
\end{longtable}

It is straightforward to check that these final expressions do satisfy the degree axiom~\eqref{degAxiom}. Moreover, one can check that, starting with the stable envelopes associated with the second chamber in the table, the stable envelopes in the other chambers can be worked out it via~\eqref{stabpi=stab-}. 

Finally, one can work out the explicit forms of the $\mathsf{R}$-matrices when moving from one chamber to the other. There are $30$ such matrices. Using the composition property~\eqref{composR}, these can be constructed from the following $6$ basic matrices:
\begingroup
\allowdisplaybreaks[1]
\begin{align}\label{RmatT*P2}
 \mathsf{R}_{(1\,2\,3),(2\,1\,3)}\,=\,\mathsf{R}_{(3\,1\,2),(3\,2\,1)}\,
    &=\,
    \frac{1}{u\,-\,\hbar}
    \begin{pmatrix}
        u & -\,\hbar & 0\\
        -\,\hbar & u & 0\\
        0 & 0 & u\,-\,\hbar
    \end{pmatrix}~,
    \notag\\[6pt]
    \mathsf{R}_{(1\,2\,3),(1\,3\,2)}\,=\,\mathsf{R}_{(2\,3\,1),(3\,2\,1)}
    &\,=\,
    \frac{1}{v\,-\,\hbar}
    \begin{pmatrix}
        v\,-\,\hbar & 0 & 0\\
        0 & v & -\,\hbar\\
        0 & -\,\hbar & v
    \end{pmatrix}~,
    \notag\\[6pt]
    \mathsf{R}_{(1\,3\,2),(3\,1\,2)}\,=\,\mathsf{R}_{(2\,1\,3),(2\,3\,1)}
    &\,=\,
    \frac{1}{w\,-\,\hbar}
    \begin{pmatrix}
        w & 0 & -\,\hbar\\
        0 & w\,-\,\hbar & 0\\
        -\,\hbar & 0 & w
    \end{pmatrix}~.
\end{align}
\endgroup
where, for compactness, we defined:
\begin{equation}
    u\,\equiv\,m_1\,-\,m_2~,
    \qquad
    v\,\equiv\,m_2\,-\,m_3~,
    \qquad
    w\,\equiv\,m_1\,-\,m_3\,=\,u\,+\,v~.
\end{equation}
Here, $w$ denotes a mass difference, rather than a fixed-point label.

In particular, the two minimal paths from $\mathfrak C_+$ to
$\mathfrak C_-$ yield:
\begin{equation}\label{YBE}
  \mathsf{R}_{(1\,2\,3),(2\,1\,3)}\,\circ\, \mathsf{R}_{(2\,1\,3),(2\,3\,1)}\,\circ\, \mathsf{R}_{(2\,3\,1),(3\,2\,1)}\,
    =\,
    \mathsf{R}_{(1\,2\,3),(1\,3\,2)}\,\circ\,\mathsf{R}_{(1\,3\,2),(3\,1\,2)}\,\circ\,\mathsf{R}_{(3\,1\,2),(3\,2\,1)}~,
\end{equation}
which both are equal to $\mathsf{R}_{(1\,2\,3),(3\,2\,1)}$.

\subsubsection{Deriving the stable envelope formula for the more general case}
Let us now give a derivation of the formula for the stable envelope~\eqref{stab-w}. Take $w = (w_1\,w_2\,\cdots\,w_k\,\cdots)\in {\rm W}^{(k;n)}$. The way we will derive the formula is by using the abelian/non-abelian correspondence. For a recent review, see~\cite{Brunner:2026fxy}.

\medskip
\noindent
\textbf{First step of the abelianization.}
As a starting point, consider the following space $Y$:
\begin{equation}
    Y \,\coloneqq\, \underbrace{T^*\bP^{n-1} \,\times\, \cdots\,\times \,T^* \bP^{n-1}}_{k\,{\rm times}}~.
\end{equation}
Symmetry group $T=(\bC^\times)^{n}\times \bC_{\hbar}^\times$ acts diagonally on the $k$ copies of $T^*\bP^{n-1}$ with equivariant parameter $m_i,\hbar$. Recall that for a single copy of $T^{*}\bP^{n-1}$, the normal bundle of the Schubert cell indexed by permutation $w' = (w'_1\,\cdots)$ is given by:\footnote{Here, we take the chamber to be $\mathfrak{C}_-$.}
\begin{equation}\label{conormal schubert class of T*P}
    e_{T}({\cal N}_{w'_1}) \,=\, \prod_{\alpha=1}^{w'_1-1}(\sigma\,-\,m_\alpha)\,\prod_{\gamma=w'_1+1}^{n}(\sigma\,-\,m_\gamma\,-\,\hbar)~,
\end{equation}
where $\sigma$ is the first Chern class of tautological bundle of $\bP^{n-1}$, ${\cal N}_{w'_1}$ is the normal bundle of the Schubert cell $C_{w'_1}$, $\prod_{\alpha=1}^{w'_1-1}(\sigma-m_\alpha)$ comes from the component of ${\cal N}_{w'_1}$ along the base $\bP^{n-1}$ and $\prod_{\gamma=w'_1+1}^{n}(\sigma-m_\gamma-\hbar)$ is the contribution of the cotangent fibers. With this in mind, let us introduce the following cohomology class:
\begin{equation}
    f(\sigma,m,\hbar) \,:=\, \prod_{a=1}^{k}e_T({\cal N}_{w_a})(\sigma_a) \,\in\, H_T{^\bullet}(Y)~,
\end{equation}
where $\sigma_a$ is the hyperplane class in the $a$-th $T^*\bP^{n-1}$ copy.

\medskip
\noindent
\textbf{Second step of the abelianization.} For simplicity, we first ignore the cotangent fiber and focus on the ${\rm Gr}(k,n)$. Recall that this admits the following GIT quotient:
\begin{equation}
    {\rm Gr}(k,n) \,= \,{\sf Hom}(\bC^k,\bC^n)/{GL}(k;\bC)~.
\end{equation}
Taking the maximal $(\bC^\times)^k\subseteq {GL}(k;\bC)$, we can consider the following GIT quotient space:
\begin{equation}
 X\,\coloneqq\,   {\sf Hom}(\bC^k,\bC^n)/(\bC^\times)^k \, =\, \bigoplus_{i=1}^k  {\sf Hom}(\bC,\bC^n)/\bC^\times \,=\, \underbrace{\bP^{n-1}\,\times\, \cdots\,\times\, \bP^{n-1}}_{k\,{\rm times}}~,
\end{equation}
where the second equality follows from the diagonal action of the maximal torus on $\bC^k$. Both ${\rm Gr}(k,n)$ and $X$ admit an $A\equiv (\bC^{*})^n$ action which is compatible with the following map:
\begin{equation}\label{map from X to Gr}
    {GL}(k;\bC)/(\bC^\times)^k \,\hookrightarrow \,X\, \to\, {\rm Gr}(k,n)~.
\end{equation}

The Schubert class $[C_w]$ is the attracting $A$-orbit of (neighborhood) of the $T$-fixed point ${\sf p}_w$, and:
\begin{equation}
    [C_w] \,=\, e_A(\widetilde{{\cal N}}_{w})\, \in\, H^\bullet_{A}({\rm Gr}(k,n))~.
\end{equation}
Here, recall that $\widetilde{\cal N}_w$ is the normal bundle to $C_w\subseteq {\rm Gr}(k,n)$.
From~\eqref{map from X to Gr}, we see that a fixed point can be lifted to a fixed point $\tilde{\sf p}_w\in X$ and the normal bundle to its attracting $A$-orbit is:
\begin{equation}\label{decomposition of the normal bundle}
    \widetilde{\cal N}_{w} \,\oplus\, V~,
\end{equation}
where $V$ is the summand in the tangent space along ${GL}(k;\bC)\times(\bC^\times)^k$. The tangent space of ${GL}(k;\bC)/(\bC^\times)^k$ is labeled by the roots of root system $\Delta$ of $\mathfrak{sl}(k;\bC)$ and the elements of the Cartan space $(\bC^\times)^k$ are labeled by $\sigma_1,\cdots,\sigma_k$. The root system can be decomposed into:
\begin{equation}
    \Delta \,=\, \Delta_{+} \,\sqcup\, \Delta_-~,
\end{equation}
 where $\Delta_-$ labels attracting orbits and $\Delta_+$ labels the normal bundle to it. Hence, we have $V = \Delta_+$.

From this, we naively have:
\begin{equation}
``e_{A}(\widetilde{\cal N}_w)" \,=\,  \frac{\prod_{a=1}^{k}\prod_{\alpha_a=1}^{w_a-1}(\sigma_a\,-\,m_{\alpha_a})}{\prod_{\alpha \in \Delta_+}\alpha(\sigma)}\,=\, \frac{\prod_{a=1}^{k}\prod_{\alpha_a=1}^{w_a-1}(\sigma_a\,-\,m_{\alpha_a})}{\prod_{1\leq a <b\leq k} (\sigma_a\,-\,\sigma_b)}~.
\end{equation}
We call it naively since we have not taken the Weyl group $W\equiv S_k$ action, which arises from the global structure of the fibration~\eqref{map from X to Gr}.\footnote{Another way of viewing it,  the lift of a fixed point ${\sf p}_w \in {\rm Gr}(k,n)$ has $|W| = k!$ fixed points in $X$ in one to one correspondence with the Weyl group elements. Pushing these fixed points from $X$ down to ${\rm Gr}(k,n)$ corresponds to taking the $S_k$ invariant part.} The Weyl group acts as permutation group $S_k$ on $\sigma_1,\cdots,\sigma_k$ and the final result of $e_{A}(\widetilde{\cal N}_{w})$ is given by taking the $S_k$-invariant part of above formula:
\begin{equation}
    e_{A}(\widetilde{\cal N}_{w}) \,=\,\sum_{\pi\in S_k} \frac{\prod_{a=1}^{k}\prod_{\alpha_a=1}^{w_a-1}(\sigma_{\pi(a)}\,-\,m_{\alpha_a})}{\prod_{1\leq a <b\leq k} (\sigma_{\pi(a)}\,-\,\sigma_{\pi(b)})} \,=\, \operatorname*{\mathsf{Sym}}_{\sigma_\bullet}\left[ \frac{\prod_{a=1}^{k}\prod_{\alpha_a=1}^{w_a-1}(\sigma_a\,-\,m_{\alpha_a})}{\prod_{1\leq a <b\leq k} (\sigma_a\,-\,\sigma_b)}\right]~,
\end{equation}
which is none other than the double Schubert polynomial~\eqref{doubleSchubert} representing the Schubert class $[C_w]\in H^\bullet_A({\rm Gr}(k,n))$.

We can easily generalize the discussion to cotangent bundles. We have instead:
\begin{equation}\label{map from Y to TGr}
     T^*({GL}(k;\bC)/(\bC^\times)^k) \,\hookrightarrow \,Y\, \longrightarrow\, T^*{\rm Gr}(k,n)~,
\end{equation}
with only two modifications:
\begin{itemize}
    \item The symmetry group is enlarged from $A = (\bC^\times)^n$ to $T = A\times\bC_\hbar^\times$, with an additional $\bC_{\hbar}^\times$ rotating the fiber.
    \item The space $V$ in~\eqref{decomposition of the normal bundle} now becomes $\Delta_{+}\oplus \left(\Delta_{-}\otimes \bC_{\hbar}\right)$.
\end{itemize}

Hence, similar to the discussion of ${\rm Gr}(k,n)$. For the conormal bundle of Schubert class $[C_w]$ (i.e. attractor manifold) inside $T^*{\rm Gr}(k,n)$, we have:
\begin{equation}
    e_T({\cal N}_{\mathcal{C}_w}) \,=\, \operatorname*{\mathsf{Sym}}_{\sigma_\bullet}\left[ \frac{ \prod_{a=1}^{k}e_T(\mathcal{N}_{\mathcal{C}_{w_a}})}{ e_{T}(\Delta_{+}\oplus (\Delta_{-}\otimes \bC_{\hbar}))}\right]\,=\, \operatorname*{\mathsf{Sym}}_{\sigma_\bullet}\left[\frac{f(\sigma,m,\hbar)}{\prod_{1\leq a<b\leq k}\left(\sigma_b\,-\,\sigma_a\right)\,\left(\sigma_a\,-\,\sigma_b\,-\,\hbar\right)}\right]~,
\end{equation}
which is none other than the stable envelope~\eqref{stab-w} up to a choice of polarization.

\medskip
\noindent
\textbf{Quantum cup product involving the stable envelopes.} 
Finally, we turn to the quantum cup products. In particular, we consider the quantum multiplication of the first Chern class $D_\lambda = c_1(L_\lambda)$ with the stable envelope of $T^*{\rm Gr}(k,n)$, where $L_\lambda$ is the line bundle
associated with a character $\lambda$ of the Grassmannian
parabolic subgroup. We choose the ${GL}(n;\mathbb C)$ weight $\lambda=(1^k,0^{n-k})$, so that $L_\lambda=\det \mathcal S$ and $D_\lambda=\sum_{a=1}^k\sigma_a$. From the general formula of~\cite[Theorem~(1.1)]{SU2016362}, for a permutation $w = (w_1\,w_2\,\cdots\,w_k\,\cdots)\in {\rm W}^{(k;n)}$ we learn that:\footnote{\cite[Theorem~(1.1)]{SU2016362} gives the action in the positive chamber. Duality of the opposite stable bases and self-adjointness of divisor multiplication give the negative-chamber formula.}
\begin{equation}\label{eq:Su_T^astGr}
\begin{aligned}
&D_\lambda\,\cup_\sfq\,\stab_{-,w}
=\left(\sum_{i=1}^{k}m_{w_i} -\hbar\min(k,n-k)\frac{\sfq}{1-\sfq}
\right)\stab_{-,w}
\\
&-\hbar \sum_{\substack{1\leq a\leq k<b\leq n\\ w_a<w_b}}
(-1)^{\mathrm{sign}(w)-\mathrm{sign}(w_{ab})}
\stab_{-,w_{ab}}
-\hbar\frac{\sfq}{1-\sfq}
\sum_{1\leq a\leq k<b\leq n}
(-1)^{\mathrm{sign}(w)-\mathrm{sign}(w_{ab})}
\stab_{-,w_{ab}}\,,
\end{aligned}
\end{equation}
where $w_{ab}$ denotes the permutation obtained by exchanging $w_a$ and $w_b$
in $w$ and then sorting the first $k$ and the remaining $n-k$ entries in increasing order separately. In Section~\ref{sec:quantumProd}, we will check this formula from the 2d A-model for a few examples.

\subsection{K-theoretic stable envelopes}
Our discussion in the above subsection can be generalized to a K-theoretic setting. Fixing a chamber $\mathfrak{C}_\pi$ as before, one can define a K-theoretic stable envelope map:
\begin{equation}
    \kstab_{\mathfrak{C}_\pi}\,:\,K_T(X^A_{k,n})\,\longrightarrow\,K_T(X_{k,n})~,
\end{equation}
satisfying a particular set of axioms. This has been done in~\cite{rimanyi2015trigonometric} for the cotangent bundle of any partial flag manifold. We will not go through the details of these axioms and how one can use them to compute an explicit expression for $\kstab_{\mathfrak{C}}$. We invite the reader to consult~\cite[Section 3]{rimanyi2015trigonometric} for the details. Here, we will just review the final result and discuss its form in some special limits.

\subsubsection{Chern character of the K-theoretic stable envelope}
For any fixed point $\sfp_w\in X_{k,n}^A$, we will denote the $T$-equivariant Chern character of its image by $\kstab_{\pi, w}$. More explicitly:
\begin{equation}
    \kstab_{\pi, w}\,\equiv\, {\rm ch}_T(\kstab_{\mathfrak{C}_\pi}(\sfp_w))\,\in\,H_T^\bullet(X_{k,n})~.
\end{equation}
\paragraph{Remark} Due to the loss of grading structure in K-theory, the degree axiom for cohomological stable envelopes~\eqref{degAxiom} is replaced by the inclusion of Newton polygons~\cite[Subsection 9.1.8]{Okounkov:2015spn}. This introduces the slope dependence (i.e., choice of a fractional line bundle) for K-theoretic envelopes. In the stable envelope considered in this paper, we take the line bundle to be trivial.\footnote{K-theoretic stable envelope corresponding to a trivial line bundle may no longer be unique, but they are all legitimate.}

In the construction of $K$-theoretic envelope, 
For the identity permutation (which we also denote by $+$), the explicit form of this Chern character is written as:
\begin{align}\label{kstab+w}
      \kstab_{+,w}(x,y,{\sf y})\,=\,(-\mathsf{y})^{-\varepsilon(w)}\,\operatorname*{\mathsf{Sym}}_{x_\bullet}\left[\frac{\prod_{a=1}^k\left[\prod_{\alpha_a=1}^{w_{a}-1}\left(1\,-\,\frac{x_a}{y_{\alpha_a}\,{\sf y}}\right)\,\prod_{\gamma_a=w_{a}+1}^n\left(1\,-\,\frac{x_a}{y_{\gamma_a}}\right)\right]}{\prod_{a<b}\left(1\,-\,\frac{x_a}{x_b}\right)\,\left(1\,-\,\frac{x_b}{x_a\,\mathsf{y}}\right)}\right]~,
\end{align}
where the symmetrization operator is defined in~\eqref{sym-op}. We recall here that $x_a\equiv e^{\boldsymbol{\beta}\sigma_a}$ are the K-theoretic Chern roots of the tautological vector bundle $\mathcal{S}$ over Gr$(k,n)$, $y_i$ and ${\sf y}$ are the equivariant characters of the torus $T$ action. The overall factor is added by hand to match the defect calculations in the next section with $\varepsilon(w)$ defined as follows:
\begin{equation}\label{vareps(w)}
    \varepsilon(w)\,\coloneqq\,\sum_{a=1}^{k} (w_a\,-\, a^2) \,-\,k\,w_1~.
\end{equation}
In contrast, in the opposite chamber, this takes the following form:
\begin{align}\label{kstab-w}
      \kstab_{-,w}(x,y,{\sf y})\,=\,(-\mathsf{y})^{-\varepsilon(w)}\,\operatorname*{\mathsf{Sym}}_{x_\bullet}\left[\frac{\prod_{a=1}^k\left[\prod_{\alpha_a=1}^{w_{a}-1}\left(1\,-\,\frac{x_a}{y_{\alpha_a}}\right)\,\prod_{\gamma_a=w_{a}+1}^n\left(1\,-\,\frac{x_a}{y_{\gamma_a}\,{\sf y}}\right)\right]}{\prod_{a<b}\left(1\,-\,\frac{x_b}{x_a}\right)\,\left(1\,-\,\frac{x_a}{x_b\,\mathsf{y}}\right)}\right]~.
\end{align}
Here $(-\mathsf{y})^{-\epsilon(w)}$ is related to the polarization choice in K-theoretic stable envelopes~\cite{Okounkov:2015spn}.

Specializing to the case $T^*\bP^{n-1}$, the Chern characters~\eqref{kstab+w} and~\eqref{kstab-w} become:
\begin{equation}\label{kstabw1}
    \begin{split}
        &\kstab_{+,w_1}(x,y,{\sf y})\,=\,(-\mathsf y)\,\prod_{\alpha=1}^{w_1-1}\left(1\,-\,\frac{x}{y_{\alpha}\,{\sf y}}\right)\,\prod_{\gamma=w_{1}+1}^n\left(1\,-\,\frac{x}{y_{\gamma}}\right)~,\\
        &\kstab_{-,w_1}(x,y,{\sf y})\,=\,(-\mathsf y)\,\prod_{\alpha=1}^{w_1-1}\left(1\,-\,\frac{x}{y_{\alpha}}\right)\,\prod_{\gamma=w_{1}+1}^n\left(1\,-\,\frac{x}{y_{\gamma}\,{\sf y}}\right)~.
    \end{split}
\end{equation}

\medskip
\noindent
\textbf{The Grassmannian limit.} Let us look at the limit where we take $\hbar\to\infty$ (i.e., ${\sf y}\rightarrow \infty$). In this case, the K-theoretic stable envelope~\eqref{kstab-w} becomes:
\begin{equation}
   (-{\sf y})^{\varepsilon(w)}\, \kstab_{-,w}(x,y,{\sf y})\,=\,\mathfrak{G}_w(x,y)\,+\,O({\sf y}^{-1})~,
\end{equation}
where, the double Grothendieck polynomial $\mathfrak{G}_w(x,y)$ is given by:\footnote{These polynomials can also be worked out in terms of a determinant formula~\cite[Equation (3.69)]{Closset:2023bdr}, or in terms of difference operators as in~\cite[Equation (A.26)]{Closset:2025akk}}
\begin{equation}
    \mathfrak{G}_w(x,y)\,=\,\operatorname*{\mathsf{Sym}}_{x_\bullet}\left[\frac{\prod_{a=1}^k\,\prod_{\alpha_a=1}^{w_{a}-1}\left(1\,-\,\frac{x_a}{y_{\alpha_a}}\right)}{\prod_{a<b}\left(1\,-\,\frac{x_a}{x_b}\right)}\right]~.
\end{equation}
This is the $A$-equivariant Chern character of the structure sheaf $[\cO_w]\in K_T({\rm Gr}(k,n))$ on the Schubert subvariety indexed by the Grassmannian permutation $w$. 

\medskip
\noindent
\textbf{The cohomological limit.} Another interesting limit for us to consider is the cohomological one. In our setting, this is the $\boldsymbol{\beta}\rightarrow 0$ limit. Applying this to~\eqref{kstab+w} and~\eqref{kstab-w}, we find:
\begin{equation}
\begin{split}
    &\kstab_{+,w}(x,y,{\sf y})\,=\,(-\boldsymbol{\beta})^{k(n-k)}\,\stab_{+,w}(\sigma, m, \hbar)\,+\,\cdots~,\\
    &\kstab_{-,w}(x,y,{\sf y})\,=\,(-\boldsymbol{\beta})^{k(n-k)}\,\stab_{-,w}(\sigma, m, \hbar)\,+\,\cdots~,\\
\end{split}
\end{equation}
where the cohomological counterparts are defined in~\eqref{stab+w} and~\eqref{stab-w}, respectively.

\subsubsection{K-theoretic \texorpdfstring{$\sf KR$}{R}-matrix}
Similar to the discussion we had in the cohomological case~\eqref{stabpi=stab-}, starting with the chamber $\mathfrak{C}_-$, one can work out the explicit form of the $\kstab_{\pi,w}$ for any $\pi\in S_n$ and $w\in {\rm W}^{(k;n)}$. More explicitly, we have:
\begin{equation}\label{kstabpi=kstab-}
    \kstab_{\pi, w}(x, y_1,\cdots,y_n, {\sf y})\,=\,\kstab_{-,w^\pi}(x, y_{\pi w_0(1)},\cdots, y_{\pi w_0(n)}, {\sf y})~.
\end{equation}

More generally, when moving from one chamber to another, the K-theoretic stable maps are connected via the K-theoretic R-matrix:
\begin{equation}
    {\sf KR}_{\pi,\pi'}\,\coloneqq\,\kstab_{\pi}^{-1}\,\circ\,\kstab_{\pi'}\,\in\,{\sf End}(K_T(X_{k,n}^A))\,\otimes\,\mathbb{C}(y,{\sf y})~,
\end{equation}
This R-matrix satisfies the same properties as those for ${\sf R}$ discussed earlier.

\medskip
\noindent
\textbf{Example: for the $T^*\mathbb{P}^1$ model.} As a first example, let us look at the case $X_{1,2}$. In this case, we have two chambers $\mathfrak{C}_\pm$ and the corresponding K-theoretic stable envelopes are given in~\eqref{kstabw1}. From this, we find that the K-theoretic R-matrices are of the form:
\begin{equation}
\begin{split}
        &\mathsf{KR}_{+,-}(y,{\sf y})\,=\,\frac{1}{y_2\,-\,y_1\,\mathsf y}
    \begin{pmatrix}
        y_2\,-\,y_1 & y_1\,(1\,-\,\mathsf y)\\
        y_2\,(1\,-\,\mathsf y) & \mathsf y\,(y_2\,-\,y_1)
    \end{pmatrix}~,\\
    &\mathsf{KR}_{-,+}(y,\mathsf y)\,=\,\mathsf{KR}_{+,-}(y,\mathsf y)^{-1}\,=\,\frac{1}{\mathsf y\, y_2\,-\,y_1}
    \begin{pmatrix}
        \mathsf y\,(y_2\,-\,y_1) & -\,y_1\,(1\,-\,\mathsf y)\\
        -\,y_2\,(1\,-\,\mathsf y) & y_2\,-\,y_1
    \end{pmatrix}~.
\end{split}
\end{equation}
It is straightforward to check that, in the cohomological limit $\boldsymbol{\beta}\rightarrow 0$, these matrices reduce to the $\mathsf R$ matrices listed in~\eqref{RmatT*P1}.

\medskip
\noindent
\textbf{Example: for the $T^*\bP^2$ model.} In contrast, for the case $X_{1,3}$, we get the following basic K-theoretic $R$-matrices:
\begin{align}
    \mathsf{KR}_{(1\,2\,3),(2\,1\,3)}\,=\,\mathsf{KR}_{(3\,1\,2),(3\,2\,1)}\,
    &=\,
    \frac{1}{y_2\,-\,y_1\,\mathsf y}
\begin{pmatrix}
y_2\,-\,y_1 & y_2(1\,-\,\mathsf y) & 0\\
y_1(1\,-\,\mathsf y) & \mathsf y(y_2\,-\,y_1) & 0\\
0 & 0 & y_2\,-\,y_1\,\mathsf y
\end{pmatrix},
    \\[6pt]
    \mathsf{KR}_{(1\,2\,3),(1\,3\,2)}\,=\,\mathsf{KR}_{(2\,3\,1),(3\,2\,1)}
    &\,=\,
   \frac{1}{y_3\,-\,y_2\,\mathsf y}
\begin{pmatrix}
y_3\,-\,y_2\,\mathsf y & 0 & 0\\
0 & y_3\,-\,y_2 & y_3(1\,-\,\mathsf y)\\
0 & y_2(1\,-\,\mathsf y) & \mathsf y(y_3\,-\,y_2)
\end{pmatrix}~,
    \\[6pt]
    \mathsf{KR}_{(1\,3\,2),(3\,1\,2)}\,=\,\mathsf{KR}_{(2\,1\,3),(2\,3\,1)}
    &\,=\,
    \frac{1}{y_3\,-\,y_1\,\mathsf y}
\begin{pmatrix}
y_3\,-\,y_1 & 0 & y_3(1\,-\,\mathsf y)\\
0 & y_3\,-\,y_1\,\mathsf y & 0\\
y_1(1\,-\,\mathsf y) & 0 & \mathsf y(y_3\,-\,y_1)
\end{pmatrix}~.
\end{align}
In particular, the two minimal paths from $\mathfrak C_+$ to
$\mathfrak C_-$ yield:
\begin{multline}
  \mathsf{KR}_{(1\,2\,3),(2\,1\,3)}\,\circ\, \mathsf{KR}_{(2\,1\,3),(2\,3\,1)}\,\circ\, \mathsf{KR}_{(2\,3\,1),(3\,2\,1)}\,
    =\,
    \mathsf{KR}_{(1\,2\,3),(1\,3\,2)}\,\circ\,\mathsf{KR}_{(1\,3\,2),(3\,1\,2)}\,\circ\,\mathsf{KR}_{(3\,1\,2),(3\,2\,1)}~,
\end{multline}
which both are equal to $\mathsf{KR}_{(1\,2\,3),(3\,2\,1)}$. One can also check that, in the cohomological limit, these reduce to~\eqref{RmatT*P2}.

\paragraph{Remark}Similar to the cohomological case, the K-theoretic R-matrices correspond to those arising from $XXZ$ spin chain, see~\cite{Pushkar:2016qvw,Okounkov:2015spn} for some details about it.

\section{Stable envelopes as BPS defects}\label{sec:BPSdefects}

In this section, we propose a construction for the stable envelopes in terms of BPS defects living in the twisted chiral ring of the GLSM. These defects are defined in terms of lower-dimensional supersymmetric quiver gauge theories that are coupled to the fields of the bulk GLSM at the point of insertion. In particular, for the cohomological stable envelopes, these will be BPS \textit{point} defects defined as  0d-2d coupled systems. In contrast, the K-theoretic ones will be BPS \textit{line} defects defined as 1d-3d coupled systems. We also provide a set of evidence supporting our proposal.

\begin{figure}[t]
    \centering
    \includegraphics[width=1\linewidth]{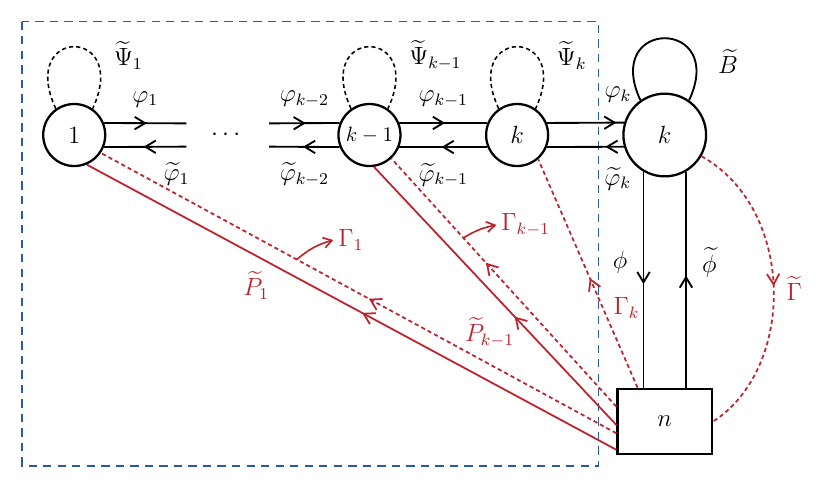}
    \caption{The BPS defect corresponding to $\stab_{-,w}\in H^{2k(n-k)}_T(X_{k,n})$ as a 0d-2d coupled system. The quiver gauge theory in the blue box is the 0d gauge theory coupled to the 2d GLSM represented by the vertical quiver on the r.h.s. The details of the matter contents are given in the main text.}
    \label{fig:BPSDefectDefn}
\end{figure}
\subsection{Stable envelopes as BPS point defects in 2d GLSM}
Fix a Grassmannian permutation $(w_1\,\cdots\,w_k\,\cdots)\in S_n$ and, as before, let us denote the corresponding Schubert cell in the chamber $\mathfrak{C}_-$ by $C_w\equiv C_{-,w}$. In this subsection, we propose a construction for the BPS point defect in the 2d twisted chiral ring that corresponds to $\stab_{-,w}\in H_T(X_{k,n},\bZ)$. The defect will be presented by a 0d-2d coupled system where we have a 0d quiver gauge theory coupled to the fields of the bulk 2d GLSM at the point of insertion. The support of our proposal will be twofold: first, we will argue that the defect restricts the Higgs branch geometry to be the conormal bundle of Schubert cell $C_w$. Second, we will show that the partition function of the 0d gauge theory matches with the form of explicit form for the stable envelope given in~\eqref{stab-w}.

\subsubsection{The proposal}

Our proposal for the defect is given in Figure~\ref{fig:BPSDefectDefn}. Let us elaborate more on the details of the defect. This defect is defined as a 0d $\mathcal{N}=(0,2)$ quiver gauge theory with gauge group:
\begin{equation}
    U(1)\,\times\,U(2)\,\times\,\cdots\,\times\,U(k)~,
\end{equation}
and the following field content:
\begin{itemize}
    \item At each gauge node $U(a)$ along the quiver, there is an $\mathcal{N}=(0,2)$ vector multiplet $\mathcal{V}^{(a)}_{\rm 0d} = (s^{(a)}, \cdots)$ where $s^{(a)}$ is a complex scalar field and the ellipses stand for the gauge field and the fermionic super-partners. Moreover, there is an adjoint Fermi multiplet $\widetilde{\Psi}_a$ whose corresponding $\widetilde{\mathsf{J}}^a$-term will be given shortly. 
    \item Between each two consecutive gauge nodes $U(a)\times U(a+1)$, we have two bifundamental $\mathcal{N}=(0,2)$ chiral multiplets $(\varphi_{a},\cdots)$ and $(\widetilde{\varphi}_a,\cdots)$ respectively living in the representations $(\square_a,\overline{\square}_{a+1})$ and $(\overline{\square}_a, \square_{a+1})$ of the two gauge groups. Between the defect $U(k)_{\rm d}$ and bulk $U(k)_{\rm b}$ gauge groups, we have the $\mathcal{N}=(0,2)$ chiral multiplets $(\varphi_k,\cdots)$ and $(\widetilde{\varphi}_k,\cdots)$ respectively living in the representations $(\square_{k,\rm d},\overline{\square}_{k,\rm b})$ and $(\overline{\square}_{k,\rm d},{\square}_{k,\rm b})$.
    
    These bifundamental chiral multiplets are coupled to the adjoint Fermi multiplets $\widetilde{\Psi}_a$ via $\widetilde{\mathsf{J}}^a\widetilde{\Psi}_a$ with:
    \begin{equation}\label{tildeJterm}
        \widetilde{\mathsf{J}}^a\,\coloneqq\,{\varphi}_a\,\cdot\,\widetilde{\varphi}_a\,-\,\widetilde{\varphi}_{a-1}\,\cdot\,{\varphi}_{a-1}~, \qquad a\,=\,1\,,\,\cdots\,,\,k~.
    \end{equation}
    The product is the matrix product where we view $\varphi_a$ as an $a\times (a+1)$ matrix and $\widetilde{\varphi}_a$ as an $(a+1)\times a$ matrix. 
    \item At each gauge node $U(a)$ along the defect, we have $M_a$ antifundamental $\mathcal{N}=(0,2)$ Fermi multiplets $\Gamma_{a,\alpha^{(a)}}$ that are positively charged under a particular set of $U(1)^{M_a}\subseteq U(n)$ of the bulk flavor symmetry group. The numbers $M_a$ are given by~\cite[Equation (3.63)]{Closset:2023bdr},~\cite[Subsection 5.3]{Closset:2026bnk}:
    \begin{equation}
    M_a \,\coloneqq\,w_{k-a+1}\,-\,w_{k-a}~,\qquad a\,=\,1\,,\,\cdots\,,\,k~,
    \end{equation}
    with the understanding that $w_0\equiv 1$. The range of the index $\alpha^{(a)}$ is given as follows:
    \begin{equation}\label{alpha-a-range}
        \alpha^{(a)}\,\in\,\{w_{k-a}\,,\,w_{k-a}+1\,,\,\cdots\,,\,w_{k-a+1}-1\}~.
    \end{equation}
     The Fermi multiplet $\Gamma_{a,\alpha^{(a)}}$ gives rise to the the following 0d $J$-term:
     \begin{equation}\label{j-a-alpha-a-term}
        \Gamma_{a,\alpha^{(a)}}\,J^a_{\alpha^{(a)}}~, \qquad J^{a}_{\alpha^{(a)}}\,\equiv\,\varphi_{a\rightarrow k}\,\cdot\,\phi_{\alpha^{(a)}}~, 
    \end{equation}    
    where, for compactness, we introduced the notation $\varphi_{a\rightarrow k}\equiv \varphi_a\,\cdot\,\varphi_{a+1}\,\cdot\,\cdots\,\cdot\,\varphi_{k}$ with the product being matrix multiplication.
    \item For $a=1,\cdots, k-1$, we have $\widetilde{M}_a$ $\mathcal{N}=(0,2)$ chiral multiplets $\widetilde{P}_{a,\beta^{(a)}}$ in the antifundamental representation of the $U(a)$ defect gauge group and positively charged under a particular $U(1)^{\widetilde{M}_a}\subset U(n)$ subgroup of the bulk flavor symmetry. The number of those multiplets is given by:
    \begin{equation}
        \widetilde{M}_a\,\coloneqq\,w_{k-a+1}\,-\,w_{k-a}~, \qquad a\,=\,1\,,\,\cdots\,,\,k-1~.
    \end{equation}
    The range of the index $\beta^{(a)}$ is taken as follows:
    \begin{equation}\label{beta-a-range}
        \beta^{(a)}\,\in\,\{w_{k-a}+1\,,\,\cdots\,,\,w_{k-a+1}\}~.
    \end{equation}
    \item At the $U(k)$ bulk gauge node, we also have $n-w_1$ fundamental $\mathcal{N}=(0,2)$ Fermi multiplets $\widetilde{\Gamma}_\beta$ that are negatively charged under $U(1)^{n-w_1}\subset U(n)$ subgroup of the bulk flavor symmetry group. More explicitly, the index $\beta$ has the following range:
    \begin{equation}\label{beta-range}
        \beta\,\in\,\{w_1+1\,,\,w_1+2\,,\,\cdots\,,\,n\}~.
    \end{equation}
    These Fermi multiplets give rise to two sets of 0d superpotentials:
    \begin{equation}\label{tilde-j-a-beta}
        \begin{split}
            &\widetilde{J}_{\beta^{(a)}}^a\,\coloneqq\,\widetilde\phi_{\beta^{(a)}}\,-\,\widetilde{P}_{a,\beta^{(a)}}\,\cdot\,\varphi_{a\rightarrow k}~, \qquad \forall a\,=\,1\,,\,\cdots\,,\,k-1~,\\
            &\widetilde{J}^k_{\beta^{(k)}}\,\coloneqq\,\widetilde{\phi}_{\beta^{(k)}}~, \qquad \qquad\qquad\qquad\quad~~\forall\beta^{(k)}\,=\,w_k\,,\,\cdots\,,\,n~.
        \end{split}
    \end{equation}
    Here, $\widetilde{\phi}$ is viewed as an $n\times k$ matrix and $\widetilde{\phi}_{i}$ denotes its $i$-th row. 
    \end{itemize}

\medskip
\noindent
\textbf{On the $U(1)_\hbar$ charges.} Additionally, the defect fields listed above are also charged under the bulk $U(1)_\hbar$ symmetry group. The charges of the defect fields under the symmetry $U(1)_\hbar$ are summarized in Table~\ref{tab:defectCharges}.

\begin{table}[t]
    \centering
        \renewcommand{\arraystretch}{1.2}
    \begin{tabular}{|c||c|c|c|c|c|c|}
    \hline
        &$\varphi_a$ & $\widetilde{\varphi}_a$ & $\widetilde{\Psi}_a$ & $\widetilde{P}_{a,\beta^{(a)}}$ & $\Gamma_{a,\alpha^{(a)}}$ &$\widetilde{\Gamma}_\beta$\\
        \hline
        \hline
        $U(1)_\hbar$&~$0$~&~$-1$~&~$+1$~&~$+1$~&~0~&~$-1$~\\
\hline
    \end{tabular}
    \caption{The $U(1)_\hbar$ charges of the fields along the BPS defect in Figure~\ref{fig:BPSDefectDefn}.}
    \label{tab:defectCharges}
\end{table}

\subsubsection{Transferring the defects across the walls}
As we discussed earlier around~\eqref{stabpi=stab-}, one can obtain the explicit form of the stable envelope $\stab_{\pi,w}$ in any other chamber $\pi\in S_n$ from its form in the chamber $\mathfrak{C}_-$~\eqref{stab-w}. With this in mind, we can extend our proposal of the BPS defects given in Figure~\ref{fig:BPSDefectDefn} to any chamber $\mathfrak{C}_\pi$. More explicitly, fix a Grassmannian permutation $w\in{\rm W}^{(k;n)}$. The BPS defect corresponding to $\stab_{\pi,w}$ can be constructed as follows:
\begin{itemize}
    \item Consider the BPS defect as in Figure~\ref{fig:BPSDefectDefn} associated with the Grassmannian permutation $w^\pi\in{\rm W}^{(k;n)}$ where $w^\pi$ is the Grassmannian representative of the class $[w_0\pi^{-1}w]$. 
    \item On the indexing sets $\alpha^{(a)}$~\eqref{alpha-a-range}, $\beta^{(a)}$~\eqref{beta-a-range}, and $\beta$~\eqref{beta-range} associated with the permutation $w^\pi$, apply the action of $\pi w_0$. 
\end{itemize}
To summarize, the BPS defect corresponding to $\stab_{\pi,w}$ is given in Figure~\ref{fig:BPSDefectDefn} with the following indexing sets of the defect Fermi multiplets $\Gamma$~\eqref{alpha-a-range}, chiral fields $\widetilde{P}$~\eqref{beta-a-range}, and Fermi multiplets $\widetilde{\Gamma}$~\eqref{beta-range}:
\begin{equation}
    \begin{split}
        \alpha^{(a)}\,&\in\,\{(\pi w_0)(w^\pi_{k-a})\,,\,\cdots\,,\,(\pi w_0)(w^\pi_{k-a+1}\,-\,1)\}~, \qquad a \,=\,1\,,\,\cdots\,,\,k~,\\
        \beta^{(a)}\,&\in\,\{(\pi w_0)(w^\pi_{k-a}\,+\,1)\,,\,\cdots\,,\,(\pi w_0)(w^\pi_{k-a+1})\}~, \qquad a\,=\,1\,,\,\cdots\,,\,k\,-\,1~,\\
        \beta\,&\in\,\{(\pi w_0)(w^\pi_1\,+\,1)\,,\,(\pi w_0)(w_1^\pi\,+\,2)\,,\,\cdots\,,\,(\pi w_0)(n)\}~.
    \end{split}
\end{equation}

\subsubsection{The underlying geometry of the coupled system}
As the first evidence supporting our proposal, let us see in detail how the insertion of the defect at some point restricts the bulk geometry to that of the conormal bundle $\mathcal{C}_{-,w}\subset X_{k,n}$. For that purpose, let us start with the defect with the ``tilded" fields removed. Taking $\zeta^{\rm d}_a$ to be the FI parameter associated with $U(1)\subseteq U(a)$ along the defect, the $D$-term equations for the defect $\varphi$ fields are given by:
\begin{equation}
    \varphi_a\,\cdot\,\varphi_a^\dagger\,-\,\varphi_{a-1}^\dagger\,\cdot\,\varphi_{a-1}\,=\,\zeta_a^{\rm d}\,1_{a\times a}~,\qquad a\,=\,1\,,\,\cdots\,,\,k~.
\end{equation}
As was shown in~\cite[Subsubsection 3.3.2]{Closset:2023bdr}, solving these equations recursively along with the $J$-term equations~\eqref{j-a-alpha-a-term} restricts the form of the bulk fundamental field $\phi$ to be of the form $\phi_w$ given in~\eqref{phiwMatrix}. For this to be the case, we assume the following stability condition:
\begin{equation}\label{stabilityConditions}
    0\,<\,\zeta^{\rm d}_{1}\,<\,\zeta^{\rm d}_{2}\,<\,\cdots\,<\,\zeta^{\rm d}_{k}~.
\end{equation}
Moreover, the forms of the defect fields get restricted as follows:
\begin{equation}\label{varphiatok}
    \varphi_{a\rightarrow k}\,=\,\left(0_{a\times (k-a)}\,\mid\,{\cal D}_a\right)~, \qquad a\,=\,1\,,\,\cdots\,,\,k~,
\end{equation}
where, $\mathcal{D}_a$ is a diagonal $a\times a$ matrix.

\medskip
\noindent
\textbf{Going to the conormal bundle.} The bulk moment map equation restricts the antichiral field $\widetilde{\phi}_w$ to live in the cotangent direction of the Schubert cell $T^*_{C_w}{\rm Gr}(k,n)$. But to get the conormal bundle $\mathcal{C}_{-,w}$, one still needs to impose the conormality constraint. This is where the defect fields $\widetilde{P}_a$ and $\widetilde{\Gamma}$ come into the game. Let us now look at the $J$-term constraints~\eqref{tilde-j-a-beta} coming from the Fermi multiplets $\widetilde{\Gamma}_\beta$:
\begin{itemize}
    \item For $a = 1,\cdots, k-1$, the superpotential $\widetilde{J}_{\beta^{(a)}}^a$ restricts the form of the $\beta^{(a)}$-th row of $\widetilde\phi_w$ as follows:
    \begin{equation}
        (\widetilde\phi_w)_{\beta^{(a)}}\,=\,\widetilde{P}_{a,\beta^{(a)}}\,\cdot\,\varphi_{a\rightarrow k}\,=\,(\underbrace{0\,\cdots\,0}_{k-a}\,\underbrace{\star\,\cdots\,\star}_a)~, \qquad \forall\beta^{(a)}~,
    \end{equation}
    where we used the explicit frozen form of $\varphi_{a\rightarrow k}$~\eqref{varphiatok}.
    \item For $\beta = \beta^{(k)}$, the superpotential $\widetilde J_{\beta^{(k)}}^k$ simply sets the $\beta^{(k)}$-th row of $\widetilde\phi_w$ to zero:
    \begin{equation}
        (\widetilde\phi_w)_{\beta^{(k)}}\,=\,(\underbrace{0\,\cdots\,0}_{k})~, \qquad \forall \beta^{(k)} \,=\,w_k\,,\,\cdots\,,\,n~.
    \end{equation}
    \item For $\beta^{(0)} = 1, \cdots, w_1$, there are no constraints on the form of the $\beta^{(0)}$-th row of $\widetilde\phi_w$, therefore, these rows are left to be free:
    \begin{equation}
        (\widetilde\phi_w)_{\beta^{(0)}}\,=\,(\underbrace{\star\,\cdots\,\star})~,\qquad \forall \beta^{(0)} \,= \,1\,,\, \cdots\,,\,w_1~.
    \end{equation}
\end{itemize}

Now, as a last step, we look at the moment map constraint of the bulk theory; $\phi_w\widetilde\phi_w=0$. Taking the form of $\phi_w$ in~\eqref{phiwMatrix}, we see that this constraint kills the pivotal rows (i.e., $i=w_a$ rows) in $\widetilde\phi_w$. Hence, combining all these observations, we see that the final form of the bulk $\widetilde{\phi}_w$ upon coupling with the BPS defect coincides with that of the conormal bundle $\mathcal{C}_{-,w}$ of the Schubert cell $C_w$ as given in~\eqref{tildePhiwMatrix}.

\medskip
\noindent
\textbf{The defect chiral fields $\widetilde{P}_a$ are now massive.} Note that, upon freezing the form of the $\varphi_{a\rightarrow k}$ as in~\eqref{varphiatok}, the superpotential $\widetilde J^{a}_{\beta^{(a)}}$ in~\eqref{tilde-j-a-beta} can be viewed as a mass term for the chiral multiplet $\widetilde P_{a,\beta^{(a)}}$ forcing them to be integrated out in the IR theory of the defect. This is why, for instance, these chiral multiplets do not give rise to new dimensions on the restricted geometry of the Higgs branch.  We will see this effect at the level of the partition function calculations below. 

\medskip
\noindent
\textbf{But what about $\widetilde\varphi$ and $\widetilde\Psi$?} As was pointed out in~\cite{Gu:2025tda}, the defect construction of~\cite{Closset:2023bdr} for the Schubert cells in the Grassmannian base in fact does not only see the Schubert cell, but rather, it sees its Bott--Samelson resolution in the complete flag manifold Fl$(n)$ spanned by $\varphi$. In the same spirit, our BPS defect here does not see the conormal bundle class $[\mathcal{C}_{-,w}]\in H_T^\bullet(X_{k,n})$ but it sees its resolution class $[\widetilde{\mathcal C}_{-,w}]\in H_T^{k(n-k)}(T^*{\rm Fl}(n))$ where now the $\widetilde{\varphi}$ and the constraints~\eqref{tildeJterm} give rise to the cotangent fibers to Fl$(n)$. See~\cite{RimanyiTarasovVarchenko2014} for more details on this resolution.

\medskip    
\noindent
\textbf{Dimension counting.} With the above explicit construction, let us now compute the dimension of the underlying geometry seen by the defect. We have:\footnote{The chiral multiplets increase the dimension, meanwhile the Fermis reduce it since they give extra constraints. Moreover, the gauge symmetries reduce the dimension since they are redundancies of the theory.}
\begin{equation}\label{dimtotal}
    \dim \,=\,\underbrace{2\,k\,(n\,-\,k)}_{T^*{\rm Gr}(k,n)}\,-\,\dim({\rm gauge})\,+\,\dim\varphi\,-\,\dim \Gamma\,+\,\dim \widetilde{\varphi}\,-\,\dim\widetilde{\Gamma}\,+\,\dim \widetilde{P}\,-\,\dim{\widetilde\Psi}~,
\end{equation}
where:
\begin{equation}
\begin{split}
      &\dim ({\rm gauge})\,=\,\sum_{a=1}^k\,a^2~,\\
      &\dim\varphi\,=\,\sum_{a=1}^{k-1} a\,(a\,+\,1)\,+\,k^2\,=\,\sum_{a=1}^ka^2\,+\,\sum_{a=1}^{k-1}a~,\\
      &\dim\Gamma\,=\,\sum_{a=1}^k a\,M_a\,=\,\sum_{a=1}^{k-1}a\,(w_{k-a+1}\,-\,w_{k-a})\,+\,k\,(w_1\,-\,1)\,=\,\sum_{a=1}^kw_a\,-\,k~,
      \\
      &\dim\widetilde{\varphi}\,=\,\dim\varphi\,=\,\sum_{a=1}^ka^2\,+\,\sum_{a=1}^{k-1}a~,\\
      &\dim\widetilde{\Gamma}\,=\,k(n\,-\,w_1)~,\\
      &\dim \,\widetilde{P}\,=\,\sum_{a=1}^{k-1}a\,\widetilde{M}_a\,=\,\sum_{a=1}^{k-1}a\,\left(w_{k+1-a}\,-\,w_{k-a}\right)~,\\
      &\dim \widetilde{\Psi}\,=\,\sum_{a=1}^ka^2~.
\end{split}
\end{equation}

Note that:
\begin{equation}
    \dim  \varphi\,-\,\dim \Gamma\,-\,\dim ({\rm gauge})\,=\,-\,\sum_{a=1}^k(w_a\,-\,a)\,=\,-\,\ell(w)~.
\end{equation}
Thus:
\begin{equation}
    \dim {\rm Gr}(k,n)\,+\,\dim  \varphi\,-\,\dim \Gamma\,-\,\dim ({\rm gauge})\,=\,\dim  X_w~,
\end{equation}
with $X_w$ being the Schubert subvariety associated with the permutation $w$. Moreover, we have:
\begin{equation}
    \begin{split}
        -\,\dim \widetilde{\Gamma}\,+\,\dim \widetilde{P}\,=\,-\,kn\,+\,\sum_{a=1}^kw_a~.
    \end{split}
\end{equation}
 Thus, 
\begin{equation}
     \dim  \widetilde{\varphi}\,-\,\dim \widetilde{\Gamma}\,+\,\dim  \widetilde{P}\,-\,\dim \widetilde{\Psi}\,=\,\sum_{a=1}^{k-1} a\,+\,\sum_{a=1}^kw_a\,-\,kn~.
\end{equation}
But, since:
\begin{equation}
    \sum_{a=1}^{k-1} a\,=\,\frac{k\,(k\,-\,1)}{2}\,=\,k^2\,-\,\frac{k\,(k\,+\,1)}{2}\,=\,k^2\,-\,\sum_{a=1}^ka~,
\end{equation}
we have:
\begin{equation}
    \dim  \widetilde{\varphi}\,-\,\dim \widetilde{\Gamma}\,+\,\dim  \widetilde{P}\,-\,\dim \widetilde{\Psi}\,=\,\sum_{a=1}^k(w_a-a)\,-\,k\,(n\,-\,k)\,=\,\ell(w)\,-\,k\,(n\,-\,k)~.
\end{equation}

Putting all this together, we see that the dimension seen by the defect~\eqref{dimtotal} is $k(n-k)$, which is indeed the dimension of the conormal bundle to the Schubert cell $C_w$.

\begin{figure}
    \centering
    \includegraphics[width=0.75\linewidth]{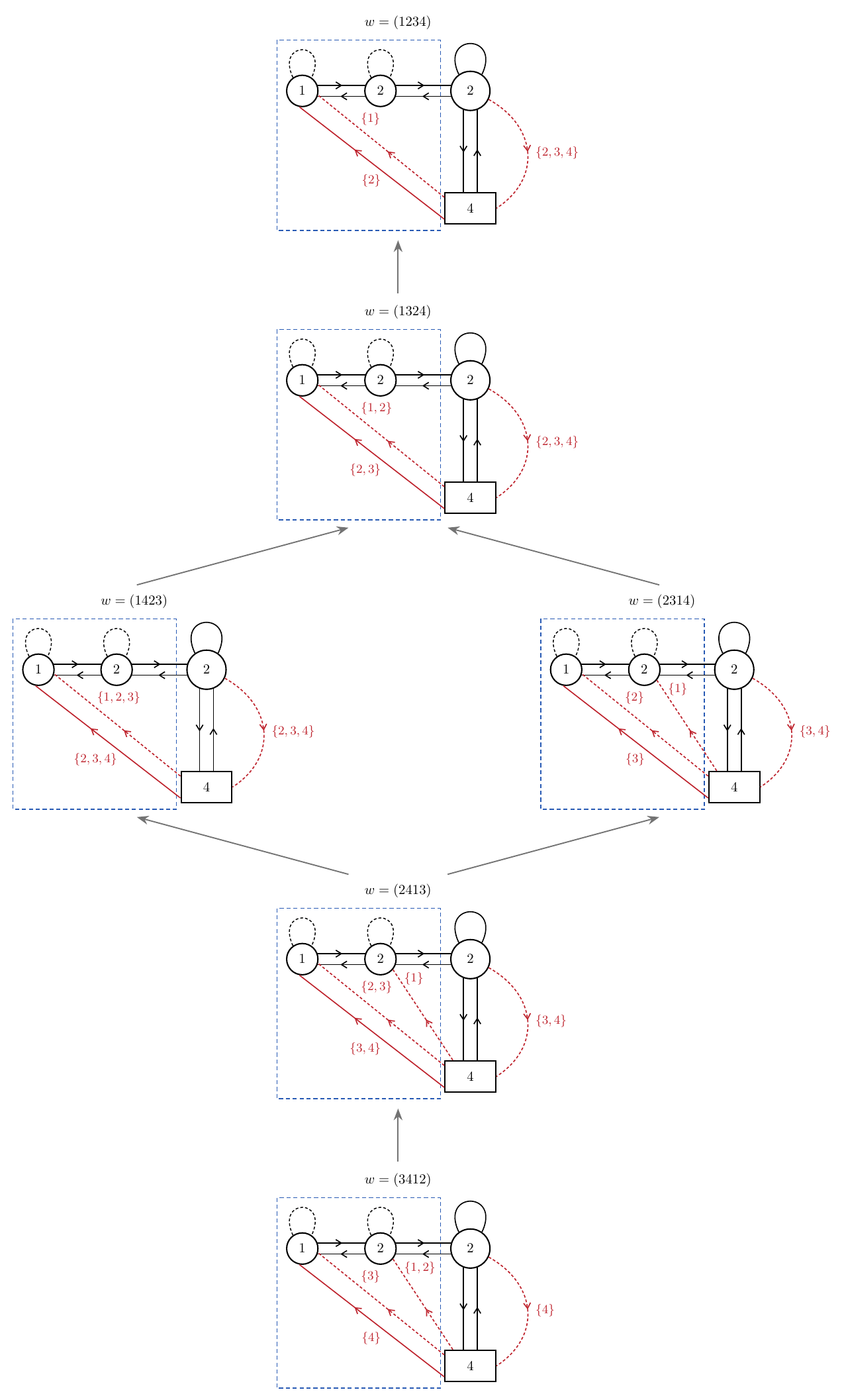}
    \caption{The BPS defects for the $6$ stable envelopes in $T^*{\rm Gr}(2,4)$.}
    \label{fig:defectsGr24}
\end{figure}

In Figure~\ref{fig:defectsGr24}, we add the final form of the proposed BPS defects for the stable envelopes in $T^*{\rm Gr}(2,4)$.

\subsubsection{The partition function of the defect}
As the second supporting evidence for our proposal, we now go through the details of computing the partition function of the 0d-2d coupled system. For a fixed Grassmannian permutation $w\in S_n$, and upon supersymmetric localization, the partition function of the system defined in Figure~\ref{fig:BPSDefectDefn} is given as the following JK residue:
\begin{equation}\label{Z0dw}
    Z_w^{\rm 0d}(\sigma, m, \hbar) \,=\,\oint_{\rm JK}\prod_{a=1}^{k}\left(\frac{ds^{(a)}}{a!\,(2\pi i)^a}\right)\,\Delta^{\rm 0d}\,Z^{\rm 0d}_{\varphi}\,Z_{\Gamma}^{\rm 0d}\,Z_{\widetilde{\varphi}}^{\rm 0d}\,Z_{\widetilde\Psi}^{\rm 0d}\,Z^{\rm 0d}_{\widetilde P}\,Z_{\widetilde{\Gamma}}^{\rm 0d}~.
\end{equation}
It is understood here that $ds^{(a)}\,\equiv\,ds^{(a)}_1\cdots ds_a^{(a)}$ with $(s_1^{(a)}, \cdots, s_a^{(a)})$ being the eigenvalues of the $U(a)$ adjoint complex scalar field $s^{(a)}$ which lives in the vector multiplet $\mathcal{V}_{\rm 0d}^{(a)}$. The different components in the residue integrand are explicitly given by:
\begin{equation}\label{Z0dwcomp}
    \begin{split}
         &\Delta^{\rm 0d}(s)\,=\,\prod_{a=1}^k\,\Delta_a^{\rm 0d}(s)\,\coloneqq\,\prod_{a=1}^k\,\prod_{\substack{\mathsf{a},\mathsf{b}=1\\\mathsf{a}\neq \mathsf{b}}}^a\,\left(s_\mathsf{a}^{(a)}\,-\,s_\mathsf{b}^{(a)}\right)~,\\
          &Z^{\rm 0d}_{\varphi}(s)\,\coloneqq\,\prod_{a=1}^{k-1}\,\left[\,\prod_{\mathsf{a}=1}^{a}\,\prod_{\mathsf{b}=1}^{a+1}\left(s_\mathsf{a}^{(a)}\,-\,s_{\mathsf{b}}^{(a+1)}\right)^{-1}\,\right]\,\prod_{\mathsf{a}=1}^k\prod_{a=1}^k\left(s_\mathsf{a}^{(k)}\,-\,\sigma_a\right)^{-1}~,\\
          & Z_{\Gamma}^{\rm 0d}(s,m)\,\coloneqq\,\prod_{a=1}^{k}\prod_{\mathsf{a}=1}^a\prod_{\alpha^{(a)}=w_{k-a}}^{w_{k-a+1}-1}\left(s^{(a)}_{\mathsf{a}}\,-\,m_{\alpha^{(a)}}\right)~,\\
        &Z^{\rm 0d}_{\widetilde\varphi}(s,\hbar)\,\coloneqq\,\prod_{a=1}^{k-1}\,\left[\,\prod_{\mathsf{a}=1}^{a}\,\prod_{\mathsf{b}=1}^{a+1}\left(s_{\mathsf{b}}^{(a+1)}\,-\,s_{\mathsf{a}}^{(a)}\,-\,\hbar\right)^{-1}\,\right]\,\prod_{a=1}^k\prod_{\mathsf{a}=1}^k\,\left(\sigma_a\,-\,s^{(k)}_{\mathsf{a}}\,-\,\hbar\right)^{-1}~,\\
        &Z_{\widetilde \Psi}^{\rm 0d}(s,\hbar)\,\coloneqq\,\prod_{a=1}^k\,\left[\hbar^a\,\prod_{\substack{\mathsf{a},\mathsf{b}=1\\\mathsf{a}\neq \mathsf{b}}}^a\,\left(s_{\mathsf{a}}^{(a)}\,-\,s_{\mathsf{b}}^{(a)}\,+\,\hbar\right)\right]~,\\
        &Z_{\widetilde P}^{\rm 0d}(s,m,\hbar)\,\coloneqq\,\prod_{a=1}^{k-1}\,\prod_{\mathsf{a}=1}^a\,\prod_{\beta^{(a)}=w_{k-a}+1}^{w_{k-a+1}}\,\left(-\,s_{\mathsf{a}}^{(a)}\,+\,m_{\beta^{(a)}}\,+\,\hbar\right)^{-1}~,\\
        &Z_{\widetilde\Gamma}^{\rm 0d}(s,m,\hbar)\,\coloneqq\,\prod_{{a}=1}^k\,\prod_{\beta=w_1+1}^n\left(\sigma_a\,-\,m_\beta\,-\,\hbar\right)~.
    \end{split}
\end{equation}
The contribution in the first line comes from the 0d `W-bosons' in each gauge vector multiplet along the defect. 

The residue integral is to be performed from left to right along the defect.\footnote{This is just a choice of convention.} As was argued in~\cite{Closset:2023bdr} (see also~\cite{Gu:2025tda,Closset:2025akk,Closset:2026bnk}), with the stability condition~\eqref{stabilityConditions}, the JK residue prescription instructs us to pick up only the poles coming from the chiral bifundamentals $\varphi$ and ignore the ones that we get from $\widetilde{\varphi}$ and $\widetilde{P}_a$.

Our claim here is that the final result of the partition function~\eqref{Z0dw} coincides with the cohomological stable envelope $\stab_{-,w}(\sigma, m, \hbar)$:
\begin{equation}\label{Z0dw=Stabw}
    Z^{\rm 0d}_w(\sigma, m,\hbar)\,=\,\stab_{-,w}(\sigma, m, \hbar)~, \qquad \forall w\,\in\,{\rm W}^{(k;n)}~.
\end{equation}
As a warm-up to proving our claim, we will start with the defects in $T^*\bP^{n-1}$ GLSM. Then, we will prove the claim in full generality. 

\begin{figure}
    \centering
    \includegraphics[width=0.3\linewidth]{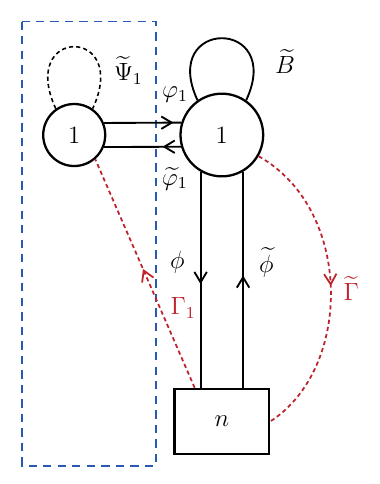}
    \caption{The BPS defect representing the stable envelope $\stab_{-,w}$ in $T^*{\bP}^{n-1}$.}
    \label{fig:BPSDefectTPn}
\end{figure}

\medskip
\noindent
\textbf{The BPS point defects in $T^*\mathbb{P}^{n-1}$.} In this case, there are $n$ possible defects, each indexed by $w = (w_1\cdots)\in {\rm W}^{(1;n)}$ with $w_1 = 1, \cdots, n$. In what follows, we will simply write such a permutation as $w_1$. For such a permutation, our proposed defect in Figure~\ref{fig:BPSDefectDefn} reduces to the one shown in Figure~\ref{fig:BPSDefectTPn} with:
\begin{equation}\label{Gamma1tildeGamma1}
    \Gamma_{1}\,=\,\{\Gamma_{1,1}\,,\,\cdots\,,\,\Gamma_{1,w_1-1}\}~,\qquad \widetilde{\Gamma}_1\,=\,\{\widetilde{\Gamma}_{1,w_1+1}\,,\,\cdots\,,\,\widetilde{\Gamma}_{1,n}\} ~.
\end{equation}

For this defect, the 0d partition function~\eqref{Z0dw} takes the form:
\begin{equation}
    Z_{w_1}^{\rm 0d}(\sigma, m, \hbar)\,=\,\hbar\,\oint_{\rm JK}\,\frac{ds}{2\pi i}\,\frac{\prod_{\alpha=1}^{w_1-1}(s\,-\,m_\alpha)\,\prod_{\beta=w_1+1}^n(\sigma\,-\,m_\beta\,-\,\hbar)}{(s\,-\,\sigma)\,(\,\sigma\,-\,s\,-\,\hbar)}~.
\end{equation}
Here, the overall factor $\hbar$ comes from the Fermi multiplet $\widetilde{\Psi}_1$ and, for ease of notation, $s\equiv s_1^{(1)}$. Since we choose a positive FI parameter $\zeta_1^{\rm d} >0$, we perform the integral by taking only the residue at $s=\sigma$:
\begin{equation}
    Z_{w_1}^{\rm 0d}(\sigma, m, \hbar)\,=\,(-1)\,\times\,\prod_{\alpha=1}^{w_1-1}(\sigma\,-\,m_\alpha)\,\prod_{\beta=w_1+1}^n(\sigma\,-\,m_\beta\,-\,\hbar)~.
\end{equation}
Which matches~\eqref{stab-w1} keeping in mind that $\ell(w) = w_1-1$.

\medskip
\noindent
\textbf{The more general case.}
Let us now prove our claim in full generality by recursively performing the integrals starting with the one at $U(1)$ and moving towards the one associated with $U(k)$ defect gauge node. Let us denote the end result of the residue integral performed at the $U(a)$ node by $\mathcal{Z}^{\rm 0d}_a$. In particular:
\begin{equation}\label{claimI}
    Z_{w}^{\rm 0d}(\sigma, m, \hbar)\,=\,\mathcal{Z}^{\rm 0d}_k(\sigma, m, \hbar)~.
\end{equation}

Starting from the left of the quiver, we have:
\begin{itemize}
    \item At $a=1$, we get:
\begin{equation}
    \mathcal{Z}^{\rm 0d}_1\,=\,\oint_{\rm JK} \frac{ds_1^{(1)}}{2\pi i}\,\frac{\hbar\,\times\,\prod_{\alpha^{(1)}}\left(s_1^{(1)}\,-\,m_{\alpha^{(1)}}\right)}{\prod_{\mathsf{a}=1}^2\left(s_1^{(1)}\,-\,s_{\mathsf{a}}^{(2)}\right)\,\left(s_{\mathsf a}^{(2)}\,-\,s_1^{(1)}\,-\,\hbar\right)\,\prod_{\beta^{(1)}}\left(-\,s_1^{(1)}\,+\,m_{\beta^{(1)}}\,+\,\hbar\right)}~,
\end{equation}
where the range for $\alpha^{(1)}$ and $\beta^{(1)}$ are defined in~\eqref{alpha-a-range} and~\eqref{beta-a-range} for $a=1$, respectively. The JK residue prescription instructs us to pick only the poles $s_1^{(1)} = s_1^{(2)}, s_2^{(2)}$. Doing so, we get:
\begin{equation}\label{Z1}
    \mathcal{Z}^{\rm 0d}_1\,=\,\operatorname*{\mathsf{Sym}}_{s^{(2)}_\bullet}\,\left[\frac{(-1)\,\prod_{\alpha^{(1)}}\left(s_1^{(2)}\,-\,m_{\alpha^{(1)}}\right)}{\prod_{\mathsf{a}<\mathsf{b}}\left(s_{\mathsf{a}}^{(2)}\,-\,s_{\mathsf{b}}^{(2)}\right)\,\left(s_{\mathsf{b}}^{(2)}\,-\,s_{\mathsf{a}}^{(2)}\,-\,\hbar\right)\,\prod_{\beta^{(1)}}\left(-\,s_1^{(2)}\,+\,m_{\beta^{(1)}}\,+\,\hbar\right)}\right]~.
\end{equation}
The symmetrization operator appearing on the r.h.s. is to be performed on both the $U(2)$ Coulomb branch parameters $s_1^{(2)},s_2^{(2)}$.
\item Substituting the above result in the $U(2)$ integral, we get:
\begin{multline}\label{eq:Z20d}
\mathcal{Z}^{\rm 0d}_2\,=\,\oint_{\rm JK}\frac{ds^{(2)}}{2\,(2\pi i)^2}\,\frac{\hbar^2\,\prod_{\mathsf{a}\neq\mathsf{b}}\left(s_{\mathsf{a}}^{(2)}\,-\,s_{\mathsf b}^{(2)}\right)\,\left(s_{\mathsf{a}}^{(2)}\,-\,s_{\mathsf b}^{(2)}\,+\,\hbar\right)}{\prod_{\mathsf{a}=1}^2\prod_{\mathsf{b}=1}^3\left(s_{\mathsf{a}}^{(2)}\,-\,s_{\mathsf{b}}^{(3)}\right)\,\left(s_{\mathsf{b}}^{(3)}\,-\,s_{\mathsf{a}}^{(2)}\,-\,\hbar\right)}\,\times\\\times\,\frac{\prod_{\mathsf{a}=1}^2\prod_{\alpha^{(2)}}\left(s_{\mathsf{a}}^{(2)}\,-\,m_{\alpha^{(2)}}\right)}{\prod_{\mathsf{a}=1}^2\prod_{\beta^{(2)}}\left(-\,s_{\mathsf{a}}^{(2)}\,+\,m_{\beta^{(2)}}\,+\,\hbar\right)}\,\times\,\mathcal{Z}^{\rm 0d}_1~.
\end{multline}
 Performing the JK residue with the choice of FI parameters $\zeta_2 > \zeta_1 >0$ amounts to taking the following poles:\footnote{Note that one needs to also consider the $S_2$ permutation of those poles. But since $\mathcal{Z}^{\rm 0d}_2$ is symmetric w.r.t. the $U(2)$ Coulomb branch parameters, then, these will give the same results as above and this will cancel the $2$ appearing in the measure of the residue integral. Moreover, the apparent poles $\prod_{\mathsf{a}<\mathsf{b}}\left(s_{\mathsf{a}}^{(2)}\,-\,s_{\mathsf{b}}^{(2)}\right)\,\left(s_{\mathsf{b}}^{(2)}\,-\,s_{\mathsf{a}}^{(2)}\,-\,\hbar\right)$ in $\mathcal{Z}^{\rm 0d}_1$ are canceled by the numerator in the first line of~\eqref{eq:Z20d}. These observations will also apply in the calculations of $\mathcal{Z}_a^{\rm 0d}$ below.}
\begin{align}\label{polesU(2)}
    (s_1^{(2)}\,,\,s_2^{(2)})\,=\,(s_1^{(3)}\,,\,s_2^{(3)})~,~(s_2^{(3)}\,,\,s_3^{(3)})~,~(s_1^{(3)}\,,\,s_3^{(3)})~.
\end{align}

Let us focus on the first pole. In this case, we get:
\begin{equation}
\mathcal{Z}^{\rm 0d}_2\,=\,\frac{(-1)^2\,\prod_{\mathsf{a}=1}^2\left[\prod_{\alpha^{(2)}}\left(s^{(3)}_{\mathsf{a}}\,-\,m_{\alpha^{(2)}}\right)\right]\,\times\,\mathcal{Z}^{\rm 0d}_1\mid_{s_\mathsf{a}^{(2)}\mapsto s_{\mathsf{a}}^{{(3)}}}}{\prod_{\mathsf{a}<3}\left(s_{\mathsf{a}}^{(3)}\,-\,s_3^{(3)}\right)\,\left(s_3^{(3)}\,-\,s_{\mathsf{a}}^{(3)}\,-\,\hbar\right)\,\prod_{\mathsf{a}=1}^2\prod_{\beta^{(2)}}\left(-\,s_{\mathsf a}^{(3)}\,+\,m_{\beta^{(2)}}\,+\,\hbar\right)}\,+\,\cdots~,
\end{equation}
where the ellipses stand for the other two terms coming from the other two poles in~\eqref{polesU(2)}.

Plugging in the form of $\mathcal{Z}^{\rm 0d}_1$ as given in~\eqref{Z1}, we see that the full expression for $\mathcal{Z}^{\rm 0d}_2$ will consist of $6$ terms which can be written as follows:
\begin{equation}\label{Z2}
    \mathcal{Z}^{\rm 0d}_2\,=\,\operatorname*{\mathsf{Sym}}_{s^{(3)}_\bullet}\left[\frac{(-1)^{1+2}\,\prod_{a'=1}^{2}\prod_{\mathsf{a}=1}^{a'}\prod_{\alpha^{(a')}}\left(s_{\mathsf{a}}^{(3)}\,-\,m_{\alpha^{(a')}}\right)}{\widetilde{\Delta}^{\rm 0d}_3(s^{(3)},\hbar)\,\prod_{a'=1}^{2}\prod_{\mathsf{a}=1}^{a'}\prod_{\beta^{(a')}}\left(-\,s_{\mathsf{a}}^{(3)}\,+\,m_{\beta^{(a')}}\,+\,\hbar\right)}\right]~,
\end{equation}
where, 
\begin{equation}
    \widetilde{\Delta}^{\rm 0d}_a(s^{(a)},\hbar)\,\coloneqq\,\prod_{\substack{\mathsf{a},\mathsf{b}=1\\\mathsf{a}<\mathsf{b}}}^a\,\left(s_{\mathsf a}^{(a)}\,-\,s_{\mathsf{b}}^{(a)}\right)\,\left(s_{\mathsf b}^{(a)}\,-\,s_{\mathsf a}^{(a)}\,-\,\hbar\right)~.
\end{equation}

\item From the final forms of the first two residue integrals~\eqref{Z1} and~\eqref{Z2}, one can extrapolate the final form of the residue integral at the $U(a)$ gauge node:
\begin{equation}\label{Za}
    \mathcal{Z}^{\rm 0d}_a\,=\,\operatorname*{\mathsf{Sym}}_{s^{(a+1)}_\bullet}\left[\frac{(-1)^{\binom{a+1}{2}}\,\prod_{a'=1}^{a}\prod_{\mathsf{a}=1}^{a'}\prod_{\alpha^{(a')}}\left(s_{\mathsf{a}}^{(a+1)}\,-\,m_{\alpha^{(a')}}\right)}{\widetilde{\Delta}^{\rm 0d}_{a+1}(s^{(a+1)},\hbar)\,\prod_{a'=1}^{a}\prod_{\mathsf{a}=1}^{a'}\prod_{\beta^{(a')}}\left(-\,s_{\mathsf{a}}^{(a+1)}\,+\,m_{\beta^{(a')}}\,+\,\hbar\right)}\right]~.
\end{equation}
In particular, at the $U(k-1)$ node, we get the following final result:
\begin{equation}\label{Zk-1}
    \mathcal{Z}^{\rm 0d}_{k-1}\,=\,\operatorname*{\mathsf{Sym}}_{s^{(k)}_\bullet}\left[\frac{(-1)^{\binom{k}{2}}\,\prod_{a'=1}^{k-1}\prod_{\mathsf{a}=1}^{a'}\prod_{\alpha^{(a')}}\left(s_{\mathsf{a}}^{(k)}\,-\,m_{\alpha^{(a')}}\right)}{\widetilde{\Delta}^{\rm 0d}_{k}(s^{(k)},\hbar)\,\prod_{a'=1}^{k-1}\prod_{\mathsf{a}=1}^{a'}\prod_{\beta^{(a')}}\left(-\,s_{\mathsf{a}}^{(k)}\,+\,m_{\beta^{(a')}}\,+\,\hbar\right)}\right]~.
\end{equation}

\item Let us now perform the final residue integral which is the one associated with the $U(k)$ defect gauge group. We have:
\begin{multline}\label{eq:Zk0d}
    \mathcal{Z}^{\rm 0d}_k\,=\,\oint_{\rm JK}\frac{ds^{(k)}}{k!\,(2\pi i)^k}\,\frac{\hbar^k\,\prod_{\mathsf{a}\neq\mathsf{b}}\left(s_{\mathsf{a}}^{(k)}\,-\,s_{\mathsf{b}}^{(k)}\right)\,\left(s_{\mathsf{a}}^{(k)}\,-\,s_{\mathsf{b}}^{(k)}\,+\,\hbar\right)}{\prod_{\mathsf{a},a=1}^k\left(s_{\mathsf{a}}^{(k)}\,-\,\sigma_{a}\right)\,\left(\sigma_{a}\,-\,s_{\mathsf{a}}^{(k)}\,-\,\hbar\right)}\,\times\\\times\,\left[\prod_{{\sf a}=1}^k\prod_{\alpha^{(k)}}\left(s_{{\sf a}}^{(k)}\,-\,m_{\alpha^{(k)}}\right)\,\prod_{{a}=1}^k\prod_{\beta}\left(\sigma_a\,-\,m_\beta\,-\,\hbar\right)\right]\,\times\,\mathcal{Z}^{\rm 0d}_{k-1}~,
\end{multline}
where the range of the index $\beta$ appearing in the second line is given in~\eqref{beta-range}. Up to $S_k$ permutation, there is only one pole to be picked, which is:
\begin{equation}
    (s_1^{(k)}\,,\,s_2^{(k)}\,,\,\cdots\,,\,s_k^{(k)})\,=\,(\sigma_1\,,\,\sigma_2\,,\,\cdots\,,\,\sigma_k)~.
\end{equation}
Taking the residue at this point, we end up with the following expression:
\begin{equation}
    \mathcal{Z}^{\rm 0d}_k\,=\,(-1)^k\,\times\,\prod_{a=1}^k\left[\prod_{\alpha^{(k)}}\left(\sigma_a\,-\,m_{\alpha^{(k)}}\right)\,\prod_{\beta}\left(\sigma_a\,-\,m_\beta\,-\,\hbar\right)\right]\,\times\,\mathcal{Z}^{\rm 0d}_{k-1}\mid_{s_a^{(k)}\mapsto \sigma_a}~.
\end{equation}
Plugging the final expression for $\mathcal{Z}^{\rm 0d}_{k-1}$ as given in~\eqref{Zk-1}, we get:
\begin{equation}
    \mathcal{Z}^{\rm 0d}_k\,=\,(-1)^{\eps(w)}\,\operatorname*{\mathsf{Sym}}_{\sigma_\bullet}\left[\frac{\prod_{a=1}^k\left[\prod_{\mathsf{a}=1}^a\prod_{\alpha^{(a)}}\left(\sigma_{\mathsf{a}}\,-\,m_{\alpha^{(a)}}\right)\,\prod_\beta\left(\sigma_a\,-\,m_\beta\,-\,\hbar\right)\right]}{\widetilde{\Delta}^{\rm 0d}_k(\sigma, \hbar)\,\prod_{a=1}^{k-1}\,\prod_{\mathsf{a}=1}^a\prod_{\beta^{(a)}}\left(\sigma_{\mathsf{a}}\,-\,m_{\beta^{(a)}}\,-\,\hbar\right)}\right]~,
\end{equation}
with $\eps(w)$ being an overall sign of the form~\eqref{eps(w)}
\end{itemize}

To further simplify this expression, let us note that, using the explicit range of $\alpha^{(a)}$ and $\beta^{(a)}$ given in~\eqref{alpha-a-range} and~\eqref{beta-a-range}, we have:
\begin{equation}\label{rel0d}
\begin{split}
&\prod_{a=1}^k\prod_{\mathsf{a}=1}^a\prod_{\alpha^{(a)}=w_{k-a}}^{w_{k-a+1}-1}\,\left(\sigma_{\mathsf{a}}\,-\,m_{\alpha^{(a)}}\right)\,=\,\prod_{a=1}^k\,\prod_{\alpha_a = 1}^{w_{k-a+1}-1}\left(\sigma_a\,-\,m_{\alpha_a}\right)~,\\
    &\prod_{a=1}^{k-1}\,\prod_{\mathsf{a}=1}^a\prod_{\beta^{(a)}=w_{k-a}+1}^{w_{k-a+1}}\left(\sigma_{\mathsf{a}}\,-\,m_{\beta^{(a)}}\,-\,\hbar\right)\,=\,\prod_{a=1}^{k-1}\prod_{\beta_a=w_1+1}^{w_{k-a+1}}\,\left(\sigma_a\,-\,m_{\beta_a}\,-\,\hbar\right)~.
\end{split}
\end{equation}
From this, we see that $\mathcal{Z}^{\rm 0d}_k$ above, simplifies as follows:
\begin{equation}
    \mathcal{Z}^{\rm 0d}_k\,=\,(-1)^{\eps(w)}\,\operatorname*{\mathsf{Sym}}_{\sigma_\bullet}\left[\frac{\prod_{a=1}^k\left[\prod_{\alpha_a=1}^{w_{k-a+1}-1}\left(\sigma_a\,-\,m_{\alpha_a}\right)\prod_{\gamma_a=w_{k-a+1}+1}^{n}\left(\sigma_a\,-\,m_{\gamma_a}\,-\,\hbar\right)\right]}{\prod_{a<b}\left(\sigma_a\,-\,\sigma_b\right)\,\left(\sigma_b\,-\,\sigma_a\,-\,\hbar\right)}\right]~.
\end{equation}
The cancellation between the contribution of $\widetilde{P}_a$ against those coming from some of the Fermis $\widetilde{\Gamma}_{\beta^{(a)}}$ is equivalent to our earlier statement about the chiral fields being massive and that they get integrated out in the IR. 

Since the final expression is symmetrized in the bulk Coulomb branch parameters $\sigma_a$, then we can write the above expression as:
\begin{equation}
    Z_w^{\rm 0d}(\sigma,m,\hbar)\,=\, (-1)^{\eps(w)}\,\operatorname*{\mathsf{Sym}}_{\sigma_\bullet}\left[\frac{\prod_{a=1}^k\left[\prod_{\alpha_a=1}^{w_{a}-1}\left(\sigma_a\,-\,m_{\alpha_a}\right)\prod_{\gamma_a=w_{a}+1}^{n}\left(\sigma_a\,-\,m_{\gamma_a}\,-\,\hbar\right)\right]}{\prod_{a<b}\left(\sigma_b\,-\,\sigma_a\right)\,\left(\sigma_a\,-\,\sigma_b\,-\,\hbar\right)}\right]~.
\end{equation}
This matches~\eqref{stab-w} and hence proves the claim~\eqref{Z0dw=Stabw}.

\subsection{K-theoretic stable envelopes as BPS line defects in 3d GLSM }
In this subsection, we generalize our construction above to the 3d GLSM case. In this case, we construct the K-theoretical stable envelope $\kstab_{-,w}\in K_T(X_{k,n})$ as a BPS line defect wrapping $\mathbb{S}^1_{\boldsymbol\beta}$ fiber at some point on the 2d space slice. This line defect is the same as the one we introduced in Figure~\ref{fig:BPSDefectDefn} but now viewed as a 1d-3d coupled system, where, on the defect, we have 1d $\mathcal{N}=(0,2)$ supersymmetric quiver gauge theory. In what follows, we will compute the 1d flavored Witten index of this SQM system and show that it coincides with the Chern character~\eqref{kstab-w}. For that goal, we will use the supersymmetric localization results of~\cite{Hori:2014tda}. See also~\cite{Closset:2023bdr,Gu:2025tda,Closset:2025akk,Closset:2026bnk} for directly related calculations.

\subsubsection{Fixing the 1d Chern--Simons levels}
As an ingredient that enters into the definition of the quiver SQM, one needs to fix the CS levels of the 1d gauge groups along the quiver. For a more recent review on 1d CS levels, see~\cite{Closset:2025zyl}. Recall that, for each gauge group $U(a)$, one can write down only one CS term, which is the one associated with $U(1)\subseteq U(a)$. Denoting the 1d CS level for $U(a)$ by $\kappa_a$, we will fix them as follows:\footnote{In the conventions of~\cite{Closset:2025zyl}, these are the integer CS levels that they denote by $Q$. In particular, these are the levels that will appear in the 1d index calculation below.}
\begin{equation}\label{1dCS}
\kappa_a(w) \,=\,\begin{cases}
     w_{k-a}\,-\,w_{k-a+1}~, \qquad &a\,=\,1\,,\,\cdots\,,\,k-1~, \\
     0~, \qquad &a\,=\,k~.
\end{cases}
\end{equation}

\medskip
\noindent
\textbf{In  the $\sf y\to\infty$ limit.} In this limit, the chiral multiplets $\widetilde{P}_a$ become massive and get integrated out. Since these fields are in the antifundamental representation of $U(a)$, this process shifts the 1d CS level $\kappa_a$ to zero. Similarly, the Fermi multiplets $\widetilde\Gamma$ also become massive in this limit. But since they are singlets under the defect gauge groups, they do not affect the 1d CS levels when integrated out. Additionally, one needs to keep track of the shifts coming from the massive $\widetilde{\varphi}_a$ and $\widetilde{\Psi}_a$. It is straightforward to see that the net shift remains zero. We will see this integrating out process at the level of the 1d index shortly.

As we explained earlier, in this limit, the cotangent fibers of the Higgs branch $X_{k,n}$ get integrated out and one ends up with the compact base Gr$(k,n)$. Following the above arguments, our BPS line defect reduces to the one introduced in~\cite{Closset:2023bdr} with the same choice of 1d CS levels at each gauge for the Schubert classes in the K-theory ring of Gr$(k,n)$.

\subsubsection{1d flavored indices as the Chern characters}
Similar to what we did in the previous subsection for the 0d-2d coupled system, let us now compute the flavored Witten indices of the 1d-3d coupled systems, which we claim to correspond to the K-theoretic stable envelopes in $T^*{\rm Gr}(k,n)$. Viewing the theory in Figure~\ref{fig:BPSDefectDefn} as a 1d-3d coupled system, and upon supersymmetric localization, the corresponding 1d index is given by the following JK residue integral:
\begin{equation}
    Z^{\rm 1d}_w(x, y, \mathsf{y}) \,=\,\oint_{\rm JK}\prod_{a=1}^{k}\left(\frac{(-1)^a\,dz^{(a)}}{a!\,(2\pi i)^a\,\det(z^{(a)})}\right)\,\Delta^{\rm 1d}\,Z^{\rm 1d}_{\rm CS}\,Z^{\rm 1d}_{\varphi}\,Z_{\Gamma}^{\rm 1d}\,Z_{\widetilde{\varphi}}^{\rm 1d}\,Z_{\widetilde\Psi}^{\rm 1d}\,Z^{\rm 1d}_{\widetilde P}\,Z_{\widetilde{\Gamma}}^{\rm 1d}~.
\end{equation}
It is understood here that $dz^{(a)}\,\equiv\,dz^{(a)}_1\cdots dz_a^{(a)}$ and $\det z^{(a)}\,\equiv z_1^{(a)}\cdots z_a^{(a)}$ which is the product of the exponentiated Coulomb parameters in the gauge vector multiplet $\mathcal{V}_{\rm 1d}^{(a)}$.

The different contributions appearing in the integrand are given explicitly by:
\begingroup
\allowdisplaybreaks[1]
\begin{align}
    & \Delta^{\rm 1d}(z)\,\coloneqq\,\prod_{a=1}^k\,\Delta_a^{\rm 1d}(z)\,\coloneqq\,\prod_{a=1}^k\,\prod_{\substack{\mathsf{a},\mathsf{b}=1\\\mathsf{a}\neq \mathsf{b}}}^a\,\left(1\,-\,\frac{z_{\mathsf a}^{(a)}}{z_{\mathsf b}^{(a)}}\right)~,\qquad Z^{\rm 1d}_{\rm CS}(z)\,\coloneqq\,\prod_{a=1}^{k-1}\,(\det z^{(a)})^{\kappa_a}~,\\
    &Z^{\rm 1d}_{\varphi}(z)\,\coloneqq\,\prod_{a=1}^{k-1}\,\left[\,\prod_{\mathsf{a}=1}^{a}\,\prod_{\mathsf{b}=1}^{a+1}\left(1\,-\,\frac{z_{\mathsf{a}}^{(a)}}{z_{\mathsf{b}}^{(a+1)}}\right)^{-1}\,\right]\,\prod_{\mathsf{a}=1}^k\prod_{a=1}^k\left(1\,-\,\frac{z_{\mathsf a}^{(k)}}{x_a}\right)^{-1}~,\\
     &Z_\Gamma^{\rm 1d}(z,y)\,\coloneqq\,\prod_{a=1}^{k}\,\prod_{\mathsf{a}=1}^a\,\prod_{\alpha^{(a)}=w_{k-a}}^{w_{k-a+1}-1}\,\left(1\,-\,\frac{z_{\mathsf a}^{(a)}}{y_{\alpha^{(a)}}}\right)~,\\
        &Z^{\rm 1d}_{\widetilde\varphi}(z,\mathsf{y})\,\coloneqq\,\prod_{a=1}^{k-1}\,\left[\,\prod_{\mathsf{a}=1}^{a}\,\prod_{\mathsf{b}=1}^{a+1}\left(1\,-\,\frac{z_{\mathsf{b}}^{(a+1)}}{z_{\mathsf a}^{(a)}\,\mathsf{y}}\right)^{-1}\,\right]\,\prod_{\mathsf{a}=1}^k\prod_{a=1}^k\left(1\,-\,\frac{x_a}{z_{\mathsf{a}}^{(k)}\,\mathsf{y}}\right)^{-1}~,\\
        &Z_{\widetilde\Psi}^{\rm 1d}(z,\mathsf{y})\,\coloneqq\,\prod_{a=1}^k\,\left[(1\,-\,\mathsf{y})^a\,\prod_{\substack{\mathsf{a},\mathsf{b}=1\\\mathsf{a}\neq \mathsf{b}}}^a\,\left(1\,-\,\frac{z_{\mathsf{a}}^{(a)}\,\mathsf{y}}{z_{\mathsf b}^{(a)}}\right)\right]~,\\
           &Z_{\widetilde P}^{\rm 1d}(z,y,\mathsf{y})\,\coloneqq\,\prod_{a=1}^{k-1}\,\prod_{\mathsf{a}=1}^a\,\prod_{\beta^{(a)}=w_{k-a}+1}^{w_{k-a+1}}\,\left(1\,-\,\frac{y_{\beta^{(a)}}\,\mathsf{y}}{z_{\mathsf a}^{(a)}}\right)^{-1}~,\\
        &  Z_{\widetilde\Gamma}^{\rm 1d}(z,y,\mathsf{y})\,\coloneqq\,\prod_{{a}=1}^k\,\prod_{\beta=w_1+1}^n\,\left(1\,-\,\frac{x_{ a}}{y_\beta\,\mathsf{y}}\right)~.
\end{align}
\endgroup
Here, $\mathsf{y}\equiv e^{\boldsymbol{\beta}\hbar}$. The contribution in the first line comes from the 1d `W-bosons' in each vector multiplet along the quiver defect.

As in the 0d-2d case, we take the stability condition~\eqref{stabilityConditions} where now $\zeta_a^{\rm d}$ is viewed as the 1d FI parameter for $U(1)\subseteq U(a)$ gauge node. The JK residue prescription hence instructs us to pick the poles coming from the 1d chiral fields $\varphi_a$. We perform the integrals from left to right along the quiver. More explicitly, we claim that the final result coincides with the Chern character of the K-theoretic stable envelope~\eqref{kstab-w} associated with the same permutation $w\in {\rm W}^{(k;n)}$ defining the defect:
\begin{equation}
    Z^{\rm 1d}_w(x, y, \mathsf{y})\,=\,\kstab_w(x,y,\mathsf{y})~.
\end{equation}

As a warm-up, we will consider the $T^*\bP^{n-1}$ GLSM case and then generalize it to any $X_{k,n}$.

\medskip
\noindent
\textbf{The BPS line defects in $T^*\mathbb{P}^{n-1}$.} Looking at the abelian case where the Higgs branch of the 3d GLSM is $T^*\mathbb{P}^{n-1}$, the BPS line defect associated with the permutation $w=(w_1\,\cdots)\in {\rm W}^{(1;n)}$ is given in Figure~\ref{fig:BPSDefectTPn} viewed now as a 1d-3d coupled system with $\Gamma_1$ and $\widetilde{\Gamma}_1$ as given in~\eqref{Gamma1tildeGamma1}. The 1d flavored Witten index of the system is then given by:
\begin{equation}
    Z_{w_1}^{\rm 1d}(x, y, \mathsf{y})\,=\,(1\,-\,\mathsf{y})\,\oint_{\rm JK}\frac{-\,dz}{2\pi i \,z}\,\frac{\prod_{\alpha=1}^{w_1-1}\left(1\,-\,\frac{z}{y_\alpha}\right)\,\prod_{\beta=w_1+1}^n\left(1\,-\,\frac{x}{y_\beta\,\mathsf{y}}\right)}{\left(1\,-\,\frac{z}{x}\right)\,\left(1\,-\,\frac{x}{z\,\mathsf{y}}\right)}~,
\end{equation}
with $z\equiv z_1^{(1)}$. Performing the residue integral at $z=x$, we end up with:
\begin{equation}
     Z_{w_1}^{\rm 1d}(x, y, \mathsf{y})\,=\,-\,\mathsf{y}\,\prod_{\alpha=1}^{w_1-1}\left(1\,-\,\frac{x}{y_\alpha}\right)\,\prod_{\beta=w_1+1}^n\left(1\,-\,\frac{x}{y_\beta\,\mathsf{y}}\right)~,
\end{equation}
matching with the corresponding K-theoretic stable envelope $\kstab_{w_1}(x,y,\mathsf{y})$ given in~\eqref{kstab-w}.

\medskip
\noindent
\textbf{The more general case.}
Let us now compute the 1d flavored Witten index of the 1d-3d coupled system in the general case. Fix $w\in {\rm W}^{(k;n)}$. Let us denote the final result of the residue integral at the $U(a)$ defect node by $\mathcal{Z}_a^{\rm 1d}$ and let us perform these integrals recursively. In particular:
\begin{equation}
    Z_{w}^{\rm 1d}(x,y,\mathsf{y})\,=\,\mathcal{Z}_k^{\rm 1d}(x,y,\mathsf{y})~.
\end{equation}

We have:
\begin{itemize}
\item At the step $a=1$, we have:
\begin{equation}
    \mathcal{Z}_1^{\rm 1d}\,=\,-\,(1\,-\,\mathsf{y})\,\oint_{\rm JK}\frac{dz_1^{(1)}}{2\pi i \,z_1^{(1)}}\,\frac{\left(z^{(1)}_1\right)^{\kappa_1}\,\prod_{\alpha^{(1)}}\left(1\,-\,\frac{z_1^{(1)}}{y_{\alpha^{(1)}}}\right)}{\prod_{\mathsf{a}=1}^2\left(1\,-\,\frac{z_1^{(1)}}{z_{\mathsf{a}}^{(2)}}\right)\,\left(1\,-\,\frac{z_{\mathsf{a}}^{(2)}}{z_1^{(1)}\,\mathsf{y}}\right)\,\prod_{\beta^{(1)}}\left(1\,-\,\frac{y_{\beta^{(1)}}\,\mathsf{y}}{z_1^{(1)}}\right)}~.
\end{equation}
As discussed above, the JK residue prescription instructs us to pick the poles $z_1^{(1)}=z^{(2)}_1,z_2^{(2)}$. Doing so, we get:
\begin{equation}\label{Z11d}
    \mathcal{Z}_1^{\rm 1d}\, =\,-\,\mathsf{y}\,\operatorname*{\mathsf{Sym}}_{z_\bullet^{(2)}}
    \left[\frac{\left(z_1^{(2)}\right)^{\kappa_1}\,\prod_{\alpha^{(1)}}\left(1-\frac{z_1^{(2)}}{y_{\alpha^{(1)}}}\right)}{\left(1-\frac{z_1^{(2)}}{z_2^{(2)}}\right)\,\left(1-\frac{z_2^{(2)}}{z_1^{(2)}\,\mathsf{y}}\right)\,\prod_{\beta^{(1)}}\left(1-\frac{y_{\beta^{(1)}}\,\mathsf{y}}{z_1^{(2)}}\right)}\right]~.
\end{equation}
Here, the symmetrization operator acts on both 1d Coulomb branch variables $z_1^{(2)},z_2^{(2)}$ and comes from the two terms that we get from the two distinct residues.
\item Let us now move to the next node. In this case, we have the following residue integral:
\begin{multline}
    \mathcal{Z}_2^{\rm 1d}\,=\,(1\,-\,\mathsf{y})^2\,\oint_{\rm JK}\frac{dz_1^{(2)}\,dz_2^{(2)}\,(\det z^{(2)})^{\kappa_2}}{2\,(2\pi i)^2\,(-z_1^{(2)})(-z_2^{(2)})}\,\frac{\prod_{\mathsf{a}\neq\mathsf{b}}\left(1\,-\,\frac{z_{\mathsf{a}}^{(2)}}{z_{\mathsf{b}}^{(2)}}\right)\,\left(1\,-\,\frac{z_{\mathsf{a}}^{(2)}\,\mathsf{y}}{z_{\mathsf{b}}^{(2)}}\right)}{\prod_{\mathsf{a}=1}^2\prod_{\mathsf{b}=1}^3\left(1\,-\,\frac{z_{\mathsf{a}}^{(2)}}{z_{\mathsf{b}}^{(3)}}\right)\,\left(1\,-\,\frac{z_{\mathsf{b}}^{(3)}}{z_{\mathsf{a}}^{(2)}\,\mathsf{y}}\right)}\,\times\\\times\,\prod_{\mathsf{a}=1}^2\frac{\prod_{\alpha^{(2)}}\left(1\,-\,\frac{z_{\mathsf{a}}^{(2)}}{y_{\alpha^{(2)}}}\right)}{\prod_{\beta^{(2)}}\left(1\,-\,\frac{y_{\beta^{(2)}}\,\mathsf{y}}{z_{\mathsf{a}}^{(2)}}\right)}\,\times\,\mathcal{Z}_1^{\rm 1d}~,
\end{multline}
where $\mathcal{Z}_1^{\rm 1d}$ is given in~\eqref{Z11d}. With our choice of FI parameters $\zeta_2>\zeta_1>0$, this integral is performed by taking residues at the following three poles (up to the $S_2$ Weyl action which is taken care of by the $1/2$ factor appearing in the measure):
\begin{equation}\label{polesZ21d}
\left(z_1^{(2)}\,,\,z_2^{(2)}\right)\,=\,\left(z_1^{(3)}\,,\,z_2^{(3)}\right)~,~\left(z_2^{(3)}\,,\,z_3^{(3)}\right)~,~\left(z_1^{(3)}\,,\,z_3^{(3)}\right)~.
\end{equation}
Doing so, we end up with the following:
\begin{equation}
    \mathcal{Z}_2^{\rm 1d}\,=\,\frac{\mathsf{y}^{4}\,(z_1^{(3)}\,z_2^{(3)})^{\kappa_2}}{\prod_{\mathsf{a}< 3}\left(1\,-\,\frac{z_{\mathsf{a}}^{(3)}}{z_{3}^{(3)}}\right)\,\left(1\,-\,\frac{z_3^{(3)}}{z_{\mathsf a}^{(3)}\,\mathsf{y}}\right)}\,\times\,\prod_{\mathsf{a}=1}^2\frac{\prod_{\alpha^{(2)}}\left(1\,-\,\frac{z_{\mathsf{a}}^{(3)}}{y_{\alpha^{(2)}}}\right)}{\prod_{\beta^{(2)}}\left(1\,-\,\frac{y_{\beta^{(2)}}\,\mathsf{y}}{z_{\mathsf{a}}^{(3)}}\right)}\,\times\,\mathcal{Z}_1^{\rm 1d}\mid_{z_{\mathsf a}^{(2)}\mapsto z_{\mathsf{a}}^{(3)}}\,+\,\cdots~,
\end{equation}
where the ellipses stand for the terms coming from the other two poles in~\eqref{polesZ21d}.

Plugging in the explicit form of $\mathcal{Z}_1^{\rm 1d}$~\eqref{Z11d}, we get six terms that can be written more compactly as a symmetrization of one of them w.r.t. to the 1d $U(3)$ Coulomb branch variables:
\begin{equation}
    \mathcal{Z}_2^{\rm 1d}\,=\,\prod_{{a'}=1}^2(-\mathsf{y})^{{a'}^2}\,\times\,\operatorname*{\mathsf{Sym}}_{z_\bullet^{(3)}}\left[\frac{1}{\widetilde{\Delta}_3^{\rm 1d}(z^{(3)},\mathsf{y})}\,\times\,\prod_{a'=1}^2\prod_{\mathsf{a}=1}^{a'}\frac{\left(z_{\sf a}^{(3)}\right)^{\kappa_{a'}}\,\prod_{\alpha^{(a')}}\left(1\,-\,\frac{z_{\mathsf{a}}^{(3)}}{y_{\alpha^{(a')}}}\right)}{\prod_{\beta^{(a')}}\left(1\,-\,\frac{y_{\beta^{(a')}}\,\mathsf{y}}{z_{\mathsf{a}}^{(3)}}\right)}\right]~,
\end{equation}
where here we introduced:
\begin{equation}
    \widetilde{\Delta}_a^{\rm 1d}(z^{(a)},\mathsf{y})\,\coloneqq\,\prod_{\mathsf{a}<\mathsf{b}}\left(1\,-\,\frac{z_{\mathsf{a}}^{(a)}}{z_{\mathsf b}^{(a)}}\right)\,\left(1\,-\,\frac{z_{\mathsf b}^{(a)}}{z_{\mathsf{a}}^{(a)}\,\mathsf{y}}\right)~.
\end{equation}
\item From the above calculations, one can extrapolate the final form of the residue integral at the $a$-th node along the defect. One finds:
\begin{multline}
    \mathcal{Z}_a^{\rm 1d}\,=\,\prod_{{a'}=1}^a(-\mathsf{y})^{{a'}^2}\,\\\times\,\operatorname*{\mathsf{Sym}}_{z_\bullet^{(a+1)}}\left[\frac{1}{\widetilde{\Delta}_{a+1}^{\rm 1d}(z^{(a+1)},\mathsf{y})}\,\times\,\prod_{a'=1}^{a}\prod_{\mathsf{a}=1}^{a'}\frac{\left(z_{\sf a}^{(a+1)}\right)^{\kappa_{a'}}\prod_{\alpha^{(a')}}\left(1\,-\,\frac{z_{\mathsf{a}}^{(a+1)}}{y_{\alpha^{(a')}}}\right)}{\prod_{\beta^{(a')}}\left(1\,-\,\frac{y_{\beta^{(a')}}\,\mathsf{y}}{z_{\mathsf{a}}^{(a+1)}}\right)}\right]~.
\end{multline}
In particular, at $a=k-1$, we have:
\begin{equation}\label{Z1dk-1}
    \mathcal{Z}_{k-1}^{\rm 1d}\,=\,\prod_{{a'}=1}^{k-1}(-\mathsf{y})^{{a'}^2}\,\times\,\operatorname*{\mathsf{Sym}}_{z_\bullet^{(k)}}\left[\frac{1}{\widetilde{\Delta}_{k}^{\rm 1d}(z^{(k)},\mathsf{y})}\,\times\,\prod_{a'=1}^{k-1}\prod_{\mathsf{a}=1}^{a'}\frac{\left(z_{\sf a}^{(k)}\right)^{\kappa_{a'}}\,\prod_{\alpha^{(a')}}\left(1\,-\,\frac{z_{\mathsf{a}}^{(k)}}{y_{\alpha^{(a')}}}\right)}{\prod_{\beta^{(a')}}\left(1\,-\,\frac{y_{\beta^{(a')}}\,\mathsf{y}}{z_{\mathsf{a}}^{(k)}}\right)}\right]~.
\end{equation}
\item Let us now come to the last gauge node. The residue integral that we have is of the following form:
\begin{multline}
    Z_w^{\rm 1d}\,=\,\mathcal{Z}_k^{\rm 1d}\,=\,\oint_{\rm JK}\,\frac{dz^{(k)}}{k!\,(2\pi i)^k\,\det(-z^{(k)})}\,\frac{(1\,-\,\mathsf{y})^k\,\prod_{\mathsf{a}\neq\mathsf{b}}\left(1\,-\,\frac{z_{\mathsf{a}}^{(k)}}{z_{\mathsf b}^{(k)}}\right)\,\left(1\,-\,\frac{z_{\mathsf{a}}^{(k)}\,\mathsf{y}}{z_{\mathsf b}^{(k)}}\right)}{\prod_{\mathsf{a},a}\left(1\,-\,\frac{z_{\mathsf a}^{(k)}}{x_a}\right)\,\left(1\,-\,\frac{x_a}{z_{\mathsf a}^{(k)}\,\mathsf{y}}\right)}\,\times\\\times\,\left[\prod_{\mathsf{a}=1}^k\prod_{\alpha^{(k)}}\left(1\,-\,\frac{z_{\mathsf{a}}^{(k)}}{y_{\alpha^{(k)}}}\right)\,\prod_{{a}=1}^k\prod_{\beta}\left(1\,-\,\frac{x_{ a}}{y_\beta\,\mathsf{y}}\right)\right]\,\times\,\mathcal{Z}_{k-1}^{\rm 1d}~.
\end{multline}
Up to the action of the permutation group $S_k$, there is only one single pole to take:
\begin{equation}
    \left(z_1^{(k)}\,,\,z_2^{(k)}\,,\,\cdots\,,\,z_k^{(k)}\right)\,=\,(x_1\,,\,x_2\,,\,\cdots\,,\,x_k)~.
\end{equation}
The residue at this pole gives us:
\begin{equation}
    {Z}_w^{\rm 1d}\,=\,(-\mathsf{y})^{k^2}\,\times\,\prod_{{a}=1}^k\left[\prod_{\alpha^{(k)}}\left(1\,-\,\frac{x_a}{y_{\alpha^{(k)}}}\right)\,\prod_{\beta}\left(1\,-\,\frac{x_a}{y_\beta\,\mathsf{y}}\right)\right]\,\times\,\mathcal{Z}_{k-1}^{\rm 1d}\mid_{z_{a}^{(k)}\mapsto x_a}~.
\end{equation}
Substituting the result for $a=k-1$~\eqref{Z1dk-1}, we end up with the following expression for the final integral:
\begin{equation}\label{Z1dw}
    Z_w^{\rm 1d}\,=\,(-\,{\sf y})^{-\varepsilon(w)}\operatorname*{\mathsf{Sym}}_{x_\bullet}\left[\frac{\prod_{a=1}^k\left[\prod_{\mathsf{a}=1}^{a}\prod_{\alpha^{(a)}}\left(1\,-\,\frac{x_{\mathsf{a}}}{y_{\alpha^{(a)}}}\right)\,\prod_{\beta}\left(1\,-\,\frac{x_a}{y_{\beta}\,\mathsf{y}}\right)\right]}{\widetilde{\Delta}_k^{\rm 1d}(x,\mathsf{y})\,\prod_{a=1}^{k-1}\prod_{\mathsf{a}=1}^a\prod_{\beta^{(a)}}\left(1\,-\,\frac{x_{\mathsf{a}}}{y_{\beta^{(a)}}\,\mathsf{y}}\right)}\right]~,
\end{equation}
where the overall factor is given in~\eqref{vareps(w)}.

From the denominator on the r.h.s. of~\eqref{Z1dw}, we got a term that perfectly cancels out the final contribution of the 1d gauge CS terms.\footnote{Additionally, there is an overall factor in terms of the $SU(n)$ flavor symmetry fugacities:
\protect\begin{equation*}
    \prod_{a=1}^k\,\prod_{\beta^{(a)}}\,y_{\beta^{(a)}}^{-a}~.
\end{equation*}
One can remove this term by turning on 1d flavored CS levels at each gauge node. Note also that this term is important when we take the $\hbar\rightarrow \infty$ limit. In this case, we get 1d flavored CS terms that cancel out this term.}

This is equivalent to our earlier statement below~\eqref{1dCS} about integrating out the cotangent fibers of $X_{k,n}$ and reproducing the Grothendieck line defects that were constructed in~\cite{Closset:2023bdr}.
\end{itemize}

The final expression can be further simplified by using the following two relations analogous to~\eqref{rel0d}. We end up with the following expression:
\begin{multline}
    Z_w^{\rm 1d}(x,y,\mathsf{y})\,=\,(-\mathsf{y})^{-\varepsilon(w)}\,\operatorname*{\mathsf{Sym}}_{x_\bullet}\left[\frac{\prod_{a=1}^k\left[\prod_{\alpha^{(a)}=1}^{w_{k-a+1}-1}\left(1\,-\,\frac{x_a}{y_{\alpha_a}}\right)\,\prod_{\gamma^{(a)}=w_{k-a+1}+1}^n\left(1\,-\,\frac{x_a}{y_{\gamma_a}\,\mathsf{y}}\right)\right]}{\prod_{a<b}\left(1\,-\,\frac{x_a}{x_b}\right)\,\left(1\,-\,\frac{x_b}{x_a\,\mathsf{y}}\right)}\right]~.
\end{multline}

Since the final form is symmetric in the 3d bulk Coulomb branch parameters $x_a$, then we can choose to write it as follows:
\begin{equation}
      Z_w^{\rm 1d}(x,y,\mathsf{y})\,=\,(-\mathsf{y})^{-\varepsilon(w)}\,\operatorname*{\mathsf{Sym}}_{x_\bullet}\left[\frac{\prod_{a=1}^k\left[\prod_{\alpha^{(a)}=1}^{w_{a}-1}\left(1\,-\,\frac{x_a}{y_{\alpha_a}}\right)\,\prod_{\gamma^{(a)}=w_{a}+1}^n\left(1\,-\,\frac{x_a}{y_{\gamma_a}\,\mathsf{y}}\right)\right]}{\prod_{a<b}\left(1\,-\,\frac{x_b}{x_a}\right)\,\left(1\,-\,\frac{x_a}{x_b\,\mathsf{y}}\right)}\right]~,
\end{equation}
which indeed matches the expression for the K-theoretic stable envelope $\kstab_{-,w}(x,y,\mathsf{y})$ given in~\eqref{kstab-w}.

\subsubsection{What about the CSM class?}
\begin{figure}[t]
    \centering
    \includegraphics[width=1\linewidth]{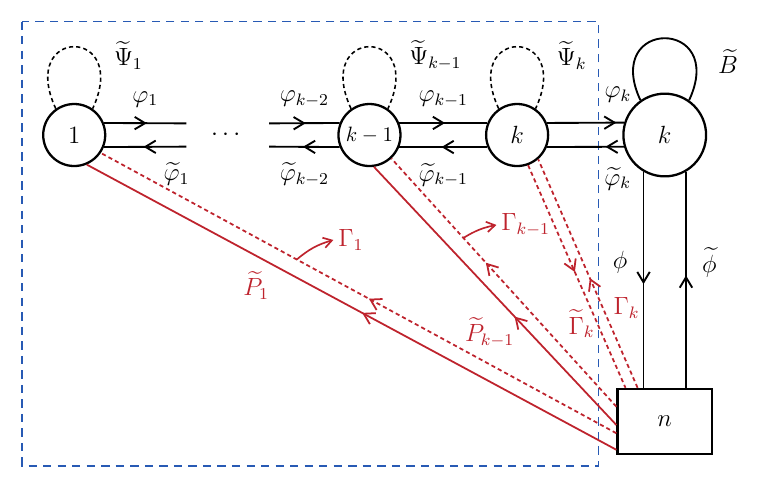}
    \caption{An alternative BPS defect proposal realizing the CSM class in the equivariant cohomology of $X_{k,n}$ and motivic Chern classes in the corresponding equivariant K-theory.}
    \label{fig:CSMdefect}
\end{figure}
In this subsection, we want to discuss an alternative BPS defect realization. Instead of introducing $(0,2)$ Fermi multiplets neutral under defect gauge groups and bulk-defect superpotentials involving $\widetilde \phi$, we can introduce $n-w_1$ Fermi multiplets (with the same flavor quantum numbers) in the fundamental representation of the $U(k)_{\rm d}$ defect gauge group, see Figure~\ref{fig:CSMdefect}. It can be checked that this defect also  gives the same stable envelopes . Additionally, it does not have a natural bulk-defect superpotential term preserving all the flavor symmetries. The existence of this alternative BPS defect construction provides the physical incarnation of the following mathematical fact:
\paragraph{`Stable envelopes = Chern--Schwartz--MacPherson (CSM) classes'.} Consider the embedding $\iota:  {\rm Gr}(k,n) \to T^*{\rm Gr}(k,n)$, the pullback of the stable envelope $\iota^*\stab_{-,w}(\sigma,m,\hbar) \in H_{T}^*({\rm Gr}(k,n))$ gives the equivariant Chern--Schwarz--Macpherson (CSM) class of corresponding Schubert cells~\cite{RimanyiVarchenko2015,Aluffi_2023}. 

Recall that, physically, going from $T^*{\rm Gr}(k,n)$ to ${\rm Gr}(k,n)$ means giving masses to $\widetilde B, \widetilde\phi$ and decouple them. The above fact suggests that for a BPS defect realizing stable envelopes in $T^*{\rm Gr}(k,n)$ GLSM, it should survive in the decoupling process and provide defect realization of CSM classes in the resulting ${\rm Gr}(k,n)$ GLSM. Since the alternative defect in Figure~\ref{fig:CSMdefect} does not require a bulk-defect superpotential involving bulk fields $\widetilde B,\widetilde \phi $, it naturally survives in the decoupling process and provides defect realization of the CSM class of ${\rm Gr}(k,n)$. To summarize, the defect in Figure~\ref{fig:CSMdefect} is the incarnation of ``Stable envelopes = CSM classes" from the physics side. The same discussion applies to the K-theoretic case~\cite{aluffi2024motivicchernclassesschubert}, with CSM classes being replaced by motivic Chern classes.

As a 0d-2d coupled system, the partition function calculation is similar to that done for~\eqref{Z0dw} with the small modification that the contribution of $\widetilde{\Gamma}$ in~\eqref{Z0dwcomp} is given by:
\begin{equation}
    Z_{\widetilde\Gamma}^{\rm 0d}(s,m,\hbar)\, \longmapsto\,\prod_{\mathsf{a}=1}^k\,\prod_{\beta=w_1+1}^n\left(s_{\mathsf{a}}^{(k)}\,-\,m_\beta\,-\,\hbar\right)~.
\end{equation}
Similarly, in the 1d-3d case, the 1d index calculation remains the same except for the replacement:
\begin{equation}
    Z_{\widetilde\Gamma}^{\rm 1d}(z,y,\mathsf{y})\,\longmapsto\,\prod_{{\sf a}=1}^k\,\prod_{\beta=w_1+1}^n\,\left(1\,-\,\frac{z^{(k)}_{\sf a}}{y_\beta\,\mathsf{y}}\right)~.
\end{equation}
Upon doing these changes, performing the JK residue calculations as done earlier in this section produces the same results for the cohomological and K-theoretical stable envelopes.

\section{Quantum product of stable envelopes from the A-model}\label{sec:quantumProd}
In this section, we explain how to compute products in quantum cohomology and quantum K-theory using the 2d and 3d A-models, respectively. We begin by introducing the necessary tools from computational algebraic geometry, then discuss the two product structures in turn. We illustrate our methods with three examples: $T^* \bP^1$, $T^* \bP^2$ and $T^\ast {\rm Gr}(2,4)$.

\subsection{Extending the Gr\"obner basis algorithm}\label{sec:Grobner}

We outline our strategy using computational algebraic geometry. From the 2d or 3d Bethe ideal $\CI_{\rm Bethe}$ we obtain the quotient ring, where the quantum cohomological or K product becomes the ordinary multiplication modulo $\CI_{\rm Bethe}$. Determining whether an element vanishes in this quotient ring thus reduces to checking whether a representative belongs to $\CI_{\rm Bethe}$. This is the ideal membership problem, a standard problem in computational algebraic geometry. Let:
\begin{equation}\label{eq:K_R_J}
R\,\coloneqq\,\mathbb{K}[x_1,\ldots,x_k]^{S_k}~,
\qquad
J\,\coloneqq\,\CI_{\rm Bethe} \, \subset \, R~,
\end{equation}
where $\mathbb{K}$ is the field of rational functions for all parameters. We assume that $Q\coloneqq R/J$ is finite-dimensional over $\mathbb{K}$.
The parameters are therefore treated as coefficients, while the
$x_a$ are the polynomial variables. Mathematically, our objective is nothing but determining
coefficients $c_1,\ldots,c_r\in \mathbb{K}$ such that
\begin{equation}\label{eq:membership}
[f]\,=\,\sum_{i=1}^{r}c_i\,[g_i] \quad\Longleftrightarrow\quad f\,-\,\sum_{i=1}^{r}c_i\,g_i\,\in\, J~,
\end{equation}
 where $f,g_1,\ldots,g_r$ are polynomials in $\mathbb{K}[\sigma_1,\ldots,\sigma_k]$, and we use $[f]$ to denote the image of $f$ under the quotient map. 

Fix a global monomial order and compute a Gr\"obner basis
$\mathcal{G}$ of $J$. For $P\in R$, write
$\operatorname{NF}_{\mathcal{G}}(P)$ for its fully reduced normal form.
This is the unique representative of $[P]$ such that:
\begin{equation}
P\,\in\, J
\quad\Longleftrightarrow\quad
\operatorname{NF}_{\mathcal{G}}(P)\,=\,0~.
\end{equation}
Moreover, the normal-form map is $\mathbb{K}$-linear. Thus, the desired relation
is equivalent to the polynomial identity:
\begin{equation}
\operatorname{NF}_{\mathcal{G}}(f)
\,=\,
\sum_{i=1}^{r}
c_i\,\operatorname{NF}_{\mathcal{G}}(g_i)~.
\end{equation}
Because $J$ is zero-dimensional, we can choose a finite
$\mathbb K$-basis
\begin{equation}
\mathcal{B}\,\coloneqq\,\{b_1\,,\,\ldots\,,\,b_s\}~,
\qquad s\,\equiv\,\dim_{\mathbb K} Q~.    
\end{equation}
Expand the normal forms in this basis:
\begin{equation}
    \operatorname{NF}_{\mathcal{G}}(f)
\,=\,\sum_{\mu=1}^{s}v_\mu\, b_\mu~,
\qquad
\operatorname{NF}_{\mathcal{G}}(g_i)
\,=\,\sum_{\mu=1}^{s}M_{\mu i}\,b_\mu~,
\end{equation}
where $v_\mu,M_{\mu i}\in \mathbb{K}$.

Equating coefficients of the standard monomials gives the linear system:
\begin{equation}
    M\,\cdot\,\mathbf{c}\,=\,\mathbf{v}~,
\qquad
M\,\equiv\,(M_{\mu i})\,\in\, \mathbb{K}^{s\times r}~,
\qquad
\mathbf{c}\,=\,(c_1\,,\,\ldots\,,\,c_r)^{\mathsf T}~.
\end{equation}
Hence, a solution exists precisely when the vector $v$ can be expressed as a linear combination of the column vectors of $M$.
If a solution exists, it is unique precisely when
$\operatorname{rank}_{\mathbb K} M=r$, or equivalently when the classes
$[g_1],\ldots,[g_r]$ are linearly independent over $\mathbb K$.
In other words, if these classes form a basis of $Q$, every target class has a unique expansion.

\subsubsection{Extending the formalism for the stable envelopes}
Our goal is to consider this formalism to work out quantum products of the stable envelopes in both quantum cohomology and quantum K-theory rings of $X_{k,n}$.  In practice, one major issue is the presence of a denominator in these stable envelopes. See their definitions in~\eqref{stab-w} and~\eqref{kstab-w}. 

Naively, we can multiply both sides of~\eqref{eq:Su_T^astGr} by a suitable polynomial to eliminate the denominator, but we still need to avoid the locus where it vanishes. This can be done by saturating the Bethe ideal with respect to the denominator.\footnote{For readers who are not interested in mathematical details, they can skip this part until~\eqref{eq:Vsat}.} Mathematically, Let $R=\mathbb{C}[x_1,\ldots,x_n]$ be a multivariate polynomial ring and $J\subset R$ be an ideal. Assume $d$ is a polynomial in $R$. The saturation of $J$ with respect to $d$ is defined by:
\begin{equation}
    J:d^\infty
\,\coloneqq\,
\bigcup_{N\geq 0}(J:d^N)
\,=\,
\left\{
f\,\in\, R \,\middle|\, d^Nf\,\in\, J
\text{ for some }N\,\geq\, 0
\right\}~.
\end{equation}
Thus, a polynomial belongs to $J:d^\infty$ if it is contained in $J$
after multiplication by a sufficiently large power of $d$.
Geometrically, over the complex field, the affine scheme defined by the saturated ideal is:
\begin{equation}
    V(J:d^\infty)
\,=\,
\overline{V(J)\setminus V(d)}~,
\end{equation}
where the overline means taking the Zariski closure. In other words, saturation removes from $V(J)$ all irreducible components
that are contained entirely in the hypersurface $d=0$. In general,
saturation does not simply impose the equation $d\neq 0$ inside affine
space, since $V(J:d^\infty)$ is the Zariski closure of the remaining open
part.

In our application, the Bethe ansatz equations define a zero-dimensional affine scheme over  $\mathbb K$. Its geometric points
form a finite set. Consequently, removing the points where $d$
vanishes leaves a Zariski-closed subset. Scheme-theoretically, saturation removes the entire components
supported at these points, while preserving the multiplicities and
nilpotent structure of all remaining solutions. Hence:
\begin{equation}\label{eq:Vsat}
    V(J:d^\infty)
\,=\,
V(J)\setminus V(d)
\,=\,
\{p\,\in\, V(J)\,\mid\, d(p)\,\neq\, 0\}~.
\end{equation}
We simply discard the solutions at which the denominator
vanishes.

The saturation can also be computed by introducing a new variable $t$. Namely,
\begin{equation}
J:d^\infty
\,=\,
\langle J\,,\, 1\,-\,t\,d\rangle
\,\cap\, R~.
\end{equation}
The additional equation $1-td=0$ forces $d$ to be invertible, with
$t=d^{-1}$. Therefore, one may compute a Gr\"obner basis of $\langle J, 1-td\rangle$ 
using an elimination order in which $t$ is eliminated, and then retaining only
the polynomials that do not involve $t$. This gives generators of
$J:d^\infty$.

Hence, we just need a slight modification to our previous strategy. Return back to~\eqref{eq:K_R_J} and~\eqref{eq:membership}. Assuming the least common multiple of denominators of $f$ and $g_i$ is $d$, we replace $J$ by $J: d^\infty$ and set:
\begin{equation}
F\,\equiv\,d \cdot f\,\in\, R~,
\qquad
G_i\,\equiv\,d \cdot g_i\,\in\, R
\quad (i\,=\,1\,,\,\ldots\,,\,r)~.
\end{equation}
Since multiplication by $d$ now is invertible in $Q$, we have:
\begin{equation}
[f]\,=\,\sum_{i=1}^{r}c_i\,[g_i]
\quad\Longleftrightarrow\quad
F\,-\,\sum_{i=1}^{r}c_i\,G_i\,\in\, J: d^\infty~.
\end{equation}
Then our method is applicable again.

\subsection{Quantum cohomology product}
As a first application for the algorithm developed above, we now compute the quantum products involving stable envelopes in the quantum cohomology ring of $X_{k,n}$. As we mentioned at the beginning of this section, we will do that in explicit examples for $T^*\bP^1$, $T^*\bP^2$, and $T^*{\rm Gr}(2,4)$. 

\subsubsection{Quantum cohomology products for \texorpdfstring{$T^* \bP^1$}{TP1}}\label{subsec:QHP1}
There are two permutations classes in $\rm W^{(1;2)}$: $(1\, 2)$ and $(2\, 1)$. The stable envelopes are~\eqref{stab-w1}: 
\begin{equation}
    \begin{aligned}
    \stab_{-,(1\,2)} &\,=\, m_2 \,-\, \sigma_1 \,+\,\hbar~,\\
    \stab_{-,(2\,1)} &\,=\, m_1 \,-\, \sigma_1~.\\
    \end{aligned}
\end{equation}
Looking back at~\eqref{eq:Su_T^astGr} and taking $\lambda = (1,0)$ (which corresponds to $\sigma_1$), we get:
\begin{equation}
\begin{aligned}\label{eq:T^astP1}
    \sigma_1\, \,\cup_\sfq\, \stab_{-,(1\,2)} &\,=\, m_1 \,\stab_{-,(1\,2)} \,-\,\hbar\, \stab_{-,(2\,1)}\, -\,\hbar\, \frac{\sfq}{1\,-\,\sfq}\,\Big(\stab_{-,(1\,2)} \,+\, \stab_{-,(2\,1)}\Big)~, \\
    \sigma_1\, \cup_\sfq\, \,\stab_{-,(2\,1)} &\,=\, m_2\, \stab_{-,(2\,1)} \,-\,\hbar\, \frac{\sfq}{1\,-\,\sfq}\,\Big(\stab_{-,(1\,2)} \,+\, \stab_{-,(2\,1)}\Big)~.
\end{aligned}
\end{equation}

On the other hand, the 2d GLSM gives the following BAE~\eqref{2dBAEs}:
\begin{equation}
    (\sigma_1 \,-\, m_1)(\sigma_1 \,-\, m_2) \,=\, \sfq\, (\sigma_1 \,-\, m_1\,-\,\hbar)\,(\sigma_1 \,-\, m_2\,-\,\hbar)~,
\end{equation}
from which we can easily check~\eqref{eq:T^astP1}. Furthermore, we can also compute the quantum product between stable envelopes:
\begin{equation}
    \begin{aligned}\label{eq:TP1stab}
    \stab_{-,(1\,2)}\, \,\cup_\sfq\, \stab_{-,(1\,2)} 
 &\,=\, (m_2 \,-\, m_1)\,\stab_{-,(1\,2)} \,+\, \frac{\hbar}{1\,-\,\sfq}\,\Big(\stab_{-,(1\,2)} \,+\, \stab_{-,(2\,1)}\Big)~,\\
 \stab_{-,(1\,2)}\, \cup_\sfq\, \stab_{-,(2\,1)} 
 &\,=\,  \frac{\hbar}{1\,-\,\sfq}\,\Big(\sfq\, \stab_{-,(1\,2)} \,+\, \stab_{-,(2\,1)}\Big)~,\\
        \stab_{-,(2\,1)}\, \cup_\sfq\, \stab_{-,(2\,1)} 
 &\,=\, (m_1 \,-\, m_2)\,\stab_{-,(2\,1)} \,+\, \hbar\,\frac{\sfq}{1\,-\,\sfq}\,\Big(\stab_{-,(1\,2)} \,+\, \stab_{-,(2\,1)}\Big) ~.
\end{aligned} 
\end{equation}

\subsubsection{Quantum cohomology products for \texorpdfstring{$T^* 
\bP^2$}{TP2}}
In this case, $\rm W^{(1;3)}$ contains three permutation classes, which are $(1\,2\,3), (2\,1\,3), (3\,1\,2)$. The stable envelopes are given by~\eqref{stab-w1}:
\begin{equation}
\begin{aligned}\label{eq:stabP2}
        \stab_{-,(1\,2\,3)} &\,=\, -\,(\sigma_1 \,-\, m_2\,-\,\hbar) \,(\sigma_1 \,-\, m_3\,-\,\hbar)~,\\ 
        \stab_{-,(2\,1\,3)} &\,=\, -\,(\sigma_1 \,-\, m_1)\, (\sigma_1 \,-\, m_3\,-\,\hbar)~,\\
        \stab_{-,(3\,1\,2)} &\,=\, -\,(\sigma_1 \,-\, m_1)\, (\sigma_1 \,-\, m_2)~. 
\end{aligned}
\end{equation}
In~\eqref{eq:Su_T^astGr}, we choose $\lambda = (1,0,0)$ (which, as before corresponds to $\sigma_1$). Then, we have:
\begin{equation}
    \begin{aligned}\label{eq:TP2stab}
        \sigma_1\, \cup_\sfq\, \stab_{-,(1\,2\,3)} 
  &\,=\,-\hbar\,\left(\stab_{-,(2\,1\,3)} \,+\, \stab_{-,(3\,1\,2)} \right) \,+\,  m_1\,\stab_{-,(1\,2\,3)}\\
 &- \hbar\,\frac{\sfq}{1\,-\,\sfq}\left(\stab_{-,(1\,2\,3)} \,+\, \stab_{-,(2\,1\,3)} \,+\, \stab_{-,(3\,1\,2)}\right) ~,\\
    \sigma_1\, \cup_\sfq\, \stab_{-,(2\,1\,3)} 
 &\,=\, -\hbar\,\stab_{-,(3\,1\,2)} \,+\,  m_2\,\stab_{-,(2\,1\,3)}\\ 
 &- \hbar\,\frac{\sfq}{1\,-\,\sfq}\,\left(\stab_{-,(1\,2\,3)} \,+\, \stab_{-,(2\,1\,3)} \,+\, \stab_{-,(3\,1\,2)}\right)~,\\
   \sigma_1\, \cup_\sfq\, \stab_{-,(3\,1\,2)}
 &\,=\,  m_3\,\stab_{-,(3\,1\,2)} \,- \,\hbar\,\frac{\sfq}{1\,-\,\sfq}\,\left(\stab_{-,(1\,2\,3)} \,+\, \stab_{-,(2\,1\,3)} \,+\, \stab_{-,(3\,1\,2)}\right)~.
    \end{aligned} 
\end{equation}

For this GLSM model, the BAE is given by~\eqref{2dBAEs}:\footnote{We redefine $\sfq$ to $-\sfq$ in~\eqref{eq:BAEP2}, and~\eqref{eq:BAEGr24} in order to match with~\eqref{eq:Su_T^astGr}.}
\begin{equation}\label{eq:BAEP2}
    (\sigma_1 \,-\, m_1)(\sigma_1 \,-\, m_2) (\sigma_1 \,-\, m_3)\,=\, \sfq (\sigma_1 \,-\, m_1\,-\,\hbar)(\sigma_1 \,-\, m_2\,-\,\hbar)(\sigma_1 \,-\, m_3\,-\,\hbar)\,.
\end{equation}
Using it, we can verify~\eqref{eq:TP2stab} straightforwardly. Moreover, we can also compute the quantum product between stable envelopes, which are summarized in Appendix~\ref{append:TastP2}.

\subsubsection{Quantum cohomology products for $T^* 
{\rm Gr}(2,4)$}
For $T^* 
{\rm Gr}(2,4)$, there are six permutation classes in $\rm W^{(2;4)}$:
\begin{equation}
    (1\,2\,3\,4)~,\quad (1\,3\,2\,4)~,\quad (1\,4\,2\,3)~,\quad (2\,3\,1\,4)~,\quad (2\,4\,1\,3)~,\quad (3\,4\,1\,2)~.
\end{equation}
 We have the following formulas for $\stab_{-,w}$ in $T^\ast {\rm Gr}(2,4)$~\eqref{stab-w}:
\begin{equation}
\begin{aligned}
        \stab_{-,(1\,2\,3\,4)} &\,=\, \operatorname*{\sf Sym}_{\sigma_1, \sigma_2} \left[\frac{\prod_{i=2}^4 (\hbar\,+\,m_i\,-\,\sigma_1)\,(m_1 \,-\, \sigma_2)\,\prod_{j=3}^4 (\hbar\,+\,m_j\,-\,\sigma_2)}{(\sigma_1\, -\, \sigma_2 \,-\,\hbar)\,(\sigma_2 \,-\, \sigma_1)} \right]~,\\
    \stab_{-,(1\,3\,2\,4)} &\,=\,\operatorname*{\sf Sym}_{\sigma_1, \sigma_2} \left[\frac{\prod_{i=2}^4 (\hbar\,+\,m_i\,-\,\sigma_1)\,(\hbar\,+\,m_4 \,-\, \sigma_2)\,\prod_{j=1}^2 (m_j\,-\,\sigma_2)}{-(\sigma_1 \,-\, \sigma_2 \,-\,\hbar)(\sigma_2 \,-\, \sigma_1)} \right]~,\\
    \stab_{-,(1\,4\,2\,3)} &\,=\,\operatorname*{\sf Sym}_{\sigma_1, \sigma_2} \left[\frac{\prod_{i=2}^4 (\hbar\,+\,m_i\,-\,\sigma_1)\,\prod_{j=1}^3 (m_j\,-\,\sigma_2)}{(\sigma_1 \,-\, \sigma_2 \,-\,\hbar)\,(\sigma_2 \,-\, \sigma_1)} \right]~,\\
    \stab_{-,(2\,3\,1\,4)} &\,=\,\operatorname*{\sf Sym}_{\sigma_1, \sigma_2} \left[\frac{(m_1 - \sigma_1)\prod_{i=3}^4 (\hbar+m_i-\sigma_1)\prod_{j=1}^2 (m_j-\sigma_2) (\hbar+m_4 - \sigma_2)}{(\sigma_1 - \sigma_2 -\hbar)(\sigma_2 - \sigma_1)} \right]~,\\
        \stab_{-,(2\,4\,1\,3)} &=\operatorname*{\sf Sym}_{\sigma_1, \sigma_2} \left[\frac{( m_1\, -\, \sigma_1)\,\prod_{i=3}^4 (\hbar \,+\, m_i\,-\,\sigma_1)\,\prod_{j=1}^3 (m_j\,-\,\sigma_2) }{-(\sigma_1 \,-\, \sigma_2 \,-\,\hbar)\,(\sigma_2 \,-\, \sigma_1)} \right]~,\\
            \stab_{-,(3\,4\,1\,2)} &\,=\,\operatorname*{\sf Sym}_{\sigma_1, \sigma_2} \left[\frac{(\hbar\, +\, m_4 \,-\, \sigma_1)\,\prod_{i=1}^2 (m_i \,-\,\sigma_1)\,\prod_{j=1}^3 (m_j\,-\,\sigma_2)}{(\sigma_1 \,-\, \sigma_2 \,-\,\hbar)\,(\sigma_2 \,-\, \sigma_1)} \right]~.
\end{aligned}
\end{equation}
Let us consider the quantum multiplication of $D_\lambda$ and $\stab_{-,w}$ in~\eqref{eq:Su_T^astGr}. For simplicity, we choose $\stab_{-,(1\,2\,3\,4)}$ and $\lambda = (1,1,0,0)$ which corresponds to $\sigma_1 + \sigma_2$. Then, the product boils down to the reduction of $(\sigma_1 + \sigma_2) \stab_{-,(1\,2\,3\,4)}$ in the polynomial ring quotient by the Bethe ansatz equations.

This is our first example where $\stab_{-,w}$ has a non-trivial denominator. Denote the numerator and denominator of a rational function $f$ by Num($f$) and Den($f$), respectively. From the explicit forms above, all six $\stab_{-,w}$ have the same denominator:
\begin{equation}\label{eq:DenT^astGr24}
     \text{Den}(\stab_{-,w}) \,=\, (\hbar \,+\, \sigma_1 \,-\, \sigma_2) (\hbar \,+\, \sigma_2 \,-\, \sigma_1)~,\qquad \forall w\,\in\,{\rm W}^{(2;4)}~.
\end{equation}
Moreover, we need to forbid the locus $\sigma_1 = \sigma_2$ because it corresponds to a nonphysical solution~\eqref{2dSBE}. Following our general strategy, we define the saturated ideal:
\begin{equation}
    J \,=\, \CI_{\rm Bethe} : \Big((\hbar + \sigma_1 - \sigma_2) (\hbar + \sigma_2 - \sigma_1)(\sigma_1 - \sigma_2)\Big)^\infty~,
\end{equation}
where $\CI_{\rm Bethe}$ is generated by:
\begin{equation}\label{eq:BAEGr24}
    \begin{aligned}
        (\sigma_1\,-\,\sigma_2\,-\,\hbar)\,\prod_{i=1}^4(\sigma_1 \,-\, m_i) &\,=\, \sfq \,(\sigma_1\,-\,\sigma_2\,+\,\hbar)\,
\prod_{i=1}^4(\sigma_1 \,-\, m_i \,-\,\hbar)~,\\
(\sigma_2\,-\,\sigma_1\,-\,\hbar)\,\prod_{i=1}^4(\sigma_2\, -\, m_i) &\,=\, \sfq \,(\sigma_2\,-\,\sigma_1\,+\,\hbar)\,
\prod_{i=1}^4(\sigma_2 \,-\, m_i \,-\,\hbar)~.
\end{aligned}
\end{equation}

Then we simply compute the normal form of $(\sigma_1 +\sigma_2)\text{Num}(\stab_{-,( 1\,2\,3\,4)})$ in the quotient ring
\begin{equation}
\mathbb{K}[\sigma_1,\sigma_2]^{S_2}/J\,, \qquad \mathbb K\, =\, \mathbb{C}[q,\hbar,m_1,m_2,m_3,m_4]\,.   
\end{equation}
We also compute the normal form for the numerator of all the stable envelopes and find the following linear relation:
{\small\begin{equation}
\begin{aligned}
    (\sigma_1 \,+\,\sigma_2)\, \cup_\sfq\,&\text{Num}(\stab_{-,( 1\,2\,3\,4)}) \,=\,(m_1\,+\,m_2\,-2\,\hbar\,\frac{\sfq}{1\,-\,\sfq})\,\text{Num}(\stab_{-,( 1\,2\,3\,4)})\\
    &+\,\frac{\hbar}{1\,-\,\sfq}\,\text{Num}\big(\stab_{-,( 1\,3\,2\,4)} \,-\, \stab_{-,( 1\,4\,2\,3)} \,-\, \stab_{-,( 2\,3\,1\,4)} \,+\, \stab_{-,( 2\,4\,1\,3)} \big)~,
    \end{aligned}
\end{equation}}
which matches precisely~\eqref{eq:Su_T^astGr} after dividing $\text{Den}(\stab_{-,w})$ on both sides. Furthermore, we can compute quantum products between stable envelopes, which are quite lengthy, so we omit the formulas here.

\subsection{Quantum K-theory product}

The above discussion can be further extended to the quantum K-theory case, where now we look at the 3d A-model of $X_{k,n}$. In this case, we take the quotient ring to be the 3d twisted chiral ring $\mathcal{R}^{\rm 3d}$ in~\eqref{QK=R3d}. Similar to what we did in the previous subsection, we will use the extended Gr\"obner algorithm to compute quantum-deformed tensor products involving $\kstab_{-,w}$. We will do this for the cases $T^*\bP^1$, $T^*\bP^2$, and $T^*{\rm Gr}(2,4)$, although our techniques can be applied to a more general $X_{k,n}$ case. 

\subsubsection{Quantum K-theory products for \texorpdfstring{$T^* \bP^1$}{TP1}}
There are two permutations classes in $\rm W^{(1;2)}$: $(1\, 2)$ and $(2\, 1)$. The K-theoretic stable envelopes are:
\begin{equation}
\begin{aligned}
    \kstab_{-,(1\,2)} &\,=\, -\,{\sf y}\,+\,\frac{x_1}{\,y_2} ~,\\
    \kstab_{-,(2\,1)} &\,=\,-\,{\sf y}\,+\,\frac{{\sf y}\, x_1}{y_1}~.
\end{aligned}
\end{equation}
From the 3d GLSM, we get the following BAE~\eqref{3dBAEs}:
\begin{equation}
    (x_1\,-\,y_1)\,(x_1\,-\,y_2) \,=\, q\, (x_1\,-\,{\sf y}\,y_1)\,(x_1\,-\,{\sf y}\,y_2)~.
\end{equation}
First we can determine the quantum K-theory product between $\rm det\, \mathcal S$ and K-theoretic stable envelopes:
\begin{equation}
\begin{aligned}
        x_1 \,\otimes_q\, \kstab_{-,(1\,2)} &\,=\, y_1 \,\left(\kstab_{-,(1\,2)} \,-\, \left(1\,-\,\frac{1}{\sf y}\right)\,\kstab_{-,(2\,1)}\right)\\&\, +\, q\, y_1\,\frac{(1\,-\,{\sf y})}{1\,-\,q}\left(\kstab_{-,(1\,2)} \,+\, \frac{1}{\sf y}\,\kstab_{-,(2\,1)}\right)~,\\
        x_1\, \otimes_q\, \kstab_{-,(2\,1)} &\,=\, y_2\, \kstab_{-,(2\,1)}\, +\, \frac{q\,(1\,-\,{\sf y})\,y_2}{1\,-\,q}\left({\sf y}\, \kstab_{-,(1\,2)}\, +\, \kstab_{-,(2\,1)}\right)~.
\end{aligned}
\end{equation}
The quantum products between K-theoretic stable envelopes can be similarly determined:
\begin{equation}
    \begin{aligned}
    \kstab_{-,(1\,2)} \,\otimes_q\, \kstab_{-,(1\,2)} &\,=\, \left(\frac{y_1}{\,y_2}\,-\,{\sf y}\right)\,\kstab_{-,(1\,2)}\, +\, \left(\frac{1}{\sf y}\,-\,1\right)\,\frac{y_1}{y_2}\,\kstab_{-,(2\,1)} \,-\, \frac{y_1}{y_2}\,\Delta^{(1)}_q~,\\
        \kstab_{-,(1\,2)} \,\otimes_q \,\kstab_{-,(2\,1)} &\,=\, \left(1\,-\,\sf y\right)\, \kstab_{-,(2\,1)} \,-\,\sf y\, \Delta^{(1)}_q~,\\
        \kstab_{-,(2\,1)} \,\otimes_q\, \kstab_{-,(2\,1)} &\,=\, \left(-\,\sf y\,+\,\frac{\sf{y}\, y_2}{y_1}\right)\,\kstab_{-,(2\,1)} \,-\, \frac{{\sf y^2}\,y_2}{y_1} \,\Delta^{(1)}_q~,
    \end{aligned} 
\end{equation}
where:
\begin{equation}
    \Delta^{(1)}_q \,\coloneqq \,q\,\frac{{\sf y}\,-\,1}{1\,-\,q}\,\left(\kstab_{-,(1\,2)} \,+\, \frac{1}{\sf y}\,\kstab_{-,(2\,1)}\right)~.
\end{equation}

Recall $x_i = e^{\boldsymbol\beta \sigma_i}, y_i = e^{\boldsymbol\beta m_i}, \sf {y} = e^{\boldsymbol\beta \hbar}$. In the limit where the radius of $\mathbb{S}^1_{\boldsymbol\beta}$ goes to zero, we find:
\begin{equation}
    \begin{aligned}
        \kstab_{-,(1\,2)} = -\,\boldsymbol\beta\left(m_2\,-\,\sigma\,+\,\hbar\right)\, +\, O(\boldsymbol\beta^2),\quad
        \kstab_{-,(2\,1)} = -\,\boldsymbol\beta\left(m_1\,-\,\sigma\right) \, +\, O(\boldsymbol\beta^2)\,.\\
    \end{aligned}
\end{equation}
Using the above formulas, one can show that when $\boldsymbol\beta \rightarrow 0$, the quantum K-products reduce to the quantum cup products in Section~\ref{subsec:QHP1}.

\subsubsection{Quantum K-theory products for \texorpdfstring{$T^* \bP^2$}{TP2}}
There are three permutation classes in $\rm W^{(1;3)}$: $(1\,2\,3)$, $(2\,1\,3)$ and $(3\,1\,2)$. The K-theoretic stable envelopes are~\eqref{kstabw1}:
\begin{equation}
\begin{aligned}\label{eq:KstabP2}
        \kstab_{-,(1\,2\,3)} &\,=\,\left(-\,\sf y\right)\, \left(1\,-\,\frac{x}{{\sf y}\,y_2}\right)\,\left(1\,-\,\frac{x}{{\sf y}\,y_3}\right)~,\\
        \kstab_{-,(2\,1\,3)} &\,=\,\left(-\,\sf y\right)\, \left(1\,-\,\frac{x}{y_1}\right)\,\left(1\,-\,\frac{x}{{\sf y}\,y_3}\right)~,\\
        \kstab_{-,(3\,1\,2)} &\,=\,\left(-\,\sf y\right)\, \left(1\,-\,\frac{x}{y_1}\right)\,\left(1\,-\,\frac{x}{y_2}\right)~.
\end{aligned}
\end{equation}
From 3d GLSM we get the following BAE~\eqref{3dBAEs}:
\begin{equation}
  (x_1\,-\,y_1)\,(x_1\,-\,y_2)\,(x_1\,-\,y_3) \,=\, q\, (x_1\,-\,{\sf y}\,y_1)\,(x_1\,-\,{\sf y}\,y_2)\,(x_1\,-\,{\sf y}\,y_3)~.
\end{equation}

Then we compute the quantum K-product between $\rm det\, \mathcal S$ and K-theoretic stable envelopes:
\begin{equation}
    \begin{aligned}
        x_1 \,\otimes_q\, \kstab_{-,(1\,2\,3)} &\,=\, y_1\,\kstab_{-,(1\,2\,3)} \,-\, \left(1\,-\,\frac{1}{\sf y}\right)\,y_1\, \kstab_{-,(2\,1\,3)}  \,\\&\,-\, \left(\frac{1}{\sf y}\,-\,\frac{1}{\sf y^2}\right)\,y_1\,\kstab_{-,(3\,1\,2)} \,+\, y_1 \,\Delta^{(2)}_q~,\\
        x_1 \,\otimes_q\, \kstab_{-,(2\,1\,3)} &\,=\,y_2\, \kstab_{-,(2\,1\,3)} \,-\, \left(1\,-\,\frac{1}{\sf y}\right)\,y_2\, \kstab_{-,(3\,1\,2)} \,+\, {\sf y}\,y_2\,\Delta^{(2)}_q~,\\
        x_1 \,\otimes_q\, \kstab_{-,(3\,1\,2)} &\,=\,y_3\, \kstab_{-,(3\,1\,2)}\,+\, {\sf y^2}\,y_3\,\Delta^{(2)}_q~,
    \end{aligned}
\end{equation}
where:
\begin{equation}
    \Delta^{(2)}_q \,\coloneqq\, q\,\frac{1\,-\,\sf y}{1\,-\,q}\,\left(\kstab_{-,(1\,2\,3)} \,+\, \frac{1}{\sf y}\,\kstab_{-,(2\,1\,3)} \,+\, \frac{1}{\sf y^2}\,\kstab_{-,(3\,1\,2)}\right)\,.
\end{equation}
The quantum products between K-theoretic stable envelopes can be similarly determined, which we collect in Appendix~\ref{append:TastP2}.

\subsubsection{Quantum K-theory products for \texorpdfstring{$T^* {\rm Gr}(2,4)$}{T*Gr24}}
There are six permutation classes in $\rm W^{(2;4)}$ for $T^* 
{\rm Gr}(2,4)$, 
\begin{equation}
    (1\,2\,3\,4)~,\quad (1\,3\,2\,4)~,\quad (1\,4\,2\,3)~,\quad (2\,3\,1\,4)~,\quad (2\,4\,1\,3)~,\quad (3\,4\,1\,2)~.
\end{equation}
 We have the following formulas for $\kstab_{-,w}$:
 \begin{equation}
\kstab_{-,(w_1\,w_2\,\cdots)} = \,\left(-\,\sf y\right)^{-\varepsilon(w)}\,\operatorname*{\sf Sym}_{x_1, x_2} \left[\frac{F_{w_1}(x_1)\,F_{w_2}(x_2)}{\left(1\,-\,\frac{x_2}{x_1}\right)\,\left(1\,-\,\frac{x_1}{{\sf y}\,x_2}\right)} \right]~,
 \end{equation}
 where we define:
 \begin{equation}
         F_i(x) \,\coloneqq\, \prod_{j=1}^{i-1}\, \left(1\,-\,\frac{x}{y_j}\right)\, \prod_{j=i+1}^4\left(1\,-\,\frac{x}{{\sf y}\,y_j}\right)~.
 \end{equation}
The Bethe ideal $\CI^{\rm 3d}_{\rm Bethe}$ is generated by the following BAEs~\eqref{3dBAEs}:
\begin{equation}
\begin{aligned}
    (x_1 \,-\,{\sf y}\,x_2)\,\prod_{i=1}^4(x_1 \,-\, {y}_i) &\,=\, q\, ({\sf y}\,x_1\, -\,x_2)\,\prod_{i=1}^4 (x_1 \,-\, {\sf y}\,y_i)~,\\
    (x_2 \,-\,{\sf y}\,x_1)\,\prod_{i=1}^4(x_2\, -\, { y}_i) &\,=\, q \,({\sf y}\,x_2\, -\,x_1)\,\prod_{i=1}^4 (x_2 \,-\, {\sf y}\,y_i)~.
\end{aligned}
\end{equation}
Let us consider the quantum K-product between $\rm det\, \mathcal S$ and $\kstab_{-,w}$. We make use of the strategy outlined at the beginning of this section. From the expressions of $\kstab_{-,w}$, the common denominator can be chosen to be $(x_1 \,-\, {\sf y}\,x_2)(x_2 \,-\, {\sf y}\,x_1)$. Furthermore, we need to forbid the non-physical locus where $x_1 = x_2$ or $x_1 x_2 = 0$ \eqref{eq:SBE3d}. So, we define a new ideal:
\begin{equation}
    J \,\coloneqq\, \CI^{\rm 3d}_{\rm Bethe} : \Big(x_1 x_2(x_1 - x_2)(x_1 - {\sf y}x_2)(x_2 - {\sf y}x_1)\Big)^\infty~.
\end{equation}
Then we simply compute the normal form of $(x_1 x_2)\text{Num}(\kstab_{-,( 1\,2\,3\,4)})$ in the quotient ring:
\begin{equation}
\mathbb{K}[x_1,x_2]^{S_2}/J~,\qquad \mathbb{K}\,=\,\mathbb{C}[q,{\sf y},y_1,y_2,y_3,y_4]\,,
\end{equation}
and re-express the result in terms of linear combinations of $\text{Num}(\kstab_{-,w})$. The final formula is obtained by dividing the common denominator. The upshot is:
\begingroup
\allowdisplaybreaks[1]
\begin{align}
&(x_1x_2)\otimes_q\kstab_{-,(1\,2\,3\,4)}=\frac{y_1y_2}{1-q}\Bigl[
 (1-q\mathsf y^2)\kstab_{-,(1\,2\,3\,4)}
 +\frac{(1-\mathsf y)^2(1+\mathsf y)}
 {(1-q\mathsf y)\mathsf y^4}\kstab_{-,(3\,4\,1\,2)}\notag\\*
&+\frac{(1-\mathsf y)(1-q\mathsf y^2)}{1-q\mathsf y}
 \Bigl(-\kstab_{-,(1\,3\,2\,4)}+\kstab_{-,(1\,4\,2\,3)}
 +\frac{\kstab_{-,(2\,3\,1\,4)}-\kstab_{-,(2\,4\,1\,3)}}{\mathsf y^2}
 \Bigr)\Bigr]\,,\notag\\[4pt]
&(x_1x_2)\otimes_q\kstab_{-,(1\,3\,2\,4)}=\frac{y_1y_3}{1-q}\Bigl[
 (1-q\mathsf y^2)\kstab_{-,(1\,3\,2\,4)}
 +\frac{(1-\mathsf y)^2(1+q\mathsf y)}
 {(1-q\mathsf y)\mathsf y^2}\kstab_{-,(2\,4\,1\,3)}\notag\\*
&-\frac{(1-\mathsf y)(1-q\mathsf y^2)}{1-q\mathsf y}
 \Bigl(q\kstab_{-,(1\,2\,3\,4)}+\kstab_{-,(1\,4\,2\,3)}
 +\frac{\kstab_{-,(2\,3\,1\,4)}}{\mathsf y^2}
 +\frac{\kstab_{-,(3\,4\,1\,2)}}{\mathsf y^4}
 \Bigr)\Bigr]\,,\notag\\[4pt]
&(x_1x_2)\otimes_q\kstab_{-,(1\,4\,2\,3)}=\frac{y_1y_4}{1-q}\Bigl[
 (1-q\mathsf y^2)\kstab_{-,(1\,4\,2\,3)}
 +\frac{q(1-\mathsf y)^2(1+\mathsf y)}
 {(1-q\mathsf y)\mathsf y^2}\kstab_{-,(2\,3\,1\,4)}\notag\\*
&+\frac{(1-\mathsf y)(1-q\mathsf y^2)}{1-q\mathsf y}
 \Bigl(q\kstab_{-,(1\,2\,3\,4)}-q\kstab_{-,(1\,3\,2\,4)}
 -\frac{\kstab_{-,(2\,4\,1\,3)}}{\mathsf y^2}
 +\frac{\kstab_{-,(3\,4\,1\,2)}}{\mathsf y^4}
 \Bigr)\Bigr]\,,\notag\\[4pt]
&(x_1x_2)\otimes_q\kstab_{-,(2\,3\,1\,4)}=\frac{y_2y_3}{1-q}\Bigl[
 (1-q\mathsf y^2)\kstab_{-,(2\,3\,1\,4)}
 +\frac{q\mathsf y^2(1-\mathsf y)^2(1+\mathsf y)}
 {1-q\mathsf y}\kstab_{-,(1\,4\,2\,3)}\notag\\*
&+\frac{(1-\mathsf y)(1-q\mathsf y^2)}{1-q\mathsf y}
 \Bigl(q\mathsf y^2\kstab_{-,(1\,2\,3\,4)}
 -q\mathsf y^2\kstab_{-,(1\,3\,2\,4)}
 -\kstab_{-,(2\,4\,1\,3)}
 +\frac{\kstab_{-,(3\,4\,1\,2)}}{\mathsf y^2}
 \Bigr)\Bigr]\,,\notag\\[4pt]
&(x_1x_2)\otimes_q\kstab_{-,(2\,4\,1\,3)}=\frac{y_2y_4}{1-q}\Bigl[
 (1-q\mathsf y^2)\kstab_{-,(2\,4\,1\,3)}
 +\frac{q\mathsf y^2(1-\mathsf y)^2(1+q\mathsf y)}
 {1-q\mathsf y}\kstab_{-,(1\,3\,2\,4)}\notag\\*
&-\frac{(1-\mathsf y)(1-q\mathsf y^2)}{1-q\mathsf y}
 \Bigl(q\mathsf y^2\kstab_{-,(1\,2\,3\,4)}
 +q\mathsf y^2\kstab_{-,(1\,4\,2\,3)}
 +q\kstab_{-,(2\,3\,1\,4)}
 +\frac{\kstab_{-,(3\,4\,1\,2)}}{\mathsf y^2}
 \Bigr)\Bigr]\,,\notag\\[4pt]
&(x_1x_2)\otimes_q\kstab_{-,(3\,4\,1\,2)}=\frac{y_3y_4}{1-q}\Bigl[
 (1-q\mathsf y^2)\kstab_{-,(3\,4\,1\,2)}
 +\frac{q^2\mathsf y^4(1-\mathsf y)^2(1+\mathsf y)}
 {1-q\mathsf y}\kstab_{-,(1\,2\,3\,4)}\notag\\*
&+\frac{q\mathsf y^2(1-\mathsf y)(1-q\mathsf y^2)}{1-q\mathsf y}
 \Bigl(-\mathsf y^2\kstab_{-,(1\,3\,2\,4)}
 +\mathsf y^2\kstab_{-,(1\,4\,2\,3)}
 +\kstab_{-,(2\,3\,1\,4)}-\kstab_{-,(2\,4\,1\,3)}
 \Bigr)\Bigr]\,.
\end{align}
\endgroup

 \subsubsection*{Acknowledgments} 
 We would like to thank Cyril Closset, Yehao Zhou, and Hao Zou for discussions and correspondence. OK and PC would like to thank the organizers of the GLSM conference at Mainz Institute for Theoretical Physics (MITP) where this project was initiated. ZD also thanks MITP
 for hospitality during which part of the calculations in Section~\ref{sec:Grobner} were performed. The work of PC is supported by the Cluster of Excellence PRISMA++ (EXC
2118/2, Project ID 390831469). The research of ZD is supported by a startup grant from SIMIS. OK is a Junior Research Fellow at the Philippe Meyer Institute. 

\subsubsection*{Declaration of AI use}
The authors acknowledge the use of ChatGPT and Claude for figure drawing and writing of Mathematica code for Section~\ref{sec:quantumProd}. The main proposal for the defect and all the other calculations were fully performed by the authors.
The authors take full responsibility for the final content.

\appendix
\section{Quantum products between Stable envelopes of \texorpdfstring{$T^* 
\bP^2$}{TP2}}\label{append:TastP2}
In this appendix, we collect formulas for quantum cup and K products between stable envelopes of $T^\ast \mathbb P^2$.

\subsection{Quantum cohomology products}
Using the Gr\"obner basis method, we can work out all the quantum cup products between stable envelopes defined in~\eqref{eq:stabP2}. The remaining products follow from commutativity.
\begingroup
\small
\setlength{\jot}{1pt}
\allowdisplaybreaks[1]
\begin{align}
&\stab_{-,(1\,2\,3)}\,\cup_\sfq\,\stab_{-,(1\,2\,3)}
=\notag\\*
&-\frac{1}{(1-\sfq)^2}\Big(
 \big[\big((1-\sfq)(m_1-m_2)-\hbar\big)
 \big((1-\sfq)(m_1-m_3)-\hbar\big)+2\sfq\hbar^2\big]
 \stab_{-,(1\,2\,3)}\notag\\*
&\qquad+\hbar\big((1-\sfq)(m_3-m_1)+(2+\sfq)\hbar\big)
 \stab_{-,(2\,1\,3)}\notag\\*
&\qquad+\hbar\big((1-\sfq)(m_2-m_1)+3\hbar\big)
 \stab_{-,(3\,1\,2)}\Big)\,.\notag\\[4pt]
&\stab_{-,(1\,2\,3)}\,\cup_\sfq\,\stab_{-,(2\,1\,3)}
=\notag\\*
&-\frac{1}{(1-\sfq)^2}\Big(
 \sfq\hbar\big((1-\sfq)(m_3-m_1)+3\hbar\big)
 \stab_{-,(1\,2\,3)}\notag\\*
&\qquad+\hbar\big((1-\sfq)(m_3-m_2)+(1+2\sfq)\hbar\big)
 \stab_{-,(2\,1\,3)}
 +(2+\sfq)\hbar^2\stab_{-,(3\,1\,2)}\Big)\,.\notag\\[4pt]
&\stab_{-,(1\,2\,3)}\,\cup_\sfq\,\stab_{-,(3\,1\,2)}
=\notag\\*
&-\frac{1}{(1-\sfq)^2}\Big(
 \sfq\hbar\big((1-\sfq)(m_2-m_1)+(2+\sfq)\hbar\big)
 \stab_{-,(1\,2\,3)}
 +3\sfq\hbar^2\stab_{-,(2\,1\,3)}\notag\\*
&\qquad+\hbar\big((1-\sfq)(m_2-m_3)+(1+2\sfq)\hbar\big)
 \stab_{-,(3\,1\,2)}\Big)\,.\notag\\[4pt]
&\stab_{-,(2\,1\,3)}\,\cup_\sfq\,\stab_{-,(2\,1\,3)}
=\notag\\*
&-\frac{1}{(1-\sfq)^2}\Big(
 \sfq\hbar\big((1-\sfq)(m_3-m_2)+(2+\sfq)\hbar\big)
 \stab_{-,(1\,2\,3)}\notag\\*
&\qquad+\big[\big((1-\sfq)(m_1-m_2)+\sfq\hbar\big)
 \big((1-\sfq)(m_3-m_2)+\hbar\big)+2\sfq\hbar^2\big]
 \stab_{-,(2\,1\,3)}\notag\\*
&\qquad+\hbar\big((1-\sfq)(m_1-m_2)+(1+2\sfq)\hbar\big)
 \stab_{-,(3\,1\,2)}\Big)\,.\notag\\[4pt]
&\stab_{-,(2\,1\,3)}\,\cup_\sfq\,\stab_{-,(3\,1\,2)}
=\notag\\*
&-\frac{1}{(1-\sfq)^2}\Big(
 \sfq(1+2\sfq)\hbar^2\stab_{-,(1\,2\,3)}
 +\sfq\hbar\big((1-\sfq)(m_1-m_2)+(2+\sfq)\hbar\big)
 \stab_{-,(2\,1\,3)}\notag\\*
&\qquad+\hbar\big((1-\sfq)(m_1-m_3)+3\sfq\hbar\big)
 \stab_{-,(3\,1\,2)}\Big)\,.\notag\\[4pt]
&\stab_{-,(3\,1\,2)}\,\cup_\sfq\,\stab_{-,(3\,1\,2)}
=\notag\\*
&-\frac{1}{(1-\sfq)^2}\Big(
 \sfq\hbar\big((1-\sfq)(m_2-m_3)+3\sfq\hbar\big)
 \stab_{-,(1\,2\,3)}\notag\\*
&\qquad+\sfq\hbar\big((1-\sfq)(m_1-m_3)+(1+2\sfq)\hbar\big)
 \stab_{-,(2\,1\,3)}\notag\\*
&\qquad+\big[\big((1-\sfq)(m_2-m_3)+\sfq\hbar\big)
 \big((1-\sfq)(m_1-m_3)+\sfq\hbar\big)+2\sfq\hbar^2\big]
 \stab_{-,(3\,1\,2)}\Big)\,.
\end{align}
\endgroup

\subsection{Quantum K-theory products}
Similarly, using the Gr\"obner basis method, we can work out all the
quantum K products between K-theoretic stable envelopes defined in
\eqref{eq:KstabP2}.
The remaining products follow from commutativity.
\begingroup
\small
\setlength{\jot}{1pt}
\allowdisplaybreaks[1]
\begin{align}
&\kstab_{-,(1\,2\,3)}\,\otimes_q\,\kstab_{-,(1\,2\,3)} =\notag\\*
&-\frac{1}{\sfy\,y_2y_3(1-q)^2}\Big(
        \big[\big((1-q\sfy)y_1-(1-q)\sfy\,y_2\big)
        \big((1-q\sfy)y_1-(1-q)\sfy\,y_3\big)\notag\\*
&\qquad\qquad
        +q(\sfy-1)^2y_1(y_2+y_3)\big]\kstab_{-,(1\,2\,3)}\notag\\*
&\qquad+\frac{(\sfy-1)y_1}{\sfy}
        \big((q\sfy-1)y_1+(\sfy-1)y_2+(\sfy-q)y_3\big)
        \kstab_{-,(2\,1\,3)}\notag\\*
&\qquad+\frac{(\sfy-1)y_1}{\sfy^2}
        \big((q\sfy-1)y_1+\big((2-q)\sfy-1\big)y_2+(\sfy-1)y_3\big)
        \kstab_{-,(3\,1\,2)}\Big)\,.\notag\\[4pt]
&\kstab_{-,(1\,2\,3)}\,\otimes_q\,\kstab_{-,(2\,1\,3)} =\notag\\*
&-\frac{\sfy-1}{\sfy^2y_3(1-q)^2}\Big(
        q\sfy^2\big((q\sfy-1)y_1+(\sfy-1)y_2
        +\big((2-q)\sfy-1\big)y_3\big)
        \kstab_{-,(1\,2\,3)}\notag\\*
&\qquad+\sfy\big(q(\sfy-1)y_1+(q\sfy-1)y_2+(\sfy-q)y_3\big)
        \kstab_{-,(2\,1\,3)}\notag\\*
&\qquad+(\sfy-1)(qy_1+y_2+y_3)
        \kstab_{-,(3\,1\,2)}\Big)\,.\notag\\[4pt]
&\kstab_{-,(1\,2\,3)}\,\otimes_q\,\kstab_{-,(3\,1\,2)} =\notag\\*
&-\frac{\sfy-1}{\sfy\,y_2(1-q)^2}\Big(
        q\sfy^2\big((q\sfy-1)y_1+(\sfy-q)y_2+(\sfy-1)y_3\big)
        \kstab_{-,(1\,2\,3)}\notag\\*
&\qquad+q\sfy(\sfy-1)(y_1+y_2+y_3)
        \kstab_{-,(2\,1\,3)}\notag\\*
&\qquad+\big(q(\sfy-1)y_1+(\sfy-q)y_2+(q\sfy-1)y_3\big)
        \kstab_{-,(3\,1\,2)}\Big)\,.\notag\\[4pt]
&\kstab_{-,(2\,1\,3)}\,\otimes_q\,\kstab_{-,(2\,1\,3)} =\notag\\*
&-\frac{1}{y_1y_3(1-q)^2}\Big(
        q\sfy(\sfy-1)y_2
        \big(q(\sfy-1)y_1+(q\sfy-1)y_2
        +\big((2-q)\sfy-1\big)y_3\big)
        \kstab_{-,(1\,2\,3)}\notag\\*
&\qquad+\big[\big((1-q)y_1-(1-q\sfy)y_2\big)
        \big((1-q)\sfy\,y_3-(1-q\sfy)y_2\big)\notag\\*
&\qquad\qquad
        +q(\sfy-1)^2y_2(y_1+y_3)\big]\kstab_{-,(2\,1\,3)}\notag\\*
&\qquad+\frac{(\sfy-1)y_2}{\sfy}
        \big((1-2q+q\sfy)y_1+(q\sfy-1)y_2+(\sfy-1)y_3\big)
        \kstab_{-,(3\,1\,2)}\Big)\,.\notag\\[4pt]
&\kstab_{-,(2\,1\,3)}\,\otimes_q\,\kstab_{-,(3\,1\,2)} =\notag\\*
&-\frac{\sfy-1}{y_1(1-q)^2}\Big(
        q\sfy^2(\sfy-1)(qy_1+qy_2+y_3)
        \kstab_{-,(1\,2\,3)}\notag\\*
&\qquad+q\sfy\big((\sfy-q)y_1+(q\sfy-1)y_2+(\sfy-1)y_3\big)
        \kstab_{-,(2\,1\,3)}\notag\\*
&\qquad+\big((1-2q+q\sfy)y_1+q(\sfy-1)y_2+(q\sfy-1)y_3\big)
        \kstab_{-,(3\,1\,2)}\Big)\,.\notag\\[4pt]
&\kstab_{-,(3\,1\,2)}\,\otimes_q\,\kstab_{-,(3\,1\,2)} =\notag\\*
&-\frac{\sfy}{y_1y_2(1-q)^2}\Big(
        q\sfy^2(\sfy-1)y_3
        \big(q(\sfy-1)y_1+(1-2q+q\sfy)y_2+(q\sfy-1)y_3\big)
        \kstab_{-,(1\,2\,3)}\notag\\*
&\qquad+q\sfy(\sfy-1)y_3
        \big((\sfy-q)y_1+q(\sfy-1)y_2+(q\sfy-1)y_3\big)
        \kstab_{-,(2\,1\,3)}\notag\\*
&\qquad+\big[\big((1-q)y_1-(1-q\sfy)y_3\big)
        \big((1-q)y_2-(1-q\sfy)y_3\big)\notag\\*
&\qquad\qquad
        +q(\sfy-1)^2y_3(y_1+y_2)\big]\kstab_{-,(3\,1\,2)}\Big)\,.
\end{align}
\endgroup

\renewcommand{\baselinestretch}{1.2} 
\bibliography{QGQK} \addcontentsline{toc}{section}{References}

@book{MaulikOkounkov2019,
  author        = {Davesh Maulik and Andrei Okounkov},
  title         = {Quantum Groups and Quantum Cohomology},
  series        = {Ast\'erisque},
  volume        = {408},
  publisher     = {Soci\'et\'e Math\'ematique de France},
  address       = {Paris},
  year          = {2019},
  doi           = {10.24033/ast.1074},
  isbn          = {978-2-85629-900-5},
  eprint        = {1211.1287},
  archivePrefix = {arXiv},
  primaryClass  = {math.AG}
}

@article{Brunner:2026fxy,
    author = "Brunner, Ilka and Cheng, Peng and Jockers, Hans",
    title = "{Quantum Elliptic Cohomology From Four-Dimensional Minimal Supersymmetric Gauge Theories}",
    eprint = "2609.09298",
    archivePrefix = "arXiv",
    primaryClass = "hep-th",
    reportNumber = "LMU-ASC 04/26, MITP/26-041",
    month = "9",
    year = "2026"
}

@article{Candelas:1990rm,
    author = "Candelas, Philip and De La Ossa, Xenia C. and Green, Paul S. and Parkes, Linda",
    editor = "Yau, Shing-Tung",
    title = "{A Pair of Calabi-Yau manifolds as an exactly soluble superconformal theory}",
    reportNumber = "UTTG-25-90",
    doi = "10.1016/0550-3213(91)90292-6",
    journal = "Nucl. Phys. B",
    volume = "359",
    pages = "21--74",
    year = "1991"
}

@article{Chiang:1999tz,
    author = "Chiang, T. M. and Klemm, A. and Yau, Shing-Tung and Zaslow, E.",
    title = "{Local mirror symmetry: Calculations and interpretations}",
    eprint = "hep-th/9903053",
    archivePrefix = "arXiv",
    reportNumber = "IASSNS-HEP-99-26",
    doi = "10.4310/ATMP.1999.v3.n3.a3",
    journal = "Adv. Theor. Math. Phys.",
    volume = "3",
    pages = "495--565",
    year = "1999"
}

@article{GiventalYan2021,
  author        = {Givental, Alexander and Yan, Xiaohan},
  title         = {Quantum {K}-Theory of {Grassmannians} and
                   {Non-Abelian} Localization},
  journal       = {Symmetry, Integrability and Geometry:
                   Methods and Applications},
  year          = {2021},
  volume        = {17},
  number        = {018},
  pages         = {24},
  doi           = {10.3842/SIGMA.2021.018},
  archivePrefix = {arXiv},
  eprint        = {2008.08182},
  primaryClass  = {math.AG},
  url           = {https://arxiv.org/abs/2008.08182}
}

@article{Aluffi_2023,
   title={Shadows of characteristic cycles, Verma modules, and positivity of Chern–Schwartz–MacPherson classes of Schubert cells},
   volume={172},
   ISSN={1547-7398},
   url={http://dx.doi.org/10.1215/00127094-2022-0101},
   DOI={10.1215/00127094-2022-0101},
   number={17},
   journal={Duke Mathematical Journal},
   publisher={Duke University Press},
   author={Aluffi, Paolo and Mihalcea, Leonardo C. and Schürmann, Jörg and Su, Changjian},
   year={2023},
   month=Nov, pages={3257–3320} }

@article{Bullimore:2021rnr,
    author = "Bullimore, Mathew and Zhang, Daniel",
    title = "{3d $\mathcal{N}=4$ Gauge Theories on an Elliptic Curve}",
    eprint = "2109.10907",
    archivePrefix = "arXiv",
    primaryClass = "hep-th",
    doi = "10.21468/SciPostPhys.13.1.005",
    journal = "SciPost Phys.",
    volume = "13",
    number = "1",
    pages = "005",
    year = "2022"
}

@article{aganagic2021ellipticstableenvelopes,
      title={Elliptic stable envelopes},
      author={Mina Aganagic and Andrei Okounkov},
      journal={Journal of the American Mathematical Society},
      volume={34},
      number={1},
      pages={79--133},
      year={2021},
      doi={10.1090/jams/954},
      eprint={1604.00423},
      archivePrefix={arXiv},
      primaryClass={math.AG},
      url={https://arxiv.org/abs/1604.00423},
}

@article{aluffi2024motivicchernclassesschubert,
      title={Motivic Chern classes of Schubert cells, Hecke algebras, and applications to Casselman's problem},
      author={Paolo Aluffi and Leonardo C. Mihalcea and J{\"o}rg Sch{\"u}rmann and Changjian Su},
      journal={Annales scientifiques de l'\'Ecole normale sup\'erieure},
      volume={57},
      number={1},
      pages={87--141},
      year={2024},
      doi={10.24033/asens.2571},
      eprint={1902.10101},
      archivePrefix={arXiv},
      primaryClass={math.AG},
      url={https://arxiv.org/abs/1902.10101},
}

@article{Bourget:2026hsg,
    author = "Bourget, Antoine and Cheng, Peng and Lamouret, Quentin",
    title = "{BPS Invariants for Generalized Toric Calabi-Yau Threefolds}",
    eprint = "2607.29451",
    archivePrefix = "arXiv",
    primaryClass = "hep-th",
    month = "7",
    year = "2026"
}

@article{Nekrasov:2009uh,
    author = "Nekrasov, Nikita A. and Shatashvili, Samson L.",
    editor = "Baulieu, L. and de Boer, J. and Douglas, M. R. and Rabinovici, E. and Vanhove, P. and Windey, P.",
    title = "{Supersymmetric vacua and Bethe ansatz}",
    eprint = "0901.4744",
    archivePrefix = "arXiv",
    primaryClass = "hep-th",
    reportNumber = "IHES-P-09-09, TCD-MATH-09-04, HMI-09-01, NSF-KITP-09-11",
    doi = "10.1016/j.nuclphysbps.2009.07.047",
    journal = "Nucl. Phys. B Proc. Suppl.",
    volume = "192-193",
    pages = "91--112",
    year = "2009"
}

@article{Nekrasov:2014xaa,
    author = "Nekrasov, Nikita A. and Shatashvili, Samson L.",
    title = "{Bethe/Gauge correspondence on curved spaces}",
    eprint = "1405.6046",
    archivePrefix = "arXiv",
    primaryClass = "hep-th",
    doi = "10.1007/JHEP01(2015)100",
    journal = "JHEP",
    volume = "01",
    pages = "100",
    year = "2015"
}

@article{Pushkar:2016qvw,
    author = "Pushkar, Petr P. and Smirnov, Andrey V. and Zeitlin, Anton M.",
    title = "{Baxter Q-operator from quantum K-theory}",
    eprint = "1612.08723",
    archivePrefix = "arXiv",
    primaryClass = "math.AG",
    doi = "10.1016/j.aim.2019.106919",
    journal = "Adv. Math.",
    volume = "360",
    pages = "106919",
    year = "2020"
}

@article{Dedushenko:2021mds,
    author = "Dedushenko, Mykola and Nekrasov, Nikita",
    title = "{Interfaces and quantum algebras, I: Stable envelopes}",
    eprint = "2109.10941",
    archivePrefix = "arXiv",
    primaryClass = "hep-th",
    doi = "10.1016/j.geomphys.2023.104991",
    journal = "J. Geom. Phys.",
    volume = "194",
    pages = "104991",
    year = "2023"
}

@article{Dedushenko:2023qjq,
    author = "Dedushenko, Mykola and Nekrasov, Nikita",
    title = "{Interfaces and Quantum Algebras, II: Cigar Partition Function}",
    eprint = "2306.16434",
    archivePrefix = "arXiv",
    primaryClass = "hep-th",
    month = "6",
    year = "2023"
}

@article{Closset:2023bdr,
    author = "Closset, Cyril and Khlaif, Osama",
    title = "{Grothendieck lines in 3d $ \mathcal{N} $ = 2 SQCD and the quantum K-theory of the Grassmannian}",
    eprint = "2309.06980",
    archivePrefix = "arXiv",
    primaryClass = "hep-th",
    doi = "10.1007/JHEP12(2023)082",
    journal = "JHEP",
    volume = "12",
    pages = "082",
    year = "2023"
}

@article{SU2016362,
  author        = {Su, Changjian},
  title         = "{Equivariant Quantum Cohomology of Cotangent Bundle of $G/P$}",
  journal       = {Advances in Mathematics},
  volume        = {289},
  pages         = {362--383},
  year          = {2016},
  issn          = {0001-8708},
  doi           = {10.1016/j.aim.2015.11.026},
  eprint        = {1501.07513},
  archivePrefix = {arXiv},
  primaryClass  = {math.AG},
  url           = {https://arxiv.org/abs/1501.07513}
}

@article{rimanyi2015partial,
  author        = {Rim{\'a}nyi, Rich{\'a}rd and Tarasov, Vitaly and Varchenko, Alexander},
  title         = "{Partial flag varieties, stable envelopes, and weight functions}",
  journal       = {Quantum Topology},
  volume        = {6},
  number        = {2},
  pages         = {333--364},
  year          = {2015},
  doi           = {10.4171/QT/65},
  eprint        = {1212.6240},
  archivePrefix = {arXiv},
  primaryClass  = {math.AG},
  url           = {https://arxiv.org/abs/1212.6240}
}

@article{tarasov2013combinatorial,
  author        = {Tarasov, Vitaly and Varchenko, Alexander},
  title         = "{Combinatorial Formulae for Nested Bethe Vectors}",
  journal       = {SIGMA. Symmetry, Integrability and Geometry: Methods and Applications},
  volume        = {9},
  pages         = {048},
  year          = {2013},
  doi           = {10.3842/SIGMA.2013.048},
  eprint        = {math.QA/0702277},
  archivePrefix = {arXiv},
  primaryClass  = {math.QA},
  url           = {https://arxiv.org/abs/math/0702277}
}

@article{Tarasov:1994bb,
  author        = {Tarasov, V. and Varchenko, A.},
  title         = "{Asymptotic Solutions to the Quantized Knizhnik--Zamolodchikov Equation and Bethe Vectors}",
  journal       = {St. Petersburg Mathematical Journal},
  volume        = {6},
  number        = {2},
  pages         = {275--301},
  year          = {1995},
  eprint        = {hep-th/9406060},
  archivePrefix = {arXiv},
  primaryClass  = {hep-th},
  url           = {https://arxiv.org/abs/hep-th/9406060}
}

@article{rimanyi2015trigonometric,
  author        = {Rim{\'a}nyi, Rich{\'a}rd and Tarasov, Vitaly and Varchenko, Alexander},
  title         = "{Trigonometric Weight Functions as K-Theoretic Stable Envelope Maps for the Cotangent Bundle of a Flag Variety}",
  journal       = {Journal of Geometry and Physics},
  volume        = {94},
  pages         = {81--119},
  year          = {2015},
  doi           = {10.1016/j.geomphys.2015.04.002},
  eprint        = {1411.0478},
  archivePrefix = {arXiv},
  primaryClass  = {math.AG},
  url           = {https://arxiv.org/abs/1411.0478}
}

@article{Gu:2025tda,
    author = "Gu, Wei and Mihalcea, Leonardo and Sharpe, Eric and Xu, Weihong and Zhang, Hao and Zou, Hao",
    title = "{Schubert defects in Lagrangian Grassmannians}",
    eprint = "2502.04438",
    archivePrefix = "arXiv",
    primaryClass = "hep-th",
    doi = "10.1007/JHEP06(2025)148",
    journal = "JHEP",
    volume = "06",
    pages = "148",
    year = "2025"
}

@article{Closset:2025akk,
    author = "Closset, Cyril and Gu, Wei and Khlaif, Osama and Sharpe, Eric and Zhang, Hao and Zou, Hao",
    title = "{Schubert line defects in 3d GLSMs. Part I. Complete flag manifolds and quantum Grothendieck polynomials}",
    eprint = "2512.19802",
    archivePrefix = "arXiv",
    primaryClass = "hep-th",
    doi = "10.1007/JHEP04(2026)074",
    journal = "JHEP",
    volume = "04",
    pages = "074",
    year = "2026"
}

@article{Okounkov:2015spn,
    author = "Okounkov, Andrei",
    title = "{Lectures on K-theoretic computations in enumerative geometry}",
    eprint = "1512.07363",
    archivePrefix = "arXiv",
    primaryClass = "math.AG",
    month = "12",
    year = "2015"
}

@article{Closset:2026bnk,
    author = "Closset, Cyril and Gu, Wei and Khlaif, Osama and Sharpe, Eric and Zhang, Hao and Zou, Hao",
    title = "{Schubert line defects in 3d GLSMs. Part II. Partial flag manifolds and parabolic quantum polynomials}",
    eprint = "2601.18881",
    archivePrefix = "arXiv",
    primaryClass = "hep-th",
    doi = "10.1007/JHEP04(2026)075",
    journal = "JHEP",
    volume = "04",
    pages = "075",
    year = "2026"
}

@article{Bullimore:2018jlp,
    author = "Bullimore, Mathew and Ferrari, Andrea and Kim, Heeyeon",
    title = "{Twisted indices of 3d $ \mathcal{N} $ = 4 gauge theories and enumerative geometry of quasi-maps}",
    eprint = "1812.05567",
    archivePrefix = "arXiv",
    primaryClass = "hep-th",
    doi = "10.1007/JHEP07(2019)014",
    journal = "JHEP",
    volume = "07",
    pages = "014",
    year = "2019"
}

@article{Closset:2023vos,
    author = "Closset, Cyril and Khlaif, Osama",
    title = "{Twisted indices, Bethe ideals and 3d $ \mathcal{N} $ = 2 infrared dualities}",
    eprint = "2301.10753",
    archivePrefix = "arXiv",
    primaryClass = "hep-th",
    doi = "10.1007/JHEP05(2023)148",
    journal = "JHEP",
    volume = "05",
    pages = "148",
    year = "2023"
}

@article{Closset:2023jiq,
    author = "Closset, Cyril and Khlaif, Osama",
    title = "{On the Witten index of 3d $\mathcal{N}=2$ unitary SQCD with general CS levels}",
    eprint = "2305.00534",
    archivePrefix = "arXiv",
    primaryClass = "hep-th",
    doi = "10.21468/SciPostPhys.15.3.085",
    journal = "SciPost Phys.",
    volume = "15",
    number = "3",
    pages = "085",
    year = "2023"
}

@article{Closset:2023izb,
    author = "Closset, Cyril and Khlaif, Osama",
    title = "{New results on 3d {\ensuremath{\mathscr{N}}}=2 SQCD and its 3d GLSM interpretation}",
    eprint = "2312.05076",
    archivePrefix = "arXiv",
    primaryClass = "hep-th",
    doi = "10.1142/S0217751X24460114",
    journal = "Int. J. Mod. Phys. A",
    volume = "39",
    number = "33",
    pages = "2446011",
    year = "2024"
}

@article{Huq-Kuruvilla:2025nlf,
    author = "Huq-Kuruvilla, I. and Mihalcea, L. and Sharpe, E. and Zhang, H.",
    title = "{Quantum K-theory levels in physics and math}",
    eprint = "2507.00116",
    archivePrefix = "arXiv",
    primaryClass = "hep-th",
    month = "6",
    year = "2025"
}

@article{Jockers:2019lwe,
    author = "Jockers, Hans and Mayr, Peter and Ninad, Urmi and Tabler, Alexander",
    title = "{Wilson loop algebras and quantum K-theory for Grassmannians}",
    eprint = "1911.13286",
    archivePrefix = "arXiv",
    primaryClass = "hep-th",
    reportNumber = "BONN-TH-2019-08, LMU-ASC 53/19",
    doi = "10.1007/JHEP10(2020)036",
    journal = "JHEP",
    volume = "10",
    pages = "036",
    year = "2020"
}

@article{Jockers:2021omw,
    author = "Jockers, Hans and Mayr, Peter and Ninad, Urmi and Tabler, Alexander",
    title = "{BPS indices, modularity and perturbations in quantum K-theory}",
    eprint = "2106.07670",
    archivePrefix = "arXiv",
    primaryClass = "hep-th",
    reportNumber = "BONN-TH-2021-03, LMU-ASC 14/21, MITP/21-027",
    doi = "10.1007/JHEP02(2022)044",
    journal = "JHEP",
    volume = "02",
    pages = "044",
    year = "2022"
}

@article{Closset:2024sle,
    author = "Closset, Cyril and Furrer, Elias and Khlaif, Osama",
    title = "{One-form symmetries and the 3d $\mathcal{N}=2$ $A$-model: Topologically twisted indices and CS theories}",
    eprint = "2405.18141",
    archivePrefix = "arXiv",
    primaryClass = "hep-th",
    doi = "10.21468/SciPostPhys.18.2.066",
    journal = "SciPost Phys.",
    volume = "18",
    number = "2",
    pages = "066",
    year = "2025"
}

@article{Closset:2025lqt,
    author = "Closset, Cyril and Furrer, Elias and Keyes, Adam and Khlaif, Osama",
    title = "{The 3d $A$-model and generalised symmetries, part I: Bosonic Chern-Simons theories}",
    eprint = "2501.11665",
    archivePrefix = "arXiv",
    primaryClass = "hep-th",
    doi = "10.21468/SciPostPhys.19.4.106",
    journal = "SciPost Phys.",
    volume = "19",
    number = "4",
    pages = "106",
    year = "2025"
}

@article{Benini:2015noa,
    author = "Benini, Francesco and Zaffaroni, Alberto",
    title = "{A topologically twisted index for three-dimensional supersymmetric theories}",
    eprint = "1504.03698",
    archivePrefix = "arXiv",
    primaryClass = "hep-th",
    reportNumber = "IMPERIAL-TP-2015-FB-01",
    doi = "10.1007/JHEP07(2015)127",
    journal = "JHEP",
    volume = "07",
    pages = "127",
    year = "2015"
}

@article{Benini:2016hjo,
    author = "Benini, Francesco and Zaffaroni, Alberto",
    editor = "Li, Si and Lian, Bong H. and Song, Wei and Yau, Shing-Tung",
    title = "{Supersymmetric partition functions on Riemann surfaces}",
    eprint = "1605.06120",
    archivePrefix = "arXiv",
    primaryClass = "hep-th",
    reportNumber = "SISSA-28-2016-FISI",
    journal = "Proc. Symp. Pure Math.",
    volume = "96",
    pages = "13--46",
    year = "2017"
}

@article{Closset:2016arn,
    author = "Closset, Cyril and Kim, Heeyeon",
    title = "{Comments on twisted indices in 3d supersymmetric gauge theories}",
    eprint = "1605.06531",
    archivePrefix = "arXiv",
    primaryClass = "hep-th",
    doi = "10.1007/JHEP08(2016)059",
    journal = "JHEP",
    volume = "08",
    pages = "059",
    year = "2016"
}

@article{Closset:2017zgf,
    author = "Closset, Cyril and Kim, Heeyeon and Willett, Brian",
    title = "{Supersymmetric partition functions and the three-dimensional A-twist}",
    eprint = "1701.03171",
    archivePrefix = "arXiv",
    primaryClass = "hep-th",
    reportNumber = "CERN-TH-2017-006",
    doi = "10.1007/JHEP03(2017)074",
    journal = "JHEP",
    volume = "03",
    pages = "074",
    year = "2017"
}

@article{Rozansky:1996bq,
  author        = {Rozansky, Lev and Witten, Edward},
  title         = {{Hyper-K{\"a}hler Geometry and Invariants of Three-Manifolds}},
  journal       = {Selecta Mathematica, New Series},
  volume        = {3},
  number        = {3},
  pages         = {401--458},
  year          = {1997},
  doi           = {10.1007/s000290050016},
  eprint        = {hep-th/9612216},
  archivePrefix = {arXiv},
  primaryClass  = {hep-th},
  url           = {https://arxiv.org/abs/hep-th/9612216}
}

@article{MarianOprea2007,
  author        = {Marian, Alina and Oprea, Dragos},
  title         = {Virtual intersections on the {Quot} scheme
                   and {Vafa--Intriligator} formulas},
  journal       = {Duke Mathematical Journal},
  volume        = {136},
  number        = {1},
  pages         = {81--113},
  year          = {2007},
  doi           = {10.1215/S0012-7094-07-13613-5},
  eprint        = {math/0505685},
  archivePrefix = {arXiv},
  primaryClass  = {math.AG},
  url           = {https://arxiv.org/abs/math/0505685}
}

@article{GukovPutrovVafa2017,
  author        = {Gukov, Sergei and Putrov, Pavel and Vafa, Cumrun},
  title         = {{Fivebranes and 3-manifold homology}},
  journal       = {JHEP},
  volume        = {07},
  pages         = {071},
  year          = {2017},
  doi           = {10.1007/JHEP07(2017)071},
  eprint        = {1602.05302},
  archivePrefix = {arXiv},
  primaryClass  = {hep-th},
  url           = {https://arxiv.org/abs/1602.05302}
}

@incollection{Kim2016StableQuasimaps,
  author        = {Kim, Bumsig},
  title         = {Stable quasimaps to holomorphic symplectic quotients},
  booktitle     = {{Schubert Calculus---Osaka 2012}},
  series        = {Advanced Studies in Pure Mathematics},
  volume        = {71},
  pages         = {139--160},
  publisher     = {Mathematical Society of Japan},
  address       = {Tokyo},
  year          = {2016},
  doi           = {10.2969/aspm/07110139},
  eprint        = {1005.4125},
  archivePrefix = {arXiv},
  primaryClass  = {math.AG},
  url           = {https://arxiv.org/abs/1005.4125}
}

@article{Ueda:2019qhg,
    author = "Ueda, Kazushi and Yoshida, Yutaka",
    title = "{3d $ \mathcal{N} $ = 2 Chern-Simons-matter theory, Bethe ansatz, and quantum $K$-theory of Grassmannians}",
    eprint = "1912.03792",
    archivePrefix = "arXiv",
    primaryClass = "hep-th",
    reportNumber = "IPMU19-0176",
    doi = "10.1007/JHEP08(2020)157",
    journal = "JHEP",
    volume = "08",
    pages = "157",
    year = "2020"
}

@article{Gu:2022yvj,
    author = "Gu, Wei and Mihalcea, Leonardo C. and Sharpe, Eric and Zou, Hao",
    title = "{Quantum K theory of Grassmannians, Wilson line operators, and Schur bundles}",
    eprint = "2208.01091",
    archivePrefix = "arXiv",
    primaryClass = "math.AG",
    doi = "10.1017/fms.2025.10088",
    journal = "Forum Math. Sigma",
    volume = "13",
    pages = "e140",
    year = "2025"
}

@article{Witten:1993yc,
    author = "Witten, Edward",
    editor = "Greene, B. and Yau, Shing-Tung",
    title = "{Phases of N=2 theories in two-dimensions}",
    eprint = "hep-th/9301042",
    archivePrefix = "arXiv",
    reportNumber = "IASSNS-HEP-93-3",
    doi = "10.1016/0550-3213(93)90033-L",
    journal = "Nucl. Phys. B",
    volume = "403",
    pages = "159--222",
    year = "1993"
}

@article{Morrison:1994fr,
    author = "Morrison, David R. and Plesser, M. Ronen",
    title = "{Summing the instantons: Quantum cohomology and mirror symmetry in toric varieties}",
    eprint = "hep-th/9412236",
    archivePrefix = "arXiv",
    reportNumber = "DUKE-TH-94-78, IASSNS-HEP-94-82",
    doi = "10.1016/0550-3213(95)00061-V",
    journal = "Nucl. Phys. B",
    volume = "440",
    pages = "279--354",
    year = "1995"
}

@article{Closset:2015rna,
    author = "Closset, Cyril and Cremonesi, Stefano and Park, Daniel S.",
    title = "{The equivariant A-twist and gauged linear sigma models on the two-sphere}",
    eprint = "1504.06308",
    archivePrefix = "arXiv",
    primaryClass = "hep-th",
    doi = "10.1007/JHEP06(2015)076",
    journal = "JHEP",
    volume = "06",
    pages = "076",
    year = "2015"
}

@article{Witten:1993xi,
    author = "Witten, Edward",
    title = "{The Verlinde algebra and the cohomology of the Grassmannian}",
    eprint = "hep-th/9312104",
    archivePrefix = "arXiv",
    reportNumber = "IASSNS-HEP-93-41",
    month = "12",
    year = "1993"
}

@article{Witten:1991zz,
    author = "Witten, Edward",
    editor = "Yau, Shing-Tung",
    title = "{Mirror manifolds and topological field theory}",
    eprint = "hep-th/9112056",
    archivePrefix = "arXiv",
    reportNumber = "IASSNS-HEP-91-83",
    journal = "AMS/IP Stud. Adv. Math.",
    volume = "9",
    pages = "121--160",
    year = "1998"
}

@article{Witten:1988xj,
    author = "Witten, Edward",
    title = "{Topological Sigma Models}",
    reportNumber = "IASSNS-HEP-88/7",
    doi = "10.1007/BF01466725",
    journal = "Commun. Math. Phys.",
    volume = "118",
    pages = "411",
    year = "1988"
}

@article{Vafa:1990mu,
  author       = {Vafa, Cumrun},
  title        = {Topological {Landau-Ginzburg} Models},
  journal      = {Mod. Phys. Lett. A},
  volume       = {6},
  number       = {4},
  pages        = {337--346},
  year         = {1991},
  doi          = {10.1142/S0217732391000324},
  reportNumber = {HUTP-90-A064},
  url          = {https://doi.org/10.1142/S0217732391000324}
}

@incollection{KimOhUedaYoshida2020,
  author        = {Kim, Bumsig and Oh, Jeongseok and
                   Ueda, Kazushi and Yoshida, Yutaka},
  title         = {Residue Mirror Symmetry for {Grassmannians}},
  booktitle     = {Schubert Calculus and Its Applications in
                   Combinatorics and Representation Theory},
  editor        = {Hu, Jianxun and Li, Changzheng and
                   Mihalcea, Leonardo C.},
  series        = {Springer Proceedings in Mathematics \& Statistics},
  volume        = {332},
  pages         = {307--365},
  publisher     = {Springer},
  address       = {Singapore},
  year          = {2020},
  doi           = {10.1007/978-981-15-7451-1_12},
  eprint        = {1607.08317},
  archivePrefix = {arXiv},
  primaryClass  = {math.AG},
  url           = {https://doi.org/10.1007/978-981-15-7451-1_12}
}

@article{Intriligator:1991nk,
  author        = {Intriligator, Kenneth},
  title         = {Fusion Residues},
  journal       = {Mod. Phys. Lett. A},
  volume        = {6},
  number        = {38},
  pages         = {3543--3556},
  year          = {1991},
  doi           = {10.1142/S0217732391004097},
  eprint        = {hep-th/9108005},
  archivePrefix = {arXiv},
  primaryClass  = {hep-th},
  reportNumber  = {HUTP-91-A041},
  url           = {https://doi.org/10.1142/S0217732391004097}
}

@article{Closset:2025zyl,
    author = "Closset, Cyril and Wynne, James",
    title = "{Dualities and trialities in $ \mathcal{N} $ = 2 supersymmetric gauged quantum mechanics}",
    eprint = "2512.02984",
    archivePrefix = "arXiv",
    primaryClass = "hep-th",
    doi = "10.1007/JHEP06(2026)072",
    journal = "JHEP",
    volume = "06",
    pages = "072",
    year = "2026"
}

@book{Anderson_Fulton_2023, 
title={Equivariant Cohomology in Algebraic Geometry}, 
author={Anderson, David and Fulton, William}, 
publisher={Cambridge University Press}, 
year={2023}, 
place={Cambridge}, 
series={Cambridge Studies in Advanced Mathematics}, 
collection={Cambridge Studies in Advanced Mathematics},
isbn={9781009349963},
url={https://books.google.com/books?id=yjbcEAAAQBAJ},
}

@article{Hori:2014tda,
    author = "Hori, Kentaro and Kim, Heeyeon and Yi, Piljin",
    title = "{Witten Index and Wall Crossing}",
    eprint = "1407.2567",
    archivePrefix = "arXiv",
    primaryClass = "hep-th",
    reportNumber = "KIAS-P14039",
    doi = "10.1007/JHEP01(2015)124",
    journal = "JHEP",
    volume = "01",
    pages = "124",
    year = "2015"
}

@article{RimanyiVarchenko2015,
  author        = {Rim{\'a}nyi, Rich{\'a}rd and Varchenko, Alexander},
  title         = {Equivariant {Chern-Schwartz-MacPherson} Classes in Partial Flag Varieties: Interpolation and Formulae},
  eprint        = {1509.09315},
  archivePrefix = {arXiv},
  primaryClass  = {math.AG},
  month         = sep,
  year          = {2015}
}

@article{Liu2019,
  author        = {Liu, Henry},
  title         = {Self-Duality in Quantum {K}-Theory},
  year          = {2019},
  month         = jun,
  archivePrefix = {arXiv},
  eprint        = {1906.10824},
  primaryClass  = {math.AG},
  doi           = {10.48550/arXiv.1906.10824},
}

@article{CiocanFontanineKim2017,
  author  = {Ciocan-Fontanine, Ionu{\c{t}} and Kim, Bumsig},
  title   = {Higher Genus Quasimap Wall-Crossing
             for Semipositive Targets},
  journal = {Journal of the European Mathematical Society},
  year    = {2017},
  volume  = {19},
  number  = {7},
  pages   = {2051--2102},
  doi     = {10.4171/JEMS/713},
  url     = {https://doi.org/10.4171/JEMS/713},
  eprint  = {1308.6377},
  archivePrefix = {arXiv},
  primaryClass  = {math.AG}
}

@article{CiocanFontanineKimMaulik2014,
  author  = {Ciocan-Fontanine, Ionu{\c{t}} and
             Kim, Bumsig and Maulik, Davesh},
  title   = {Stable Quasimaps to {GIT} Quotients},
  journal = {Journal of Geometry and Physics},
  year    = {2014},
  volume  = {75},
  pages   = {17--47},
  doi     = {10.1016/j.geomphys.2013.08.019},
  url     = {https://doi.org/10.1016/j.geomphys.2013.08.019},
  eprint  = {1106.3724},
  archivePrefix = {arXiv},
  primaryClass  = {math.AG}
}

@article{Givental2016,
  author  = {Givental, Alexander},
  title   = {Explicit Reconstruction in Quantum Cohomology
             and {K}-Theory},
  journal = {Annales de la Facult{\'e} des Sciences de Toulouse:
             Math{\'e}matiques},
  year    = {2016},
  volume  = {25},
  number  = {2--3},
  pages   = {419--432},
  doi     = {10.5802/afst.1500},
  url     = {https://doi.org/10.5802/afst.1500},
  eprint  = {1506.06431},
  archivePrefix = {arXiv},
  primaryClass  = {math.AG}
}

@article{Okounkov2020,
  author        = {Okounkov, Andrei},
  title         = {Nonabelian Stable Envelopes, Vertex Functions
                   with Descendents, and Integral Solutions of
                   {$q$}-Difference Equations},
  year          = {2020},
  month         = oct,
  archivePrefix = {arXiv},
  eprint        = {2010.13217},
  primaryClass  = {math.AG},
  doi           = {10.48550/arXiv.2010.13217},
}

@article{Givental2017,
  author  = {Givental, Alexander},
  title   = {Permutation-Equivariant Quantum {K}-Theory I.
             Definitions. Elementary {K}-Theory of
             {$\overline{\mathcal{M}}_{0,n}/S_n$}},
  journal = {Moscow Mathematical Journal},
  year    = {2017},
  volume  = {17},
  number  = {4},
  pages   = {691--698},
  doi     = {10.17323/1609-4514-2017-17-4-691-698},
  url     = {https://doi.org/10.17323/1609-4514-2017-17-4-691-698},
  eprint  = {1508.02690},
  archivePrefix = {arXiv},
  primaryClass  = {math.AG}
}

@article{Lee2004,
  author  = {Lee, Y.-P.},
  title   = {Quantum {K}-Theory, I: Foundations},
  journal = {Duke Mathematical Journal},
  year    = {2004},
  volume  = {121},
  number  = {3},
  pages   = {389--424},
  doi     = {10.1215/S0012-7094-04-12131-1},
  url     = {https://doi.org/10.1215/S0012-7094-04-12131-1},
  eprint  = {math/0105014},
  archivePrefix = {arXiv},
  primaryClass  = {math.AG}
}

@article{RimanyiTarasovVarchenko2014,
  author  = {Rim{\'a}nyi, Rich{\'a}rd and
             Tarasov, Vitaly and
             Varchenko, Alexander},
  title   = {Cohomology Classes of Conormal Bundles
             of Schubert Varieties and Yangian
             Weight Functions},
  journal = {Mathematische Zeitschrift},
  year    = {2014},
  volume  = {277},
  number  = {3--4},
  pages   = {1085--1104},
  doi     = {10.1007/s00209-014-1295-5},
  url     = {https://doi.org/10.1007/s00209-014-1295-5},
  eprint  = {1204.4961},
  archivePrefix = {arXiv},
  primaryClass  = {math.AG}
}

@misc{Nekrasov2013,
  author       = {Nekrasov, Nikita},
  title        = {Bethe States as Defects in Gauge Theories},
  howpublished = {Simons Center for Geometry and Physics,
                  SCGP Weekly Talk},
  year         = {2013},
  month        = oct,
  url          = {http://scgp.stonybrook.edu/video_portal/video.php?id=1775},
  note         = {Research seminar, October 2013}
}

@article{Bullimore:2017lwu,
    author = "Bullimore, Mathew and Kim, Hee-Cheol and Lukowski, Tomasz",
    title = "{Expanding the Bethe/Gauge Dictionary}",
    eprint = "1708.00445",
    archivePrefix = "arXiv",
    primaryClass = "hep-th",
    doi = "10.1007/JHEP11(2017)055",
    journal = "JHEP",
    volume = "11",
    pages = "055",
    year = "2017"
}

@book{Hori:MirrorSymmetry,
  author    = {Hori, Kentaro and Katz, Sheldon and Klemm, Albrecht and Pandharipande, Rahul and Thomas, Richard and Vafa, Cumrun and  Vakil, Ravi and Zaslow, Eric},
  title     = {Mirror Symmetry},
  series    = {Clay Mathematics Monographs},
  volume    = {1},
  publisher = {American Mathematical Society},
  address   = {Providence, RI},
  year      = {2003},
  isbn      = {0-8218-2955-6}
}

@article{Cao:2026qbt,
    author = "Cao, Z. and Sharpe, E. and Zhang, H.",
    title = "{Towards quantum elliptic cohomology from GLSMs}",
    eprint = "2609.09277",
    archivePrefix = "arXiv",
    primaryClass = "hep-th",
    month = "9",
    year = "2026"
}

@article{Ishtiaque:2023acr,
    author = "Ishtiaque, Nafiz and Moosavian, Seyed Faroogh and Zhou, Yehao",
    title = "{Elliptic Stable Envelopes for Certain Non-Symplectic Varieties and Dynamical $R$-Matrices for Superspin Chains from the Bethe/Gauge Correspondence}",
    eprint = "2308.12333",
    archivePrefix = "arXiv",
    primaryClass = "hep-th",
    doi = "10.3842/SIGMA.2024.099",
    journal = "SIGMA",
    volume = "20",
    pages = "099",
    year = "2024"
}

@article{Galakhov:2022uyu,
    author = "Galakhov, Dmitry and Li, Wei and Yamazaki, Masahito",
    title = "{Gauge/Bethe correspondence from quiver BPS algebras}",
    eprint = "2206.13340",
    archivePrefix = "arXiv",
    primaryClass = "hep-th",
    doi = "10.1007/JHEP11(2022)119",
    journal = "JHEP",
    volume = "11",
    pages = "119",
    year = "2022"
}

@article{Bullimore:2020nhv,
    author = "Bullimore, Mathew and Ferrari, Andrea E. V. and Kim, Heeyeon and Xu, Guangyu",
    title = "{The twisted index and topological saddles}",
    eprint = "2007.11603",
    archivePrefix = "arXiv",
    primaryClass = "hep-th",
    doi = "10.1007/JHEP05(2022)116",
    journal = "JHEP",
    volume = "05",
    pages = "116",
    year = "2022"
}

@article{Intriligator:2013lca,
    author = "Intriligator, Kenneth and Seiberg, Nathan",
    title = "{Aspects of 3d N=2 Chern-Simons-Matter Theories}",
    eprint = "1305.1633",
    archivePrefix = "arXiv",
    primaryClass = "hep-th",
    reportNumber = "UCSD-PTH-12-17",
    doi = "10.1007/JHEP07(2013)079",
    journal = "JHEP",
    volume = "07",
    pages = "079",
    year = "2013"
}

@phdthesis{Khlaif:2025ccg,
    author = "Khlaif, Osama K. F.",
    title = "{Novel Aspects of 3D $\mathcal{N}= 2$ Chern{\textendash}Simons{\textendash}Matter theories}",
    school = "Birmingham U.",
    month = "4",
    year = "2025"
}

@article{Bullimore:2014awa,
    author = "Bullimore, Mathew and Kim, Hee-Cheol and Koroteev, Peter",
    title = "{Defects and Quantum Seiberg-Witten Geometry}",
    eprint = "1412.6081",
    archivePrefix = "arXiv",
    primaryClass = "hep-th",
    doi = "10.1007/JHEP05(2015)095",
    journal = "JHEP",
    volume = "05",
    pages = "095",
    year = "2015"
}

@article{Koroteev:2017nab,
    author = "Koroteev, Peter and Pushkar, Petr P. and Smirnov, Andrey V. and Zeitlin, Anton M.",
    title = "{Quantum K-theory of quiver varieties and many-body systems}",
    eprint = "1705.10419",
    archivePrefix = "arXiv",
    primaryClass = "math.AG",
    doi = "10.1007/s00029-021-00698-3",
    journal = "Selecta Math. (N.S.)",
    volume = "27",
    number = "5",
    pages = "87",
    year = "2021"
}

@article{Jockers:2018sfl,
    author = "Jockers, Hans and Mayr, Peter",
    title = "{A 3d Gauge Theory/Quantum K-Theory Correspondence}",
    eprint = "1808.02040",
    archivePrefix = "arXiv",
    primaryClass = "hep-th",
    reportNumber = "BONN-TH-2018-12, LMU-ASC 47/18",
    doi = "10.4310/ATMP.2020.v24.n2.a4",
    journal = "Adv. Theor. Math. Phys.",
    volume = "24",
    number = "2",
    pages = "327--457",
    year = "2020"
}

@article{Jockers:2019wjh,
    author = "Jockers, Hans and Mayr, Peter",
    title = "{Quantum K-Theory of Calabi-Yau Manifolds}",
    eprint = "1905.03548",
    archivePrefix = "arXiv",
    primaryClass = "hep-th",
    reportNumber = "BONN-TH-2019-03, LMU-ASC 19/19",
    doi = "10.1007/JHEP11(2019)011",
    journal = "JHEP",
    volume = "11",
    pages = "011",
    year = "2019"
}

@article{Bullimore:2019qnt,
    author = "Bullimore, Mathew and Ferrari, Andrea E. V. and Kim, Heeyeon",
    title = "{The 3d twisted index and wall-crossing}",
    eprint = "1912.09591",
    archivePrefix = "arXiv",
    primaryClass = "hep-th",
    doi = "10.21468/SciPostPhys.12.6.186",
    journal = "SciPost Phys.",
    volume = "12",
    number = "6",
    pages = "186",
    year = "2022"
}

@article{Gu:2020zpg,
    author = "Gu, Wei and Mihalcea, Leonardo and Sharpe, Eric and Zou, Hao",
    title = "{Quantum K theory of symplectic Grassmannians}",
    eprint = "2008.04909",
    archivePrefix = "arXiv",
    primaryClass = "hep-th",
    doi = "10.1016/j.geomphys.2022.104548",
    journal = "J. Geom. Phys.",
    volume = "177",
    pages = "104548",
    year = "2022"
}

@article{Gu:2021yek,
    author = "Gu, Wei and Pei, Du and Zhang, Ming",
    title = "{On phases of 3d $\mathcal{N}=2$ Chern-Simons-matter theories}",
    eprint = "2105.02247",
    archivePrefix = "arXiv",
    primaryClass = "hep-th",
    doi = "10.1016/j.nuclphysb.2021.115604",
    journal = "Nucl. Phys. B",
    volume = "973",
    pages = "115604",
    year = "2021"
}

@incollection{Gu:2021beo,
    author = "Gu, Wei",
    title = "{Vacuum Structures Revisited}",
    booktitle = "{2021--2022 MATRIX Annals}",
    editor = "Wood, David R. and de Gier, Jan and Praeger, Cheryl E.",
    series = "MATRIX Book Series",
    volume = "5",
    publisher = "Springer",
    address = "Cham",
    pages = "835--854",
    year = "2024",
    doi = "10.1007/978-3-031-47417-0_35",
    eprint = "2110.13156",
    archivePrefix = "arXiv",
    primaryClass = "hep-th"
}

@article{Gu:2023tcv,
    author = "Gu, Wei and Mihalcea, Leonardo and Sharpe, Eric and Xu, Weihong and Zhang, Hao and Zou, Hao",
    title = "{Quantum K theory rings of partial flag manifolds}",
    eprint = "2306.11094",
    archivePrefix = "arXiv",
    primaryClass = "hep-th",
    doi = "10.1016/j.geomphys.2024.105127",
    journal = "J. Geom. Phys.",
    volume = "198",
    pages = "105127",
    year = "2024"
}

@article{Gu:2023fpw,
    author = "Gu, Wei and Mihalcea, Leonardo C. and Sharpe, Eric and Xu, Weihong and Zhang, Hao and Zou, Hao",
    title = "{Quantum K Whitney relations for partial flag varieties}",
    eprint = "2310.03826",
    archivePrefix = "arXiv",
    primaryClass = "math.AG",
    month = "10",
    year = "2023"
}

@article{Sharpe:2024ujm,
    author = "Sharpe, Eric and Zhang, Hao",
    title = "{Decomposition squared}",
    eprint = "2405.12269",
    archivePrefix = "arXiv",
    primaryClass = "hep-th",
    doi = "10.1007/JHEP10(2024)168",
    journal = "JHEP",
    volume = "10",
    pages = "168",
    year = "2024"
}

@article{Huq-Kuruvilla:2024tsg,
    author = "Huq-Kuruvilla, Irit",
    title = "{Quantum K-Rings of Partial Flag Varieties, Coulomb Branches, and the Bethe Ansatz}",
    eprint = "2409.15575",
    archivePrefix = "arXiv",
    primaryClass = "math.AG",
    month = "9",
    year = "2024"
}

@article{AminiHuqKuruvillaMihalceaOrrXu2025,
    author = "Amini, Kamyar and Huq-Kuruvilla, Irit and Mihalcea, Leonardo C. and Orr, Daniel and Xu, Weihong",
    title = "{Toda-Type Presentations for the Quantum K Theory of Partial Flag Varieties}",
    journal = "SIGMA",
    volume = "21",
    pages = "098",
    year = "2025",
    doi = "10.3842/SIGMA.2025.098",
    eprint = "2504.07412",
    archivePrefix = "arXiv"
}

@article{GuMihalceaSharpeXuZhangZou2025,
    author = "Gu, Wei and Mihalcea, Leonardo C. and Sharpe, Eric and Xu, Weihong and Zhang, Hao and Zou, Hao",
    title = "{A Nakayama result for the quantum K theory of homogeneous spaces}",
    journal = "{\'E}pijournal G{\'e}om. Alg{\'e}brique",
    volume = "9",
    year = "2025",
    doi = "10.46298/epiga.2025.17016",
    eprint = "2507.15183",
    archivePrefix = "arXiv",
    primaryClass = "math.AG"
}

@article{Duan:2026mfq,
    author = "Duan, Zhihao and Khlaif, Osama and Zou, Hao",
    title = "{On the Schubert calculus of the quantum K-theory for partial flag manifolds: a 3d A-model perspective}",
    eprint = "2606.15261",
    archivePrefix = "arXiv",
    primaryClass = "hep-th",
    doi = "10.1007/JHEP08(2026)103",
    journal = "JHEP",
    volume = "08",
    pages = "103",
    year = "2026"
}
\bibliographystyle{ytamsalpha} 
\end{document}